\documentclass[pra,aps,eqsecnum,showpacs]{revtex4}
\usepackage{amsmath}
\usepackage{amssymb}
\usepackage{latexsym}
\usepackage{color}
\usepackage{float}
\usepackage{graphicx}

\newtheorem{theorem}{Theorem}

\newtheorem{corollary}{Corollary}

\begin{document}
\title{Operator-sum Representation for Bosonic Gaussian Channels}
\author{J. Solomon Ivan}
\email{solomon@rri.res.in}
\affiliation{Raman Research Institute, C. V. Raman Avenue, Sadashivanagar, 
Bangalore 560 080, India.}
\author{Krishnakumar Sabapathy}
\email{kkumar@imsc.res.in}
\affiliation{Centre for Quantum Science, The Institute of Mathematical
  Sciences, C.I.T Campus, Tharamani, Chennai 600 113, India.}
\author{R. Simon}
\email{simon@imsc.res.in}
\affiliation{Centre for Quantum Science, The Institute of Mathematical
  Sciences, C.I.T Campus, Tharamani, Chennai 600 113, India.}
\begin{abstract} 
Operator-sum or Kraus representations for single-mode Bosonic Gaussian
channels are developed, and several of their consequences 
 explored. The fact that the two-mode metaplectic operators acting
as unitary purification of these channels do not, in their canonical form,
mix the position and momentum variables is exploited
to present a procedure which applies uniformly to all families
in the Holevo classification. In this procedure the Kraus operators of every
quantum-limited Gaussian channel can be simply read off from
the matrix elements of a corresponding metaplectic operator.
 Kraus operators are employed to bring out, in the Fock basis, the
 manner in which the antilinear, unphysical matrix transposition 
 map when accompanied by injection of a threshold classical noise
becomes a physical channel, denoted ${\cal D}(\kappa)$ in the Holevo
classification. The matrix transposition  channels
${\cal D}(\kappa)$,  ${\cal D}(\kappa ^{-1})$ turn out to be a dual pair
in the sense that their Kraus operators are related by the adjoint
operation. The amplifier channel with amplification factor $\kappa$ 
and the beamsplitter channel with attenuation factor $\kappa^{-1}$
turn out to be mutually dual in the same sense. The action of the
quantum-limited attenuator and amplifier channels as simply scaling
maps on suitable quasi-probabilities in phase space is examined in the
Kraus picture.
 Consideration of cumulants is used to examine the issue of fixed points.
 The semigroup property of the
 amplifier and attenuator families leads  in both cases to a Zeno-like
effect arising as a consequence of interrupted evolution. In the cases of
entanglement-breaking channels a description in terms of rank one 
Kraus operators is shown to emerge quite simply. In contradistinction,
 it is shown that there is not even one finite rank operator in  the entire
linear span of  Kraus operators of the quantum-limited amplifier or
attenuator families, an assertion far stronger than the statement that
these are not entanglement breaking channels. A characterization of extremality
in terms of  
Kraus operators, originally due to Choi, is employed to show that
 all quantum-limited Gaussian channels are extremal.
 The fact that every noisy Gaussian channel can be realised as product of a pair
of quantum-limited channels is used to construct a discrete set of
linearly independent Kraus operators for noisy Gaussian channels,
including the classical noise channel, and these Kraus operators have
a particularly simple structure.

\end{abstract}
\maketitle

\section{Introduction}
Gaussian states and Gaussian channels play a major role in quantum
information processes, and this is primarily due to their potential
experimental realization within current technologies
\cite{ caves94, adesso07,
  rev2, polzik}. Indeed, the basic protocols of quantum information
processing including teleportation and dense coding have been
implemented in the quantum optical domain \cite{tele, crypto}. The feasibility 
of processing information using Gaussian channels was originally
explored in \cite{hall94, caves94}. More recently, the problem of evaluating
the classical capacity of Gaussian channels was addressed in
\cite{preskill01, hirota99, giovannetti042}, and the quantum
capacities in \cite{holevo01, wolf05, caruso061, mmwolf07, giovannetti041,
  giovannetti043}. In  
particular, the classical capacity of the attenuator channel was
evaluated in \cite{giovannetti042}, and the quantum capacity of a class of
channels was studied in \cite{mmwolf07}. A systematic study of the structure
of the family of all Gaussian channels has been carried out in
\cite{caruso06, holevo07, mmwolf08, caruso08, caruso10}; 
single-mode Gaussian channels have been classified in
\cite{caruso06, holevo07}, and the case of multimodes in
\cite{mmwolf08, caruso08, caruso10}. 

Phase space description in terms of quasiprobabilities or,
equivalently, in terms of the associated characteristic functions
underlies the very notions of Gaussian states and Gaussian
channels. Gaussian states are those with Gaussian characteristic
functions, and  Gaussian channels are those trace-preserving CP maps
which image every input Gaussian state into a Gaussian state at the
output. 

Gaussian states are fully specified by their first and second
moments. Since the first moments play no significant role in our
study, we may assume that they vanish (this can indeed be ensured
using the unitary Weyl-Heisenberg displacement operators), so that a
Gaussian state for our purpose is fully described by its covariance matrix
\cite{simon87, simon88, dutta94, krp}. The symplectic group of real linear canonical
transformations (acting through its unitary metaplectic
representation) and the Weyl-Heisenberg group of phase space translations
are the only unitary evolutions which preserve Gaussianity, and these
groups are generated by hermitian Hamiltonians which are respectively
quadratic and linear in the creation and annihilation operators
\cite{simon87, simon88, dutta94}. This suggests that more general
Gaussian channels on system A 
may be realized as Gaussianity preserving unitaries on a suitably
enlarged  system\,: 
\begin{equation}
\rho_A \rightarrow \rho_A^{\,'} = \text{Tr}_B\left(U_{AB} \,(\rho_A
  \otimes \rho_B) \, U_{AB}^{\dagger} \right). 
\label{i1}
\end{equation}  
Here $\rho_{B}$ is a Gaussian state of the ancilla B, and $U_{AB}$ is a
linear canonical transformation on the enlarged composite system
consisting of the system of interest A and
the ancilla B. That all Gaussian channels can indeed be realized in
this manner has been shown by the work of Holevo and coauthors
\cite{caruso10, caruso06, caruso08,holevo07}. 

It is clear that the most general trace-preserving linear map $\Omega$
which takes Gaussian characteristic functions to Gaussian, taking
states with vanishing first moments to ones with vanishing first
moments, are necessarily of the form $\Omega\,: \chi(\xi) \rightarrow
\chi^{\,'}(\xi ) = \chi(X\xi) \exp[-\frac{1}{2} \xi^T Y \xi] $, where
$X, Y$ are real matrices with $Y=Y^T \geq0$. And $X,Y$ need to obey an
appropriate matrix inequality to ensure that the trace-preserving map
$\Omega$ is completely positive \cite{holevo01,
    lindblad00, demoen79, mmwolf07}. For a
given $X$, the minimal $Y$, say $Y_0$, meeting this inequality
represents the threshold Gaussian noise that needs to be added to $\chi(X\xi)$
to make atonement for the failure of $X$ to be a symplectic matrix,
and thus rendering the map completely positive; if $X$ happens to be
a symplectic  matrix, then the corresponding minimal $Y_0=0$. 

Now, given a Gaussian channel $\Omega$ we can construct, `quite
cheaply', an entire family of Gaussian channels by simply preceding
and following $\Omega$ with unitary (symplectic) Gaussian channels
$U(S_1), U(S_2)$ corresponding respectively to symplectic matrices
$S_1,\,S_2$. Therefore in classifying Gaussian 
channels it is sufficient to classify these orbits or double cosets
and, further, we may identify each orbit with the `simplest' looking
representative element of that orbit (the canonical form). Since 
\begin{align}
U(S_1) \,\Omega \, U(S_2) \, :\, \chi(\xi) \rightarrow \chi(S_2XS_1\,
\xi) \exp[-\frac{1}{2} \xi^T S_1^T Y S_1 \xi],
\end{align}
the task actually reduces to enumeration of the orbits of $(X,Y)$ under the
transformation $(X,Y) \rightarrow (X^{\,'}, Y^{\,'}) = (S_2XS_1,\,
S_1^TYS_1)$.

One final point before turning to the special case of single-mode
Gaussian channels. The injection of an arbitrary amount of classical
(Gaussian) noise into the state is obviously a Gaussian channel\,:
$\chi(\xi) \rightarrow \chi(\xi) \, \exp[-\frac{a}{2} \xi^T\xi ], \,
a>0$. It is called the classical noise channel. Now, given a Gaussian
channel we may follow it up with a classical noise channel to obtain
another Gaussian channel. A Gaussian channel will be said to be {\em
  quantum-limited} if it cannot be realized as another Gaussian
channel followed by a classical noise channel. Conversely, the most general
Gaussian channel is a quantum-limited Gaussian channel followed by a classical
noise channel, and it follows that quantum-limited channels are the
primary objects which need to be classified into orbits. 

In the single-mode case where $X,Y$ are $2 \times 2$ matrices, $S_1,
S_2 \in Sp(2,R)$ can be so chosen that $X^{\,'}$ equals a multiple of
identity, a multiple of $\sigma_3$, or $(1\!\!1 + \sigma_3)/2$ while $Y^{\,'}$
equals a multiple of identity or $(1\!\!1 + \sigma_3)/2$. Thus the canonical
form of a Gaussian channel $X,Y$ is fully determined by the rank and
determinant of $X,Y$ and we have the following classification of
{\em quantum-limited bosonic  Gaussian channels} \cite{caruso06,
  holevo07}
\begin{equation}
\begin{array}{rll}
{\cal D}(\kappa;0)\,: ~~~~&~~~~X = -\kappa \sigma_3, ~~~~&~~~~Y_0=
(1+\kappa^2) 1\!\!1, \, \kappa >0;\\
{\cal C}_1(\kappa;0)\,: ~~~~&~~~~X = \kappa 1\!\!1 , ~~~~&~~~~Y_0= (1-\kappa^2)1\!\!1,
\, 0 \leq \kappa \leq 1;\\
{\cal C}_2(\kappa;0)\,: ~~~~&~~~~X = \kappa 1\!\!1, ~~~~&~~~~Y_0= (\kappa^2-1) 1\!\!1,
\, \kappa \geq 1;\\
{\cal A}_1(0)\,: ~~~~&~~~~X =0 , ~~~~&~~~~Y_0=1\!\!1;\\
{\cal A}_2(0)\,: ~~~~&~~~~X =(1\!\!1 + \sigma_3)/2 , ~~~~&~~~~Y_0=1\!\!1;\\
{\cal B}_2(0)\,: ~~~~&~~~~X = 1\!\!1, ~~~~&~~~~Y_0=0;\\
{\cal B}_1(0)\,: ~~~~&~~~~X =1\!\!1 , ~~~~&~~~~Y_0= 0.
\end{array}
\label{int3}
\end{equation}
It may be noted that the quantum-limited end of both the ${\cal B}_1$
and ${\cal B}_2$ families is the trivial identity channel.

By following the above listed quantum-limited channels by injection of
classical noise of magnitude $a$ we get respectively ${\cal
  D}(\kappa;a)$, ${\cal C}_1(\kappa;a)$, ${\cal C}_2(\kappa;a)$,
${\cal A}_1(a)$, ${\cal A}_2(a)$, and  ${\cal B}_2(a)$; the last case
${\cal B}_1(a)$ is special in that it is obtained from ${\cal B}_1(0)$
by injection of noise into {\em just one quadrature}\,: $\chi(\xi)
\rightarrow \chi(\xi) \, \exp[-\frac{a}{4} \xi^T (1\!\!1 + \sigma_3) \xi ]$. 

It is clear in the case of ${\cal D}(\kappa;0)$ that $X= -\kappa
\sigma_3$ corresponds to (scaled) phase conjugation or matrix
transposition of the density operator. And the phase conjugation is the most famous
among positive maps which are not CP \cite{peres96, horodecki96, simon00}; it is the injection of additional
classical noise of magnitude (not less than) $1+\kappa^2$, represented
by $Y_0$, that mends it into a CP map. 

The reason for the special emphasis on quantum-limited channels in our
enumeration of the Holevo classification is this\,: every noisy
Gaussian channel [except ${\cal B}_1(a)$] can be realized, as we shall
see later, as the
composite of a pair of quantum-limited channels.  And this fact proves
to be of much value to the study presented in this paper. In the
original classification of Holevo \cite{holevo07, caruso06} the families
${\cal C}_1$ and ${\cal C}_2$, which correspond respectively to
the attenuator (beamsplitter) and the amplifier (two-mode squeezing)
channels, together constituted a single family ${\cal C}$. From the
point of view of the present study, however, these two families turn
out to be qualitatively different from one another, hence we prefer to
keep them as two distinct families. 

It is well known that every trace-preserving completely positive map has an operator-sum
representation of the form 
\begin{align}
\rho \rightarrow \rho^{\,'} = \sum_{\alpha} W_{\alpha} \, \rho \,
W_{\alpha}^{\dagger}, ~~~ \sum_{\alpha} W_{\alpha}^{\dagger}
W_{\alpha} = 1\!\!1,
\end{align}
often called Kraus representation \cite{hellwig}. It may be noted, however,
that this representation appears as Theorem 4 of a much earlier
work of Sudarshan et al \cite{matthews61}. It has
been presented also by Choi \cite{choimap75}, apparently
independently. Mathematicians seem to view it as a direct and
immediate consequence of the dilation theorem of Stinespring
\cite{stinespring}. In this paper we develop and present a systematic analysis of
the operator-sum representation for single-mode bosonic Gaussian
channels. 

Knowledge of Kraus representation of a channel could prove useful for
several purposes. For instance, since the set of channels for a given
system 
is convex, it is of interest to know its extremals. And a theorem of
Choi \cite{choimap75} gives a necessary and sufficient test for extremality
of a channel in terms of Kraus operators. It is known that a
channel is entanglement breaking if and only if it can be described in
terms of a set of rank one Kraus operators \cite{horodecki03,
  holevo08ebt, shirokov05}. Further, the work of \cite{winter} and
\cite{gregoratti03,
  hayden-king} relate error correctability to the structure of the Kraus
operators of a channel. Finally, there has been considerable recent
interest in contrasting the Gaussian with nonGaussian states in respect of
robustness and degradation of bipartite entanglement under one-sided
and two-sided action by Gaussian channels \cite{agarwal09,
  carmichael10, allegra10}, and it is likely
that Kraus representation could throw light on this problem. 

The content of this paper is organised as follows. Section II presents
a general scheme for computation of Kraus operators, and this scheme applies
uniformly to all quantum-limited Gaussian channels. This scheme takes
particular advantage of the fact that the symplectic two-mode
transformation which realizes the channel in the sense of \eqref{i1}
does not couple, in the Holevo canonical form, the position variables
with the momentum variables. With the ancilla mode assumed to be in
its vacuum state initially, it turns out that the Kraus operators for
each channel can be simply read off from the matrix elements of the
appropriate two-mode metaplectic operator. 

This scheme is applied in Sections III to VI to detail the Kraus operators
of respectively the ${\cal D},\,  {\cal C}_1, \, {\cal C}_2,$ and
${\cal A}_2$ families of quantum-limited channels. The Kraus operators
in every case is found to have an extremely simple-looking sparse
structure. In each case we ask if the channel has any fixed points
(invariant states), and if there are sufficient number of rank one
operators in the linear span of the Kraus operators, a question which
is at the very root of the entanglement breaking property of the channel. 

In the case of the phase conjugation or (matrix) transposition family
${\cal D}(\kappa;0)$ in Section III we explore how the threshold noise
of magnitude $1+\kappa^2$ renders the antilinear phase conjugation
into a linear map and channel. We bring out a well-defined sense in
which the channels 
${\cal D}(\kappa;0)$ and ${\cal D}(\kappa^{-1};0)$ are dual to one
another. The case ${\cal D}(1;0)$ is self-dual and hence doubly
stochastic, but it turns out that it is not a random unitary channel,
a fact which is of relevance to the possibility or otherwise of
extending the classical Birkhoff theorem to the quantum domain
\cite{landau93}. We examine in the Kraus picture the manner in which
quantum-limited attenuator and amplifier families ${\cal C}_1$, ${\cal
C}_2$ act as simple scaling maps respectively on the diagonal weight
function and the Husimi $Q$-distribution.  
Comparing the Kraus operators of the ${\cal C}_1$ family
with those of the ${\cal C}_2$ family, we show in Section V that these
two families are dual to one another. The intersection between ${\cal
  C}_1, \, {\cal C}_2$ consists of just the identity channel, the only
self-dual or doubly stochastic channel in the union of ${\cal
  C}_1, \, {\cal C}_2$. The manner in which the semigroup structures
of the ${\cal C}_1$ and ${\cal C}_2$ families are reflected in their
respective sets of Kraus operators is brought out in Sections IV and
V, and this enables us to point to a Zeno-like effect \cite{zeno}, in
both cases, arising as consequence of interrupted evolution.  Finally, even
though the single-quadrature classical noise 
channels ${\cal B}_1(a), \, a \neq 0$ [${\cal B}_1(0)$ is the
identity channel] are not quantum-limited, we deal with them briefly
in Section VII just to bring out the fact that this case too is obedient
to the general computational scheme presented in Section II. 

In Section VIII wherein we use Choi's theorem \cite{choimap75} to study if there
are any extremals among Gaussian channels, we show that all quantum
limited channels, and these alone, are extremal. That our concern upto 
this stage of the presentation is (almost) exclusively with the
quantum-limited case gets justified by our demonstrations in Section IX
that every noisy Gaussian channel [except ${\cal B}_1(a)$] can be
realized as the composite of a pair of quantum-limited ones. This
demonstration leads, in particular, to an operator-sum representation
for all noisy channels, including the all important classical noise
channels ${\cal B}_2(a)$ [but excluding ${\cal B}_1(a)$, and only
this case], in terms of a linearly independent {\em discrete set} of
Kraus operators having very simple sparse structure.

The composition of pairs of quantum-limited channels studied in
Section IX, and conveniently summarized in Table \ref{tab1} there,
assumes that both the constituent channels are simulatneously in their
respective canonical forms. When this assumption is removed, the
situation with the composition process gets much richer. The general
case is fully classified and presented in Table \ref{tab2} of the
Appendix. 

The final Section X contains a brief summary of the principal results
and also some additional remarks.

\section{Kraus representation: Some general considerations}  
Given density operator ${\rho}^{(a)}$ describing the state of a
single-mode radiation field,  
the action of a quantum-limited Gaussian channel 
takes it to \cite{caruso06, holevo07}
\begin{eqnarray}
{\rho}^{\,'(a)} = 
\text{Tr}_{b} (U^{(ab)} \, ({\rho}^{(a)} \otimes \,
|0\rangle_{b} {}_{b} \langle 0|)\,{ U^{(ab)}}^{\dagger} ). 
\label{ph2}
\end{eqnarray}
Here $|0\rangle_{b}$ is the vacuum state of the ancilla mode $b$, and
$U^{(ab)}$  is the unitary operator corresponding to a suitable 
two-mode linear canonical transformation.  
It is convenient to perform the partial trace in the Fock basis of
mode $b$. We have 
\begin{eqnarray}
{\rho}^{\,'(a)} &=& \sum_{{\ell}} {}_{b}\langle {\ell} | U^{(ab)}  \,
({\rho}^{(a)} \otimes \, 
|0 \rangle_{b} {}_{b}\langle 0 |) \,U^{(ab) \,\dagger }|{\ell} \rangle_{b} \,  \nonumber \\
&=& \sum_{{\ell}} {}_{b}\langle {\ell} | U^{(ab)}|0 \rangle_{b} \,{\rho}^{(a)} {}_{b}\langle
 0| U^{(ab)\, \dagger} |{\ell} \rangle_{b} \, . 
\label{ph3}
\end{eqnarray}
Clearly, $ {}_{b}\langle {\ell} | U^{(ab)} |0 \rangle_{b}$ is an operator
acting on the Hilbert space  of 
mode $a$. The last expression thus leads us to the Kraus representation of the
channel \cite{hellwig}\,:
\begin{eqnarray}
\rho \rightarrow {\rho}^{\,'(a)} = \sum_{{\ell}} W_{{\ell}} \,{\rho}^{(a)}
{W}_{{\ell}}^{\dagger}, ~\,\, ~~ 
 W_{\ell} =  {}_{b}\langle {\ell} | U^{(ab)}|0 \rangle_{b}.
\label{ph4}
\end{eqnarray}
It follows that once the Fock basis matrix elements of 
$U^{(ab)}$ are known, the Kraus operators ${W}_{{\ell}}$ can 
be easily read off. Let $ \langle m_1 m_2 | U^{(ab)} |
 n_1 n_2 \rangle \equiv {C}_{n_1 n_2}^{m_1 m_2}$ be 
the matrix elements of $U^{(ab)}$ in
the two-mode Fock basis. Since the ancilla mode $b$ is assumed to be in the vacuum state, the
$W_{\ell}$'s are obtained by setting $n_2 =0 $ and $m_2 = {\ell} $\,:
\begin{equation}
W_{{\ell}} = \sum_{n_1,m_1 =0} ^{\infty} \, {C}^{m_1 {\ell}}_{n_1 0} |m_1 \rangle
\langle n_1|.
\label{ph5}
\end{equation}

Now, in evaluating ${C}^{m_1 m_2}_{n_1 n_2}$ it proves
useful to employ a resolution of identity in the position basis \cite{leonhardt94}\,:
\begin{align}
C_{n_1n_2}^{m_1m_2} &=  \langle m_1 m_2 | U^{(ab)} | n_1 n_2 \rangle \nonumber \\
&= \int_{-\infty}^{\infty} dx_1 dx_2 \langle m_1 m_2 | x_1 x_2 \rangle
\langle x_1 x_2 | U^{(ab)} | n_1 n_2 \rangle. 
\label{ph6}
\end{align}
Under conjugation by $U^{(ab)}$ the quadrature variables $q_j, p_j
~~(j=1,2)$ undergo a linear canonical transformation $S \in Sp(4,R)$, of which
$U^{(ab)}$ is the (metaplectic) unitary representation \cite{dutta94}.  
Let us assume that this canonical transformation does not mix the
position variables with the momentum variables. That is, 
\begin{eqnarray}
\left ( \begin{matrix}
  {q}_{1}^{}  \\
  {q}_{2}^{} 
 \end{matrix}
\right)
&\rightarrow {U^{(ab)}}^{\dagger} \left ( \begin{matrix}
  {q}_{1}^{}  \\
  {q}_{2}^{} 
 \end{matrix}
\right) U^{(ab)} = &
\left ( \begin{matrix}
  {q}_{1}^{\,'}  \\
  {q}_{2}^{\,'} 
 \end{matrix}
\right) = 
M \, \left( \begin{matrix}
  q_1  \\
  q_2 
 \end{matrix}
\right), \nonumber \\
\left ( \begin{matrix}
  {p}_{1}^{}  \\
  {p}_{2}^{} 
 \end{matrix}
\right)
&\rightarrow {U^{(ab)}}^{\dagger} \left ( \begin{matrix}
  {p}_{1}^{}  \\
  {p}_{2}^{} 
 \end{matrix}
\right) U^{(ab)} = &
\left ( \begin{matrix}
  {p}_{1}^{\,'}  \\
  {p}_{2}^{\,'} 
 \end{matrix}
\right) = 
({M}^{-1})^T \, \left( \begin{matrix}
  p_1  \\
  p_2 
 \end{matrix}
\right),
\label{ph7}
\end{eqnarray}
where $M$ is a real non-singular $2 \times 2$ matrix. This assumption
that our $S \in Sp(4,R)$ has the direct sum structure $S = M \oplus
(M^{-1})^T$ 
will prove to be of much value in our analysis. We have
\begin{align} 
C_{n_1n_2}^{m_1m_2}  &= \int_{-\infty}^{\infty} dx_1 dx_2 \langle m_1 m_2 | x_1 x_2 \rangle
\langle x_1 x_2 | U^{(ab)} | n_1 n_2 \rangle \nonumber \\
&= \int^{\infty}_{-\infty} dx_1 dx_2 \langle m_1m_2
| x_1 x_2 \rangle \psi_{n_1} (x_1') \psi_{n_2}(x_2')  \nonumber\\
& = \int^{\infty}_{-\infty} dx_1 dx_2  \psi^*_{m_1}(x_1)
\psi^*_{m_2}(x_2)   \psi_{n_1}(x_1')  \psi_{n_2}(x_2'),
\label{ph8}
\end{align}
where $(x_1^{\,'}, \, x_2^{\,'})$ is linearly related to $(x_1, \,
x_2)$ through $M$. 
These wavefunctions are the familiar Hermite functions, the Fock
states in the position representation. The above 
integral may be evaluated using the generating function for Hermite
polynomials \cite{leonhardt94}\,: 
\begin{eqnarray}
\psi_n (x) &=& \frac{\pi^{-1/4}}{\sqrt{2^n n!}} e^{-x^2} H_n(x) \nonumber \\
&=& \frac{\pi^{-1/4}}{\sqrt{n!}}  \frac{\partial^{n}}{\partial z^{n}} 
\exp \left( -\frac{1}{2} [ (x-z \sqrt{2} )^2 - z^2 ] \right) \Big|_{z=0}.
\label{ph9}
\end{eqnarray}
Inserting in Eq. \eqref{ph8} the generating function for each of the four wavefunctions we have
\begin{align}
C_{n_1n_2}^{m_1m_2} = \frac{1}{\sqrt{n_1! n_2! m_1! m_2! }}
\frac{\partial^{m_1} }{\partial \eta_1^{m_1}} \frac{\partial^{m_2}}
{\partial \eta_2^{m_2}} \frac{\partial^{n_1} }{\partial z_1^{n_1}}
\frac{\partial^{n_2} }{\partial z_2^{n_2}} \, F(z_1, z_2,
\eta_1, \eta_2 ) \Big|_{z_1, z_2,\eta_1, \eta_2 = 0},
\label{ph10}
\end{align}
where
\begin{align}
F(z_1, z_2,\eta_1, \eta_2 ) = \pi^{-1} & \int_{-\infty}^{\infty} dx_1 \,
dx_2 \exp \left\{ -\frac{1}{2} [ (x_1- \eta_1\sqrt{2})^2
+ (x_2- \eta_2\sqrt{2})^2 \right.\nonumber\\
&\left. + (x_1'- z_1\sqrt{2})^2+ (x_2'-
z_2\sqrt{2})^2 - \eta_1^2 - \eta_2^2 - z_1^2 - z_2^2  ] \right\}.
\label{ph11}
\end{align}
The Gaussian integration over the variables $ x_1$ and $x_2$ can be
easily carried out to obtain $F(z_1, z_2,\eta_1, \eta_2 )$, and from 
$F(z_1, z_2,\eta_1, \eta_2 )$ we may readily  obtain 
$ C_{n_1n_2}^{m_1m_2}$, and hence the Kraus operators. This is the
general scheme we will employ in what follows to obtain Kraus
representation for quantum-limited Gaussian channels of the various families.

\section{Phase conjugation  or transposition channel  ${\cal
    D(\kappa)}, ~ \kappa \geq 0$} 
We now use the above scheme to evaluate a set of Kraus operators representing the phase
conjugation channel. The metaplectic unitary operator $U^{(ab)}$ appropriate for this case induces on the quadrature 
operators of the bipartite phase space a linear canonical
transformation corresponding to the following $S \in Sp(4,R)$\cite{caruso06}\,: 
 \begin{eqnarray}
S &=& ~
\left(
\begin{matrix}
\sinh{\mu} &0 & \cosh{\mu}& 0 \\
0& -\sinh{\mu} & 0 & \cosh{\mu} \\
\cosh{\mu} & 0&\sinh{\mu}& 0\\
0& \cosh{\mu} & 0 & -\sinh{\mu} 
\end{matrix}
\right).
\label{phc1}
\end{eqnarray}
Written in detail, the phase space variables undergo, under the action
of this channel, the transformation
\begin{eqnarray}
\left ( \begin{matrix}
  {q}_{1}^{}  \\
  {q}_{2}^{} 
 \end{matrix}
\right)
&\rightarrow&
\left ( \begin{matrix}
  {q}_{1}^{\,'}  \\
  {q}_{2}^{\,'} 
 \end{matrix}
\right) = 
M \, \left( \begin{matrix}
  q_1  \\
  q_2 
 \end{matrix}
\right), \nonumber \\
\left ( \begin{matrix}
  {p}_{1}^{}  \\
  {p}_{2}^{} 
 \end{matrix}
\right)
&\rightarrow&
\left ( \begin{matrix}
  {p}_{1}^{\,'}  \\
  {p}_{2}^{\,'} 
 \end{matrix}
\right) = 
(M^{-1})^T \, \left( \begin{matrix}
  p_1  \\
  p_2 
 \end{matrix}
\right), \nonumber \\
M &=& \left(
\begin{matrix} 
-\sinh{\mu} &\cosh{\mu} \\
\cosh{\mu} &-\sinh{\mu} 
\end{matrix}
\right).
\label{phc2}
\end{eqnarray}
It is seen that the above $S$ is indeed of the form $S = M \oplus
(M^{-1})^T\in Sp(4,R)$, and 
does not mix the position variables with the momentum variables, and
so our general scheme above readily applies.

It is clear from the structure of $S$ that the parameter $\mu$ is
related to $\kappa$ in ${\cal D(\kappa)}$ 
through $\kappa = -\sinh{\mu} >0$, so that $\cosh{\mu} = \sqrt{\kappa^2
+1}$. Thus \eqref{ph11} translates, for the present case, to the
following expression\,:
\begin{eqnarray}
F(z_1, z_2,\eta_1, \eta_2 ) &=& \pi^{-1} \int_{-\infty}^{\infty} dx_1 \,
  dx_2 \exp  \Bigg\{ - \frac{1}{2} [ (x_1- \eta_1\sqrt{2})^2
 +(x_2- \eta_2\sqrt{2})^2 \Bigg. \nonumber \\ &&  + (-\kappa x_1 + \sqrt{1+\kappa^2} \, x_2- z_1\sqrt{2})^2 
 + (\sqrt{1+\kappa^2}\,x_1 - \kappa x_2-z_2\sqrt{2})^2  \nonumber \\
 && \Bigg. - \eta_1^2 - \eta_2^2 - z_1^2 - z_2^2  ] \Bigg\}.
\label{phc3}
\end{eqnarray}
Performing the Gaussian integrals in $x_1$ and $x_2$ we obtain 
\begin{align}
F(z_1, z_2,\eta_1, \eta_2 ) = ({\sqrt{1+\kappa^2}})^{-1} ~\text{exp}
&\left\{(\sqrt{1+\kappa^{-2}})^{-1}
(\eta_1\eta_2 -
  z_1 z_2) \right.\nonumber\\ 
&\left. + (\sqrt{1+\kappa^2})^{-1}(\eta_1 
  z_2 + \eta_2 z_1)   \right\}.
\label{phc4}
\end{align}

To obtain the matrix elements $C_{n_1 n_2}^{m_1m_2}$ we need to carry out the procedure
indicated in Eq. (\ref{ph10}). 
This may be done in two steps. We begin by
rewriting the function $F(z_1, z_2,\eta_1, \eta_2 )$ as
\begin{align}
F(z_1, z_2,\eta_1, \eta_2 )=  (\sqrt{\kappa^2+1})^{-1} ~\text{exp} &\left\{
    \, z_2 [ (\sqrt{1+\kappa^2})^{-1} \eta_1 - (\sqrt{1+\kappa^{-2}})^{-1} z_1] \right. \nonumber\\
&\left. + \eta_2 [ (\sqrt{1+\kappa^{-2}})^{-1} \eta_1 +
  (\sqrt{1+\kappa^2})^{-1} z_1]\right\}. 
\label{phc6}
\end{align}
Performing the $z_2$ and $\eta_2$ differentiations respectively $n_2$ and $m_2$ times
on $F(z_1, z_2,\eta_1, \eta_2 )$, we obtain 
\begin{equation}
[ (\sqrt{1+\kappa^2})^{-1} \eta_1 -  (\sqrt{1+\kappa^{-2}})^{-1} z_1]^{n_2}
 [  (\sqrt{1+\kappa^{-2}})^{-1} \eta_1 + (\sqrt{1+\kappa^2})^{-1} z_1]^{m_2}
 F \equiv G F.  
\label{phc7}
\end{equation} 
The remaining differentiations can be carried out using the Leibniz rule.
Since we finally set $z_1, z_2,\eta_1, \eta_2 =
0$, and since $F(0) = 1$, 
the only terms that could possibly survive are necessarily of the form 
\begin{equation}
\frac{\partial^{m_1} }{\partial \eta_1^{m_1}} \frac{\partial^{n_1}}
{\partial z_1^{n_1}} [  (\sqrt{1+\kappa^2})^{-1} \eta_1 - 
(\sqrt{1+\kappa^{-2}})^{-1} z_1]^{n_2} [ (\sqrt{1+\kappa^{-2}})^{-1} \eta_1 +
(\sqrt{1+\kappa^2})^{-1} z_1]^{m_2}. 
\label{phc8}
\end{equation}
To evaluate the above expression we set $ x =  (\sqrt{\kappa^2+1})^{-1}  \eta_1
- (\sqrt{1+\kappa^{-2}})^{-1} z_1 $ and $y =
(\sqrt{1+\kappa^{-2}})^{-1} \eta_1 + (\sqrt{1+\kappa^2})^{-1}  
z_1$, and compute 
\begin{equation} 
[ (\sqrt{1+\kappa^2})^{-1} \partial_x + 
(\sqrt{1+\kappa^{-2}})^{-1} \partial_y]^{m_1} ~ [
-(\sqrt{1+\kappa^{-2}})^{-1} \partial_x +
(\sqrt{1+\kappa^2})^{-1} \partial_y]^{n_1} ~ x^{n_2} \, y^{m_2} |_{x,y =0}.  
\label{phc9}
\end{equation}
Straight forward algebra leads, in view of Eq. (\ref{ph10}),  to 
\begin{align}
C_{n_1n_2}^{m_1m_2}= \frac{ (\sqrt{1+\kappa^2})^{-1} \, }{\sqrt{n_1! n_2! m_1! m_2!} }  
\sum_{j=  0}^{n_1} \, \sum_{r = 0}^{m_1} & {}^{n_1}C_{j}
{}^{m_1}C_{r}\,   (-\sqrt{1+\kappa^{-2}})^{-(m_1
  + j -r)}\,  (\sqrt{1+\kappa^2})^{-(n_1 - j + r)}
\nonumber \\ &\times (-1)^{m_1 - r}  ~n_2! m_2! \delta_{n_2, r+j} \, \delta_{m_2,
  n_1 - j + m_1 -r} \, . 
\label{phc10}
\end{align}
\noindent
The Kraus operators $W_{\ell}$, denoted $T_{\ell}(\kappa)$ in this
case, are obtained from these matrix elements by setting $n_2 
=0$ and $m_2 = \ell$. Since $n_2 =0 \Rightarrow r,j =0$, we have,
\begin{align} 
T_{\ell}(\kappa) =  (\sqrt{1+\kappa^2})^{-1} \sum_{n_1, m_1 =0}^{\infty}
\frac{ (\sqrt{1+\kappa^2})^{-n_1} (-\sqrt{1+\kappa^{-2}})^{-m_1} 
  \sqrt{\ell!}}{\sqrt{n_1 ! m_1 !}} \delta_{\ell, n_1 + m_1} (-1)^{m_1} ~ |m_1
  \rangle \langle n_1|. 
\label{phc11}
\end{align} 
\noindent
We set $ n_1 +  m_1 = \ell$ and denote $n_1 =n$, leading to 
\begin{align}
T_{\ell}(\kappa) 
 = (\sqrt{1+\kappa^2})^{-1} \sum_{n=0}^{\ell} (\sqrt{1+\kappa^2})^{-n}
 (\sqrt{1+\kappa^{-2}})^{-(\ell-n)} \sqrt{{}^{\ell}C_n}   |\ell-n
 \rangle \langle n |,   \,\,\,\,\ell=0,1,2, \cdots
\label{phc12}
\end{align} 
as our final form for the Kraus operators of the phase conjugation
channel. We note that the $T_{\ell}(\kappa)$'s are real and manifestly
trace-orthogonal\,: $\text{tr}(
T_{\ell}(\kappa)^{\dagger}T_{\ell^{\,'}}(\kappa))=0$ if $\ell \neq
\ell^{\,'}$. 

\subsection{The dual of ${\cal D}(\kappa)$} 
As is well known (and also obvious), if a set of Kraus operators $\{
W_{\ell} \}$ describes the 
completely positive map $\Phi: {\rho} \rightarrow {\rho}^{\,'} =
\sum_{\ell}  W_{\ell} {\rho} {W_{\ell}}^{\dagger}$, then the 
dual map $\tilde{\Phi}:  {\rho} 
\rightarrow {\rho}^{\,'}= \sum_{\ell}  {W_{\ell}}^{\dagger} {\rho}
W_{\ell}$, described by the dual or adjoint set of operators $\{ W_{\ell}^{\dagger}\}$, 
is also completely positive. It is clear that the dual map
$\tilde{\Phi}$ is unital or trace-preserving according as $\Phi$ is
trace-preserving or unital. 

For the present case of ${\cal D}(\kappa)$, it is readily verified
that the Kraus operators $\{T_{\ell}(\kappa) \}$ presented in (\ref{phc12}) meet
$\sum_{\ell} {T_{\ell}}^{\dagger}(\kappa) T_{\ell}(\kappa) = 
1\!\!1$, consistent with the expected trace-preserving nature of 
${\rho} \rightarrow {\rho}^{\,'} = \sum_{\ell}  T_{\ell}(\kappa)\, {\rho}\,
{T_{\ell}}^{\dagger}(\kappa)$. But the phase conjugation channel is not unital
in general, for we have 
\begin{equation}
 \sum_{\ell} T_{\ell}(\kappa)\,{T_{\ell}}^{\dagger}(\kappa) =
 \kappa^{-2}  1\!\!1 . 
\end{equation}
We may say that it is `almost unital' to emphasise the minimal nature
of the failure\,: the
unit element is taken by the channel into a scalar multiple of
itself. However, the scale factor $\kappa^{-2} $
can not be transformed away by absorbing $ \kappa^{-1} $ into
the Kraus operators, for the Kraus operators so modified would not then
respect the  trace-preserving property of the map. 

It is thus of interest
to understand the nature of the {\em unital channel} described by the
set of Kraus operators 
$\{{T}_{\ell}(\kappa)^{\dagger} 
\}$. We have 
\begin{eqnarray}
{{T}_{\ell}(\kappa)}^{\dagger}
&=&(\sqrt{1+\kappa^2})^{-1} \sum_{n=0}^{{\ell}} (\sqrt{1+\kappa^2})^{-n}\,
(\sqrt{1+\kappa^{-2}})^{-({\ell}-n)}\, \sqrt{{}^{{\ell}}C_n}   |n
\rangle \langle {\ell}-n | \nonumber \\
&=&  (\sqrt{1+\kappa^2})^{-1}\,\sum_{n'={\ell}}^{0}
(\sqrt{1+\kappa^2})^{-({\ell}-n')} (\sqrt{1+\kappa^{-2}})^{-n'} \,
\sqrt{{}^{{\ell}}C_{{\ell}-n'}} \,  |{\ell}-n' \rangle 
\langle n' | \nonumber \\
&=&  (\sqrt{1+\kappa^2})^{-1}\,\sum_{n=0}^{{\ell}}
(\sqrt{\kappa^2+1})^{-({\ell}-n)}\, (\sqrt{1+\kappa^{-2}})^{-n} 
\sqrt{{}^{{\ell}}C_{n}}   |{\ell}-n \rangle 
\langle n | \nonumber \\
&=&   \kappa^{-1} T_{{\ell}}(\kappa^{-1}).
\label{phc14}
\end{eqnarray}
Thus the dual $\{ T_{\ell}(\kappa)^{\dagger} \}$ differs from the
original $\{T_{\ell}(\kappa) \}$ in two elementary 
aspects. 
The multiplicative factor $\kappa^{-1}$
{\em is the same} for all Kraus operators, independent of ${\ell}$. Thus
the only significant difference is change in the argument of
$T_{\ell}$, from  $\kappa$ to $\kappa^{-1}$. We conclude
that the `dual' channel whose Kraus operators are $\kappa \,T_{\ell}
(\kappa)^{\dagger}$ is the (trace-preserving) phase
conjugation channel ${\cal D}(\kappa^{-1})$. We have thus proved
\begin{theorem}\label{pth1}
While the Kraus operators $\{ T_{\ell}(\kappa) \}$ describe ${\cal
  D(\kappa)}$, the `dual' channel described by Kraus
operators $\{ \kappa  T_{\ell}(\kappa)^{\dagger} \} $ is the trace-preserving phase
conjugation channel ${\cal D}(\kappa^{-1})$ with reciprocal scale parameter. 
\end{theorem}

\subsection{Properties of the Kraus operators}
We now explore the properties of the Kraus operators of
Eq. (\ref{phc12}), connecting their explicit action in the
Fock basis to the expected transformation of the characteristic
function. The question of its fixed points is studied through the action of the channel on
the cumulants. The action of the
channel is illustrated with simple examples and, finally,
the entanglement breaking nature of the channel is made transparent by obtaining a
set of rank one Kraus operators describing the channel. 

The expected or defining action of the phase conjugation channel on the
characteristic function is \cite{caruso06}\,: 
\begin{equation}
\chi_W (\xi) \rightarrow \chi_W^{\,'}(\xi) = \chi_{W}(-\kappa \,\xi^*) ~ \exp[-(1+\kappa^2 )|\xi|^2/2]. 
\label{phc16}
\end{equation}  
It is of interest to understand how the `antilinear' phase
conjugation $(\xi \rightarrow \xi^{*})$ action of this 
channel on the characteristic function emerges from the linear
action of the Kraus operators. To this end, it is sufficient to establish such an action on the
`characteristic function' corresponding to the operators  
$ |n\rangle \langle m|$, for arbitrary pairs of integers $n, \, m \geq 0$. 
The `characteristic function' of $|n\rangle \langle m|$ is
given by \cite{cahill692}
\begin{align}
{\chi_W}_{|n\rangle \langle m|}(\xi)&\equiv \langle m|D(\xi)|n \rangle \nonumber \\ 
&=\sqrt{\frac{m!}{n!}} {(-\xi^*)}^{n-m}
{L}^{n-m}_{m}({|\xi|}^2) \, \exp[-|\xi|^2/2]\,\,\,\,{\rm for}
\,\,\,\, n \geq m,   \nonumber\\
 &=\sqrt{\frac{n!}{m!}} {(\xi)}^{m-n}
{L}^{m-n}_{n}({|\xi|}^2)\, \exp[-|\xi|^2/2 \,\,\,\,{\rm for}
\,\,\,\, n \leq m.
\label{phc17}
\end{align}
Assuming $n \geq m$, the action of the phase conjugation channel on the operator 
$| n \rangle \langle m|$ is
\begin{align}
&\sum_{{\ell}=0}^{\infty} T_{\ell}(\kappa) |n\rangle \langle m|
{T}_{{\ell}}^{\dagger}(\kappa) \nonumber \\
&= (1+\kappa^2)^{-1} \sum_{{\ell}=n}^{\infty} 
(\sqrt{1+\kappa^2})^{-(n+m)} \, (\sqrt{1+\kappa^{-2}})^{-(2{\ell} -n-m)}\, \sqrt{{}^{\ell}C_n 
 \,   {}^{{\ell}}C_m} \, | {\ell}-n \rangle \langle {\ell}-m|. 
\label{phc18}
\end{align}
Denoting $ n= m+ \delta$ and 
${\ell}-n = \lambda$, we have 
\begin{align}
\sum_{{\ell}=0}^{\infty} {T}_{\ell}(\kappa) |m+\delta\rangle \langle m| {T}_{{\ell}}^{\dagger}(\kappa) 
&= (1+\kappa^2)^{-1} (\sqrt{1+\kappa^2})^{-(2m + \delta)}
(\sqrt{1+\kappa^{-2}})^{-\delta} \nonumber \\ &\times \sum_{\lambda=0}^{\infty} 
\frac{(\lambda + m + \delta)!
(1+\kappa^{-2})^{-\lambda}}{\sqrt{(m+\delta)! m!  \lambda! (\lambda + \delta)!}}
|\lambda \rangle \langle \lambda + \delta|.
\label{phc19}
\end{align}
The manner in which ${\cal D}(\kappa)$, matrix transposition
accompanied by threshold Gaussian noise
$\exp[-(1+\kappa^2)|\xi|^2/2]$, acts as a channel may now be
appreciated. Every operator $M$ can be written in the Kronecker delta
basis $\{|j \rangle \langle \ell| \}$ as $M = \sum_{j ,\ell} c_{j\ell}
|j \rangle \langle \ell|$. The coefficient matrix $C$ associated with
$|5 \rangle \langle 3|$, for instance, is $c_{j,k} = \delta_{5j}
\delta_{\ell 3}$, with non-zero entry only at the lower-diagonal
location $(5,3)$ marked $\otimes$ in the matrix below. 
\begin{equation}
\left( \begin{array}{ccccccccccccccccc}
0&&0&&{\color{blue}\times}&&0&&0&&0&&0&&0&&\cdots\\
0&&0&&0&&{\color{blue}\times}&&0&&0&&0&&0&&\cdots\\
0&&0&&0&&0&&{\color{red}\oplus}&&0&&0&&0&&\cdots\\
0&&0&&0&&0&&0&&{\color{blue}\times}&&0&&0&&\cdots\\
0&&0&&{\color{green}\otimes}&&0&&0&&0&&{\color{blue}\times}&&0&&\cdots\\
0&&0&&0&&0&&0&&0&&0&&{\color{blue}\times}&&\cdots\\
0&&0&&0&&0&&0&&0&&0&&0&&{\color{blue}\times}\\
\vdots&&\vdots&&\vdots&&\vdots&&\vdots&&\vdots&&\vdots&&\vdots&&\vdots
\end{array} \right). \nonumber 
\end{equation}
On transposition this entry moves to the upper-diagonal location
$(3,5)$ marked $\oplus$, and the threshold noise then spreads it along
the parallel upper diagonal $(3+r, 5+r)$, $-3 \leq r < \infty$ marked
$\times$. 

Let the Weyl-ordered characteristic function $\text{tr}(D(\xi)
|m+\delta \rangle \langle m|)$ where $D(\xi) = \exp [\xi a^{\dagger} -
\xi^{*} a] $  is the displacement operator, be denoted
${\chi_W}_{|m+\delta \rangle \langle m|}(\xi)$, and that of the output
$\sum_{\ell=0}^{\infty} T_{\ell}(\kappa) |m+\delta \rangle \langle m|
T_{\ell}(\kappa)^{\dagger} $    
 be denoted ${\chi_W}^{\,'}_{|m + \delta\rangle \langle
   m|}(\xi)$. Then we have from Eq. \eqref{phc19}
\begin{align}
{\chi_W}^{\,'}_{| m+\delta \rangle \langle m| }(\xi) 
&=  (1+\kappa^2)^{-1} (\sqrt{1+\kappa^2})^{-(2m +
  \delta)} (\sqrt{1+\kappa^{-2}})^{-\delta}  \nonumber \\ &\times \sum_{\lambda
  =0}^{\infty} \frac{(\lambda + m + \delta)!
  (1+\kappa^{-2})^{-\lambda}}{\sqrt{(m+\delta)! m! 
    \lambda! (\lambda + \delta)!}} \langle \lambda + \delta |D(\xi))|
\lambda \rangle \nonumber \\
&= \frac{ (1+\kappa^2)^{-1} e^{-|\xi|^2 /2}}{\sqrt{(m+\delta)! m!}}
(\sqrt{1+\kappa^2})^{-(2m + \delta)}
(\sqrt{1+\kappa^{-2}})^{-\delta} \nonumber \\ &\times
\sum_{\lambda =0}^{\infty}  (1+\kappa^{-2})^{-\lambda}
\frac{(\lambda + m + \delta)!}{(\lambda + \delta)!} \xi^{\delta}
L_{\lambda}^{\delta} (|\xi|^2), 
\label{phc20}
\end{align}
where we used (\ref{phc17}), the Fock basis representation of the displacement
operator. While no `phase conjugation' is
manifest as yet, we expect from Eq. (\ref{phc16})
that the channel should take the characteristic function of $|m+\delta
\rangle \langle m|$ to 
\begin{align}
{\chi^{\,''}_W}_{| m+\delta \rangle \langle m|}(\xi) 
&= \langle m| D(-\kappa \xi^*) |m+\delta \rangle \,
\exp \left[-\frac{1}{2}(1+\kappa^2) |\xi|^2\right] \nonumber \\
&= \langle m+\delta | D(\kappa \xi^*) |m \rangle^{*} \,
\exp \left[-\frac{1}{2}(1+\kappa^2) |\xi|^2\right] \nonumber \\
&= \sqrt{\frac{m!}{m+\delta!}} (\kappa \xi)^{\delta} L_{m}^{\delta} (\kappa^2|\xi|^2) \,
\exp \left[-\left(\frac{1}{2}+\kappa^2\right) |\xi|^2\right] \,.
\label{phc21}
\end{align}
Thus the problem reduces to one of establishing equality of ${\chi^{\,'}_W}_{| m+\delta \rangle \langle m| }(\xi) $ in
\eqref{phc20} and ${\chi^{\,''}_W}_{| m+\delta \rangle \langle m|}(\xi)$ in
\eqref{phc21}. That is, it remains to prove 
\begin{eqnarray} 
&&\sqrt{\frac{m!}{m+\delta!}}  (\kappa \xi)^{\delta} L_{m}^{\delta} (\kappa^2 |\xi|^2) \,
\exp \left[-(1/2+ \kappa^2) |\xi|^2\right] \,
  \nonumber \\
&=&\frac{ (1+\kappa^2)^{-1} e^{-|\xi|^2
    /2}}{\sqrt{(m+\delta)! m!}} 
\sum_{\lambda =0}^{\infty}  (1+\kappa^{-2})^{-\lambda}
(\sqrt{1+\kappa^2})^{-(2m + \delta)}
(\sqrt{1+\kappa^{-2}})^{-\delta} \nonumber\\
&\times &\frac{(\lambda + m + \delta)!}{(\lambda + \delta)!} \xi^{\delta}
L_{\lambda}^{\delta} (|\xi|^2),
\label{phc22}
\end{eqnarray}
for all $m,\, \delta \geq 0$ [the case of $|m \rangle \langle
m+\delta| $ can be handled similarly].

Since the associated Laguerre functions form a complete orthonormal
set, we 
may expand the LHS of Eq. (\ref{phc22}) in the Laguerre basis. 
That is, we multiply both sides of Eq. (\ref{phc22}) by 
$ (\xi^*)^{\delta} L_{\ell}^{\delta} (|\xi|^2)  \,e^{-\frac{|\xi|^2}{2}}$ 
and evaluate the overlap integrals. We use the following
two standard results  : (i)  
orthogonality relation among Laguerres, and (ii) the overlap  between
a Laguerre and a scaled Laguerre function \cite{gradstein1}\,:
\begin{eqnarray}
&&\int_0^{\infty} e^{-|\xi|^2} \, {|\xi|}^{2\delta} L^{\delta}_n(|\xi|^2) \, L^{\delta}_m({|\xi|}^2) d{|\xi|}^2  = \,
\frac{(n+\delta)!}{n!} \delta_{n,m} .\nonumber \\
&&\int_0^{\infty} e^{-t|\xi|^2} \, |\xi|^{2\delta} L_{m}^{\delta}
(\eta^2 |\xi|^2) L^{\delta}_{\ell}(|\xi|^2) d|\xi|^2  \nonumber \\
&&= \frac{(m+{\ell}+\delta)!}{m!{\ell}!} \frac{(t-\eta^2)^m \,
(t-1)^{\ell}}{t^{m+{\ell}+\delta+1}} 
 \times F \left[ -m,-{\ell}; -m-{\ell}-\delta, 
\frac{t(t-\eta^2-1)}{(t-1)(t-\eta^2)} \right].~~~~
\label{phc23}
\end{eqnarray}
\noindent
Here $F[\cdot]$ is the hypergeometric function. In our case $t=\eta^2
+1$, which implies that the last argument of $F[\cdot]$ in
Eq. (\ref{phc23}) is zero, and thereby $F[\cdot]=1$. 
Performing the overlap integrals, we obtain for the left and right
hand sides of (\ref{phc22}) 
\begin{eqnarray}
\text{LHS}&=& \frac{(m+{\ell}+\delta)!}{{\ell}! \sqrt{(m+\delta)! m!}} \frac{{\kappa}^{2{\ell}+\delta}
}{(1+\kappa^2)^{m+{\ell}+\delta +1}} \,\,\,\,\, \nonumber, \\ 
\text{RHS}&=&
\frac{(m+{\ell}+\delta)!}{{\ell}!\sqrt{(m+\delta)!m!}} \, 
(\sqrt{1+\kappa^2})^{-(2+2m+\delta)}\,  (\sqrt{1+\kappa^{-2}})^{-(2{\ell}+ \delta)}.
\label{phc24}
\end{eqnarray} 
These two expressions obviously equal  one another for all $\ell$.
We have thus established Eq. (\ref{phc22}), and the fact that the
Kraus operators indeed effect the `completely positive phase
conjugation' operation, transforming the characteristic function as
expected in (\ref{phc16}).  
\begin{theorem}
The scaled phase conjugation transformation $\chi_W(\xi) \rightarrow {\chi^{\,'}}_W(\xi) = \chi_W(-\kappa \,\xi^*) 
~ \exp [-(1+\kappa^2 ) \frac{|\xi|^2}{2}]$ is, in view of the
threshold noise $\exp [-(1+\kappa^2)|\xi|^2/2]$ a completely
positive map, and is implemented linearly by the Kraus operators $\{ T_{\ell} (\kappa)\}$ in Eq. \eqref{phc12}. 
\label{phth1}
\end{theorem}

The phase conjugation channel has an interesting property in respect
of classicality/nonclassicality of the output states.
We may say a channel is {\em nonclassicality breaking} if the output of
the channel is classical for every input state. That is, if the
normal-ordered characteristic function $\chi^{\,'}_N(\xi)$ of the
output, related to the Weyl-ordered characteristic function
$\chi^{\,'}_W(\xi)$ of  \eqref{phc16} through $\chi^{\,'}_N(\xi) =
\chi^{\,'}_W(\xi) \exp[{|\xi|^2/2}]$, is such that its Fourier transform,
called the diagonal `weight' function $\phi(\alpha)$
\cite{sudarshan63}, is a genuine probability density. 

Now, Eq. \eqref{phc16} written in terms of the normal-ordered
characteristic function reads 
\begin{align}
\chi_N(\xi) \rightarrow \chi_N^{\,'}(\xi) &= \chi_W(-\kappa\xi^{*})
\exp[{-\kappa^2|\xi^{*}|^2}/2] \nonumber \\
&= \chi_A(-\kappa\xi^{*}), 
\label{phc24a}
\end{align} 
where $\chi_A(\xi) = \chi_N(\xi) \exp[-|\xi|^2] $ is the
antinormal-ordered characteristic function corresponding to the $Q$ or
Husimi distribution.

Under Fourier transformation this important relation \eqref{phc24a},
namely $\chi_N^{\,'} (\xi) = \chi_A(-\kappa\xi^*)$, reads that the {\em output}
diagonal weight function $\phi^{\,'}(\alpha)$ evaluated at $\alpha$ equals the
{\em input} $Q(\alpha)$ evaluated at $\kappa^{-1}\alpha^{*}$. Thus
$\phi^{\,'}(\alpha)$ is a genuine probability density for every input
state, and we have 
\begin{align}
{\cal D}(\kappa)\,:~ \phi_{\text{in}}(\alpha) \rightarrow
\phi_{\text{out}}(\alpha) = \kappa^{-2} Q_{\text{in}}(\kappa^{-1}\alpha^*).
 \end{align}
Since the $Q$-distribution of a density operator
is given by $Q(\alpha) = \langle \alpha|\rho|\alpha\rangle$, it is a
genuine probability distribution for all states including nonclassical
states. We have thus proved 
\begin{theorem}
The phase conjugation channel is a nonclassicality breaking channel.
\label{phco1}
\end{theorem}

\subsection{Fixed points}
We now study the fixed points of the phase conjugation channel through
consideration of cumulants. The characteristic function
of the $s$-ordered quasiprobabilities differ from one and another just
by a Gaussian factor, and it follows that the   
cumulants of order $>2$ of the $s$-ordered quasi-probability are
independent of the ordering parameter $s$ \cite{schack90, solong08},
and are thus intrinsic to the state. Hence it is sufficient to
work with a particular choice of $s$. We work with $s=0$, the case of
symmetric or Weyl ordering. 

Given a symmetric ordered characteristic function $\chi_W(\xi)$,
the corresponding cumulant generating function is defined as  
\begin{eqnarray}
\Gamma({\xi})= {\rm log}\, [\chi_W(\xi)]. 
\label{phc27}
\end{eqnarray}
With $\xi = \xi_1 + i \xi_2$, the cumulants are defined through
\begin{eqnarray}
\gamma_{m_1,m_2} =  \frac{\partial^{m_1}}{\partial
(i\xi_1)^{m_1}} \frac{\partial^{m_2}}{\partial
(i\xi_2)^{m_2}} \, \Gamma(\xi) |_{\xi=0}, ~~m_1,m_2=0,1,2, \cdots
\label{phc27a}
\end{eqnarray}
From Eq. (\ref{phc16}) we know that under the action of the phase
conjugation channel 
\begin{eqnarray}
{\cal D}(\kappa)\,: ~\Gamma(\xi) \rightarrow \Gamma^{\,'}(\xi)=
\text{log }[\chi^{\,'}_W(\xi)] = \text{log }[\chi_W(-\kappa \xi^*)] \, -\frac{1}{2} |\xi|^2 (1+\kappa^2).  
\label{phc27b}
\end{eqnarray}
Since the additional term on the right hand side is quadratic in $\xi$,
the cumulants $\gamma^{\,'}_{m_1,m_2}$ of $\chi^{\,'}_W(\xi)$ of order $m_1,
m_2 \neq 2$ are 
\begin{eqnarray}
\gamma^{\,'}_{m_1 m_2}= \left( \frac{\partial}{\partial(i\xi_1)} \right)^{m_1}  
\left( \frac{\partial}{\partial(i\xi_2)}
\right)^{m_2}   \, \text{log }[\chi_W(-\kappa \xi^*)] \Big{|}_{\xi=0}.
\label{phc27c}
\end{eqnarray}
Denoting $-\kappa \xi^*=t$ we have 
\begin{eqnarray}
\gamma^{\,'}_{m_1 m_2}&=& (-1)^{m_1} (\kappa)^{m_1+m_2} \left( \frac{\partial}{\partial(it_1)} \right)^{m_1}  
\left( \frac{\partial}{\partial(it_2)}
\right)^{m_2}   \, \text{log }[\chi_W(t)] \nonumber \\
&=& (-1)^{m_1} (\kappa)^{m_1+m_2} \,\gamma_{m_1m_2},
\label{phc27d}
\end{eqnarray}
Now, for a state to be invariant all its cumulants need to remain
invariant. By \eqref{phc27d}, none of the cumulants of order $m_1,m_2
>2$ are preserved for $\kappa \neq 1$. Indeed, under repeated use of
the channel the higher order cumulants monotonically increase or
decrease depending
on whether $\kappa$ is $>1$ or $<1$. In the case $\kappa=1$, the
cumulants with $m_1$ or $m_2 =2$ are not preserved because of the last
additional term on the right hand side of \eqref{phc27b}, showing that
no non-Gaussian state can be a fixed point of ${\cal D}(\kappa)$.

We are therefore left with the case $\kappa<1$ to consider. It is clear
that in this case all cumulants of order $\neq 2$ die out under
repeated use of the channel, and any initial state is driven towards a
fixed Gaussian (thermal state). A similar situation was discussed in
\cite{browne03}, where linear devices were used to drive non-Gaussian
pure states to  Gaussian states. 

We are thus led to look for fixed points among Gaussian states. The
additional last term in Eq. (\ref{phc27b})
is proportional to $|\xi|^2$. Since there
is no cross term involving the real and imaginary parts of $\xi$, it is
sufficient to look for fixed point among the thermal states, given by 
 \begin{align}
{\rho}_{\text{th}}(a_0) &= \frac{2}{a_0+1} \sum_{n=0}^{\infty} \left( \frac{a_0-1}{a_0+1}
\right)^n  |n \rangle \langle n| \nonumber \\
&=(1-x) \sum_{n=0}^{\infty} x^n |n \rangle \langle n|,
\label{phc28a}
\end{align}
where $x =(a_0-1)(a_0+1)^{-1}$, and the average photon number  
$\text{tr}\, (\rho_{\text{th}} a^{\dagger} a )= (a_0-1)/2$.  
By Eq. (\ref{phc19}) the output of the channel is
\begin{align}
{\rho}^{\,'} &= (1+\kappa^2)^{-1} (1-x) \sum_{n=0}^{\infty} \sum_{j=0}^{\infty}
{}^{n+j}C_j (1+\kappa^2)^{-j} (1+\kappa^{-2})^{-n} \, x^j \, |n \rangle\langle n|.
\label{phc29}
\end{align}
With the use of the identity
\begin{align}
\sum_{j=0}^{\infty} & {}^{n+j}C_j (x(1+\kappa^2)^{-1})^j = \frac{1}{(1-(1+\kappa^2)^{-1}x)^{n+1}},
\label{phc30}
\end{align} 
the double summation in Eq. (\ref{phc29}) reduces to a single sum, and we
have
\begin{eqnarray}
{\rho}_{\text{th}}(a_0) \rightarrow
{\rho}_{\text{th}}(a^{\,'}_0) = \frac{1-x}{1+\kappa^2-x}
\sum_{n=0}^{\infty} \left(\frac{\kappa^2}{1+\kappa^2-x} \right)^n
| n \rangle\langle n|.
\label{phc31}
\end{eqnarray}
This is a thermal state, and we see that the effect of the channel is to
change the thermal parameter $x$ as follows\,: 
\begin{align}
x \rightarrow x^{\, '} = \frac{\kappa^2}{1+\kappa^2-x}.
\label{phc31a}
\end{align}
Thus the fixed points are $\bar{x} = \kappa^2$ and $\bar{x}=1$. The first,
$\bar{x}=\kappa^2$, corresponds to a finite temperature state
$(\kappa^2<1)$, to which all other states are attracted under repeated
use of the channel. The second one corresponds to infinite
temperature, and is uninteresting for this reason and for the fact
that no state is attracted towards it. 

\begin{figure}
\includegraphics{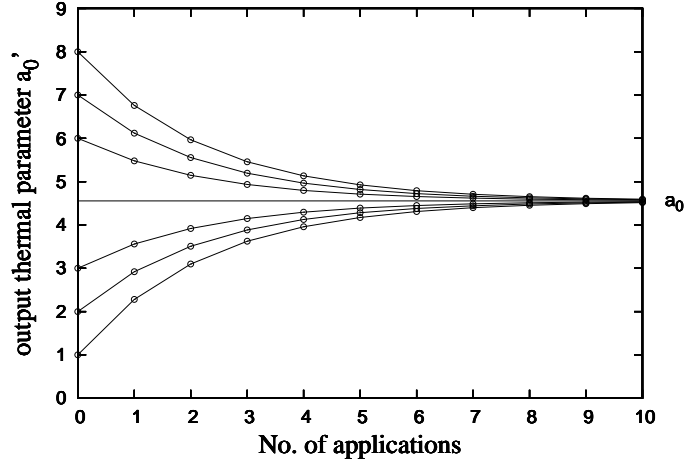}
\caption{Showing the variation of the output thermal parameter $a_0^{\,'}$ under repeated
  application of the quantum-limited phase conjugation channel ${\cal
    D}(\kappa)$ with $\kappa=0.8$ for
  various input thermal parameters. The stable thermal
parameter is $a_0=4.56$. \label{fig0}}
\end{figure}

In terms of the parameter $a_0$, the channel action reads 
\begin{eqnarray}
{\cal D}(\kappa)\,: ~{\rho}_{\text{th}}(a_0) \rightarrow
{\rho}_{\text{th}}(a^{\,'}_0), ~ a_0 \rightarrow a^{\,'}_0 = 
\kappa^2a_0 + 1+\kappa^2. 
\label{phc32}
\end{eqnarray}
\noindent
As was to be expected from \eqref{phc31a}, the recursion relation in 
Eq. (\ref{phc32}) is stable when $\kappa^2 < 1$ and unstable for $\kappa^2 \geq 1$.
For a given $\kappa < 1$, any input state is driven towards this `stable'
thermal state under repeated use of the channel, and the thermal
parameter of this attracting fixed state is $a_0 = (1+\kappa^2)(1-\kappa^{2})^{-1}$. 
This is illustrated in Fig. (\ref{fig0}). 
We may summarize as follows 
\begin{theorem}
For the phase conjugation channel\, ${\cal D}(\kappa), \,\kappa<1$, there is a unique
thermal state ${\rho}_{\rm{th}}(a_0)$ with parameter $a_0=
(1+\kappa^2)(1-\kappa^2)^{-1}$ which is left invariant by the
channel. All other states are driven towards this thermal state under
repeated use of the channel. Channels ${\cal D}(\kappa)$ for which $\kappa \geq 1$ have no fixed
points. 
\end{theorem}


As a simple illustration of the action of the phase conjugation channel, assume the input to be a Fock
state. By Eq. (\ref{phc19}), we have 
\begin{align}
|n \rangle \langle n |  &\rightarrow \sum_{\ell=0}^{\infty} T_{\ell}(\kappa)
|n\rangle\langle n|T_{\ell}(\kappa)^{\dagger} \nonumber\\
&= (1+\kappa^2)^{-1} \sum_{{\ell}=n}^{\infty} {}^{\ell}C_n
(1+\kappa^2)^{-n} (1+\kappa^{-2})^{-({\ell}-n)} |{\ell}-n\rangle 
\langle {\ell}-n |.
\label{phc25}
\end{align}
Setting $ {\ell}-n = j$, we have 
\begin{align}
\sum_{{\ell}=0}^{\infty} T_{\ell}(\kappa) |n\rangle\langle n|T_{\ell}(\kappa)^{\dagger} 
&= (1+\kappa^2)^{-1} \sum_{j=0}^{\infty} {}^{n+j}C_j (1+\kappa^2)^{-n} (1+\kappa^{-2})^{-j}
|j \rangle \langle j|. 
\label{phc26}
\end{align}
That is, a Fock state is taken to a convex sum of
{\em all} Fock states. As an immediate consequence we have\,: if a density
operator ${\rho}$ has $\langle n | 
{\rho} | n \rangle = 0$ for some $n$, then   
${\rho}$ cannot remain invariant under the action of the
channel. This is true, in particular, of any state ${\rho}$ which is
in the support of a finite number of Fock states.

As another simple example, consider the phase averaged coherent state given by 
\begin{align}
{\rho} = e^{-\lambda} \sum_{j=0}^{\infty} \, \frac{\lambda^j}{j!}
|j \rangle \langle j|. 
\label{phc32a}
\end{align}
This mixed state has Poissonian photon number distribution
[PND]. Under the action of the channel,  we have
\begin{align}
{\rho} \rightarrow {\rho}^{\,'} = e^{-\lambda} (1+\kappa^2)^{-1}  \sum_{j=0}^{\infty}\frac{\lambda^j}{j!} 
\sum_{n=0}^{\infty} {}^{n+j} C_j  (1+\kappa^2)^{-j}
(1+\kappa^{-2})^{-n}  |n \rangle \langle n|. 
\label{phc33}
\end{align}
We cannot solve consistently for parameters $\kappa$ and $\lambda$ such that
the output PND is also Poissonian. In
view of Theorem \ref{phco1}, the output is a classical state, and
hence is necessarily super-Poissonian \cite{mary97}. To summarize,
under the action of the channel ${\cal D(\kappa)}$ a thermal 
PND is taken to a thermal PND, 
whereas a Poissonian PND is taken to a super-Poissonian PND, all
moving towards the fixed thermal PND if $\kappa <1$.\\

\subsection{Entanglement breaking property}
It is known that the phase conjugating channel is entanglement
breaking \cite{holevo08ebt, shirokov05}. It is also known that every
entanglement breaking channel has a description in terms of rank one
Kraus operators \cite{horodecki03}. We 
demonstrate these aspects using our Kraus operators $\{T_{\ell}(\kappa) \}$.

The Kraus operators $T_{\ell}(\kappa)$ presented in \eqref{phc12} are not of
unit rank; indeed, rank $T_{\ell}(\kappa) = \ell +1$, $\ell = 0,1,2, \cdots$. We noted
immediately following \eqref{phc12} that $T_{\ell}(\kappa)$ are
trace-orthogonal. In the generic case, trace-orthogonality requirement
would render the Kraus operators unique, but this is not true with the
present situation. The reason is that all these trace-orthogonal
$T_{\ell}(\kappa)$'s have the same Frobenius norm: $\text{tr}\,\left(
T_{\ell}(\kappa)T_{\ell}(\kappa)^{\dagger} \right)= (1+\kappa^2)^{-1}$, independent of
$\ell$. Thus the set $\{T_{r}^{\,'} \}$ defined through $T_{r}^{\,'}(\kappa) =
\sum_{\ell} U_{\ell r} T_{\ell}(\kappa)$, for any unitary matrix
$(U_{\ell r})$
will be a set of trace-orthogonal Kraus operators describing the same
channel as the original trace-orthogonal set $\{ T_{\ell}(\kappa) \}$. 

More generally, and independent of  trace-orthogonality, the map $\rho
\rightarrow \rho^{\,'} = \sum_{\alpha} T^{\,'}_{\alpha}(\kappa) \rho
T^{\,' \dagger}_{\alpha}(\kappa)$ describes the same channel as  $\rho
\rightarrow \rho^{\,'} = \sum_{\ell} T_{\ell}(\kappa) \rho
{T}^{\dagger}_{\ell}(\kappa)$ if the matrix $U$ connecting the sets $\{T_{\ell}(\kappa)
\}$ and $\{ T^{\,'}_{\alpha} (\kappa)\}$ is an isometry \cite{choimap75, landau93}\,:
\begin{align}
T_{\alpha}^{\,'}(\kappa) &= \sum_{\alpha} U_{\ell \alpha} T_{\ell}(\kappa),
~~~~\sum_{\alpha} U_{\ell \alpha} U^{*}_{r \alpha } = \delta_{\ell r}
\nonumber \\
&\Rightarrow \sum_{\ell} T_{\ell} (\kappa)\,\rho\, T_{\ell}^{\dagger}(\kappa) =
\sum_{\alpha} T^{\,'}_{\alpha}(\kappa)\, \rho \,{T^{\,'}}^{\dagger}_{\alpha}(\kappa). 
\label{phc33a}  
\end{align}   
If the index set $\alpha$ is continuous, as in the case below, then $\sum_{\alpha}$ is to be
understood, of course, as an integral. Now, the matrix elements between
coherent states $|\alpha \rangle$ and Fock states $|k\rangle$ define
such an isometry 
\begin{align}
U_{\ell \alpha} \equiv \langle \ell | \alpha \rangle =
\exp[-|\alpha|^2/2] \frac{\alpha^{\ell}}{\sqrt{\ell !}}. 
\label{phc33b}
\end{align} 
The resulting new 
Kraus operators $T^{\,'}_{\alpha}(\kappa)$ are 
\begin{align}\label{phc39}
T^{\,'}_{\alpha}(\kappa) &=
e^{-\frac{|\alpha|^2}{2}}\sum_{{\ell}=0}^{\infty}
\frac{\alpha^{\ell}}{\sqrt{{\ell}!}} T_{\ell}(\kappa) 
\nonumber \\
&= e^{-\frac{|\alpha|^2}{2}}\sum_{{\ell}=0}^{\infty}
\frac{\alpha^{\ell}}{\sqrt{{\ell}!}}\, (\sqrt{1+\kappa^2})^{-1}\,
\sum_{n=0}^{{\ell}} \sqrt{{}^{\ell}C_n}\, (\sqrt{1+\kappa^2})^{-n}\,
(\sqrt{1+\kappa^{-2}})^{-({\ell}-n)} |{\ell}-n\rangle \langle n|
\nonumber \\ 
&= e^{-\frac{|\alpha|^2}{2}} \sum_{{\ell}=0}^{\infty}
(\sqrt{1+\kappa^2})^{-1} \sum_{n=0}^{{\ell}}
\frac{[(\sqrt{1+\kappa^2})^{-1}\alpha]^n \,
  [(\sqrt{1+\kappa^{-2}})^{-1}\alpha]^{{\ell}-n}} {\sqrt{(\ell-n)!n!}}\,|{\ell}-n\rangle \langle n|\nonumber \\
&= \frac{1}{\sqrt{1+\kappa^2}} |\alpha/\sqrt{1+\kappa^{-2}} \rangle
\langle \alpha^*/\sqrt{1+\kappa^2}|, \, \forall \alpha \in {\cal C}. 
\end{align}
It is manifest that rank $T^{\,'}_{\alpha}(\kappa) =1$ for all $\alpha \in {\cal
  C}$, the complex plane, showing that the phase conjugation channel is indeed
entanglement breaking. However $\{T^{\,'}_{\alpha}(\kappa) \}$ are not
trace-orthogonal even though $\{T_{\ell}(\kappa) \}$ from which the former are
constructed were trace-orthogonal. This is due to the fact that the
isometry $U$ defined in \eqref{phc33b} is not an unitary, which in turn
is a consequence of the overcompleteness of the coherent states.

This brings us to another aspect of ${\cal D}(\kappa)$. In terms of
these new Kraus operators the phase conjugation channel 
${\cal D(\kappa)}$ reads 
\begin{align}
\rho \rightarrow \rho^{\,'} &= \pi^{-1} \int d^2\alpha \,
T^{\,'}_{\alpha}(\kappa)\, \rho \, T^{\,'\,
  \dagger}_{\alpha}(\kappa)\nonumber \\  &= \pi^{-1}(1+\kappa^2)^{-1}\int d^2 \alpha \,
Q((\sqrt{1+\kappa^2})^{-1}\alpha^{*}) |\alpha/\sqrt{1+\kappa^{-2}}
\rangle \langle \alpha/\sqrt{1+\kappa^{-2}}|. 
\label{phc33c}
\end{align} 
Thus the diagonal weight function of the output state of the channel
is the $Q$-distribution of the input state $\rho$\,:
$\phi_{\text{out}} = \kappa^{-2} Q_{\text{in}}(\kappa^{-1}\alpha^*)$.
We may combine this
result with the earlier one on rank one Kraus operators to state
\begin{theorem}
The diagonal weight of the output of the quantum-limited phase
conjugation channel is essentially the $Q$-distribution of the input
state. The channel ${\cal D}(\kappa)$ is not only classicality breaking, but also
entanglement breaking. 
\end{theorem}

The diagonal weight of the {\em output state} at $\alpha$ is the
$Q$-distribution of the {\em input state} evaluated at 
$\kappa^{-1} \alpha^{*}$. Since $Q(\alpha) \geq 0$ for all $\alpha$ and for
any $\rho$, the channel is nonclassicality breaking. The intimate
relationship between this result and the earlier one on
nonclassicality breaking may be noted. While the former followed
directly from the behaviour of the characteristic function, the
present one required consideration of the Kraus operators.     \\

\section{Beamsplitter/attenuator channel ${\cal C}_1(\kappa), ~ 0<\kappa<1$}
The two-mode unitary operator corresponding to the beamsplitter
channel 
induces the following symplectic
transformation on the quadrature operators of  
the bipartite phase space \cite{caruso06}\,: 
\begin{eqnarray}
S &=& ~
\left(
\begin{matrix}
\cos{\theta} &0 & -\sin \theta& 0 \\
0& \cos {\theta} & 0 & -\sin{\theta} \\
\sin \theta  & 0&\cos \theta& 0\\
0& \sin{\theta} & 0 & \cos{\theta}  
\end{matrix}
\right).
\label{bs1}
\end{eqnarray}
Note that $S$ is a direct sum of identical two-dimensional rotations:
as in the case of ${\cal D}(\kappa)$, the
position and momentum operators are not mixed by this 
transformation. The position variables transform as
\begin{equation}
\left ( \begin{matrix}
  {q}_{1}^{}  \\
  {q}_{2}^{} 
 \end{matrix}
\right) \rightarrow
\left ( \begin{matrix}
  {q}_{1}^{\,'}  \\
  {q}_{2}^{\,'} 
 \end{matrix}
\right) = 
M \, \left ( \begin{matrix}
  q_1  \\
  q_2 
 \end{matrix}
\right)= 
\left(
\begin{matrix} 
\cos \theta & \sin \theta  \\
- \sin \theta & \cos \theta 
\end{matrix}
\right) ~
\left(
\begin{matrix}
q_1\\
q_2
\end{matrix}
\right)
\label{bs2}
\end{equation}
and, consequently, the momentum variables as
\begin{eqnarray}
\left ( \begin{matrix}
  {p}_{1}^{}  \\
  {p}_{2}^{} 
 \end{matrix}
\right) \rightarrow
\left ( \begin{matrix}
  {p}_{1}^{\,'}  \\
  {p}_{2}^{\,'} 
 \end{matrix}
\right) = 
({M}^{-1})^T \, \left ( \begin{matrix}
  p_1  \\
  p_2 
 \end{matrix}
\right)
= {M} \, \left ( \begin{matrix}
  p_1  \\
  p_2 
 \end{matrix}
\right).
\label{bs3}
\end{eqnarray}
It is evident from $S$ that the parameter $\kappa$ in ${\cal C}_1(\kappa)$ is related to $\theta$
through $\cos{\theta}=\kappa, \,\sin{\theta} = \sqrt{1-\kappa^2}$.
The function $F(z_1, z_2,\eta_1, \eta_2 )$ of \eqref{ph11} for the present case is given by
\begin{eqnarray}
F(z_1, z_2,\eta_1, \eta_2 ) = \exp \left[ \eta_{2} ( \sqrt{1-\kappa^2}\,
  z_1 + \kappa z_2) + \eta_{1} ( \kappa z_1 - \sqrt{1-\kappa^2}
\,  z_2 ) \right]. 
\label{bs4}
\end{eqnarray}
\noindent
As in the previous case of ${\cal D}(\kappa)$,  
the differentiation on $F(z_1, z_2,\eta_1, \eta_2 )$ can be performed
in a straight forward manner to 
obtain the matrix elements of the unitary operator \cite{saleh89}, leading to   
\begin{eqnarray}
&C_{n_1n_2}^{m_1m_2}= \frac{1}{\sqrt{n_1! n_2! m_1! m_2!} } \sum_{r =
  0}^{n_1} \, \sum_{j = 0}^{n_2} &{}^{n_1}C_{r} \,
{}^{n_2}C_{j}\,  (-1)^{n_2 - j} \, {\kappa}^{n_1  -r +j}\, {(\sqrt{1-\kappa^2})}^{r +
  n_2 -j} \nonumber \\ 
&& \times ~m_1! m_2!\, \delta_{m_2, r+j} \, \delta_{m_1, n_1 + n_2 -r -j} .
\label{bs5}
\end{eqnarray}
Now, to obtain the Kraus operators from these matrix elements we set,
as in the case of ${\cal D}(\kappa)$, $n_2=0$ and $m_2=\ell$. 
Setting  $n_2=0$ $\Rightarrow$ $j=0$, and we have
\begin{eqnarray}
B_{\ell}(\kappa)= \sum_{m=0}^{\infty} \sqrt{{}^{m+\ell} C_{\ell}}\,
(\sqrt{1-\kappa^2})^{\ell}\, {\kappa}^{m}  
| m \rangle \langle m+\ell|, \,\,\,\,\, \ell=0,1,2, \cdots
\label{bs6}
\end{eqnarray}
as the Kraus operators of the beamsplitter or quantum-limited
attenuator channel. It is easy to
see that the Kraus operators 
are real and pairwise trace-orthogonal, as in the case of ${\cal D(\kappa)}$.

\subsection{Properties of the Kraus operators}
We now explore the properties of the Kraus operators presented in
Eq. (\ref{bs6}), connecting the action of the channel on the
Fock basis to that on the characteristic
function. We firstly exhibit the fact that the beamsplitter channel simply
effects a scaling on the weight function of the diagonal
representation. We show that
vacuum is the only fixed point
of the channel. It is further shown that in any set of Kraus operators
describing the channel ${\cal C}_1(\kappa)$, there will be not even
one operator of unit rank, thus demonstrating that ${\cal
  C}_1(\kappa)$ is not an 
entanglement breaking channel. Finally, the manifestation of the semigroup structure
of the family of channels ${\cal C}_1(\kappa), \, 0<\kappa< 1$ is
brought out in the Kraus representation, as also an associated
Zeno-like effect.

Recall that the beamsplitter channel induces the following transformation 
on the characteristic function \cite{caruso06}\,: 
\begin{align}
\chi_W(\xi) \rightarrow \chi_W{\,'}(\xi) &= \chi_W(\kappa  \,\xi) ~
\exp[-(1-\kappa^2)|\xi|^2/2] \nonumber \\
&=  \chi_W(\kappa  \,\xi)\, \exp [\kappa^2 |\xi|^2/2]  \exp[-|\xi|^2/2] .
\label{bs7}
\end{align}
Thus the normal
ordered characteristic function $\chi_N(\xi)$ transforms as 
\begin{align}
\chi_N(\xi) \equiv \chi_W(\xi) \exp(|\xi|^2/2) \rightarrow \chi^{\,'}_N(\xi) = 
\chi_N(\kappa \,\xi).
\label{bs9}
\end{align}
Since $\chi_N (\xi)$ and the diagonal weight $\phi(\alpha)$ form a
Fourier transform pair, it is immediately   
seen that $\phi(\alpha)$  gets simply scaled under the action of the
 ${\cal C}_1(\kappa)$ channel\,: $\phi(\alpha) \rightarrow \phi^{\,'}(\alpha) = {\kappa}^{-2}
\phi(\kappa^{-1}\alpha)$ \cite{nair}.  

It is instructive to bring out this fact from the perspective of the Kraus
operators. Since every state ${\rho}$ can be expressed through a
diagonal `weight' $\phi(\alpha)$ as \cite{sudarshan63}
\begin{align}
{\rho} = \pi^{-1} \int d^2 \alpha \, \phi(\alpha) |\alpha \rangle
\langle \alpha|,
\label{bs10}
\end{align}   
to exhibit the action of the channel on an arbitrary state it is sufficient to
consider its action on a generic coherent state. We have
\begin{align}
| \alpha \rangle \langle \alpha | \rightarrow&
\sum_{\ell=0}^{\infty} B_{\ell}(\kappa)  |\alpha \rangle \langle \alpha|
{B}_{\ell}^{\dagger}(\kappa)\nonumber \\
= & \sum_{{\ell}=0}^{\infty} \sum_{m=0}^{\infty} \sum_{n=0}^{\infty}
\frac{(({{1-\kappa^2}})|\alpha|^2)^{\ell}}{{\ell}!}
(\kappa\alpha^*)^m(\kappa \alpha )^n 
\frac{e^{-|\alpha|^2}}{\sqrt{m!n!}} |m\rangle\langle n|,
\label{bs11}
\end{align}
where we used the fact that the operator 
\begin{align}
|m \rangle \langle n | & \rightarrow \sum_{\ell=0}^{\infty} B_{\ell}(\kappa)
|m\rangle \langle n| B_{\ell}^{\dagger}(\kappa) \nonumber \\ 
& = \sum_{{\ell}=0}^{\text{min}\{m,n \}} \sqrt{{}^m C_{\ell}\, {}^n C_{\ell}}\,
{({1-\kappa^2})}^{\ell} {\kappa}^{m+n-2{\ell}} 
|m-{\ell} \rangle \langle n-{\ell}|. 
\label{bs12}
\end{align}
Carrying out the summations in Eq. (\ref{bs11}), one finds \cite{imoto04}
\begin{align}
\sum_{{\ell}=0}^{\infty} B_{\ell}(\kappa)  |\alpha \rangle \langle
\alpha| {B}_{{\ell}}^{\dagger}(\kappa) 
= |\kappa\alpha \rangle \langle \kappa\alpha|.
\label{bs13}
\end{align}
With this the action of the channel ${\cal C}_1(\kappa)$ reads 
\begin{eqnarray}
{\rho} &\rightarrow& {\rho}^{\,'} = 
\pi^{-1}\int d^2 \alpha \, \phi (\alpha) |\kappa\alpha \rangle
\langle \kappa\alpha| \nonumber \\
&= & \pi^{-1}\kappa^{-2} \int d^2 \alpha\,  \phi (\kappa^{-1}\alpha) 
|\alpha \rangle \langle \alpha|,
\label{bs14}
\end{eqnarray} 
which means 
\begin{align}
{\cal C}_1(\kappa)\,:\, \phi(\alpha) \rightarrow \kappa^{-2}
\phi\left(\kappa^{-1} \alpha \right). 
\label{bs15}
\end{align}
We have thus proved in the Kraus representation
\begin{theorem}
The scaling  $\phi_{\rho} (\alpha)$ $\rightarrow$ ${\phi}^{\,'}_{\rho}(\alpha)
= \kappa^{-2} \phi_{\rho} (\kappa^{-1} \alpha) $, $0< \kappa < 1$, is a completely positive
map whose Kraus decomposition is given by  $\{ B_{\ell}(\kappa) \}$ of \eqref{bs6}.
\label{bsth1}
\end{theorem}
As an immediate consequence we have   
\begin{corollary}
The beamsplitter channel cannot generate or destroy nonclassicality.
\end{corollary}
\noindent
{\em Proof}\,: By definition a state is classical if and only if its diagonal
weight  function $\phi(\alpha)$
is pointwise nonnegative everywhere in the complex plane \cite{sudarshan63}. Since
a pointwise positive 
function goes to a pointwise positive function under the above scaling
transformation, it follows that a classical state (and
a classical state alone)
is taken to a classical state under the action of the (quantum-limited) attenuator channel.  

\subsection{Fixed Points}
We now examine, using cumulants, if there are any fixed points for
this channel. In view
of Eq. (\ref{bs7}), the cumulant generating function transforms as
follows under the action of this channel\,: 
\begin{eqnarray}
{\cal C}_1(\kappa)\,: ~\Gamma(\xi) \rightarrow \Gamma^{\,'}(\xi) =
{\rm log}\, [\chi_W(\kappa \xi)] - \frac{1}{2} (1-\kappa^2)\,
{|\xi|}^2 
\label{bs17a}
\end{eqnarray}
As in the previous case of ${\cal D}(\kappa)$, the cumulants of order $>2$ of $\chi^{\,'}_W(\xi)$ are
\begin{eqnarray}
\gamma^{\,'}_{m_1 m_2}&=& (\kappa)^{m_1+m_2} \left( \frac{\partial}{\partial(it_1)} \right)^{m_1}  
\left( \frac{\partial}{\partial(it_2)}
\right)^{m_2}   \, \text{log }[\chi(t)], ~~~(\kappa\xi=t)\nonumber \\
&=&  (\kappa)^{m_1+m_2} \gamma_{m_1m_2},
\label{bs17b}
\end{eqnarray} 
where $\gamma^{\,'}_{m_1 m_2} $ and $\gamma_{m_1m_2}$ are the
cumulants of respectively 
the output and the input states. Thus, with the exception of the
trivial case $\kappa=1$, action of the channel ${\cal C}_1(\kappa)$
attenuates the higher order cumulants. Therefore 
any state that is preserved is 
necessarily Gaussian. Since the additional term $\frac{1}{2} (1-\kappa^2)
{|\xi|}^2$  in
Eq. (\ref{bs17a}) does not involve a cross term between the real and imaginary
parts of $\xi$, in looking for fixed points it is sufficient to consider the action only on thermal states.
Given an input thermal state ${\rho}_{\text{th}}$ as in
\eqref{phc28a} the output, in view of Eq. \eqref{bs12}, is
\begin{align}
{\rho}_{\text{out}}= \frac{(1-x)}{1-(1-\kappa^2)x} \sum_{n=0}^{\infty}
\left(\frac{\kappa^2x}{1-(1-\kappa^2)x}\right)^n |n \rangle \langle n|,
\label{bs19}
\end{align}
where we have used the identity in Eq. \eqref{phc30}. Since
${\rho}_{\text{out}}$ is a thermal state,  
comparing Eqs. (\ref{phc28a}) and (\ref{bs19})  we have this
transformation law for the thermal parameter\,:
\begin{align} 
{\cal C}_1(\kappa)\,: ~x \rightarrow \frac{\kappa^2x}{1-(1-\kappa^2)x} \text{ or } a_0
\rightarrow \kappa^2 a_0 + (1-\kappa^2). 
\label{bs20}
\end{align}
This means that the output thermal state is always {\em strictly} `cooler' than the input thermal state.
Thus any thermal state is driven towards the ground state under repeated  use
of the channel. Indeed, an arbitrary state is driven towards the ground state.
We thus have
\begin{theorem}
For the action of the  beamsplitter channel ${\cal C}_1(\kappa)$, the
ground state is the only fixed point.
\end{theorem}

As an illustration of the action of ${\cal C}_1(\kappa)$, consider the
case of a Fock state as the input.  
By Eq. \eqref{bs12} we find
\begin{eqnarray}
\sum_{\ell} B_{\ell}(\kappa) |n\rangle \langle n| B_{\ell}^{\dagger}(\kappa) 
&=& \sum_{{\ell}=0}^{n} {}^n C_{\ell} \, ({{1-\kappa^2}})^{{\ell}} {\kappa}^{2n-2{\ell}}
|n-{\ell} \rangle \langle n-{\ell}| \nonumber \\
&=& \sum_{{\ell}=0}^{n} {}^n C_{\ell} \,  ({1-\kappa^2})^{n-{\ell}} {\kappa}^{2{\ell}}
|{\ell} \rangle \langle {\ell}|. 
\label{bs17}
\end{eqnarray} 
That is, the beamsplitter channel takes a Fock state 
$|n \rangle \langle n|$ to a convex sum 
of all Fock states with photon number {\em less than or equal} to $n$. It is
thus clear that 
any input state which is diagonal in the Fock basis is taken to a Fock diagonal
state at the output. We also note that for an arbitrary input ${\rho}$, a
Fock diagonal entry $|m \rangle \langle m|$ at the output gets contribution only from the
Fock diagonal entries $|n \rangle \langle n|$ of the input with $n
\geq m$. Putting these facts together, 
it is easy to see that no state which is in the support of a finite number
of Fock states is preserved under the beamsplitter channel, for the
strength of the highest Fock state strictly decreases. 

As a second example, consider the phase averaged
coherent state \eqref{phc32a} as the input. By Eq. \eqref{bs13}, 
we have a phase averaged coherent 
state at the output\,: a Poissonian PND at the input
is taken to another Poissonian PND at the
output. 
Unlike the phase conjugation channel, both a thermal
as well as a Poissonian PND at the input are taken at the
output to respectively a thermal and
Poissonian PND of strictly decreasing mean photon number. The fixed
point should therefore be a Poissonian or thermal
state of vanishing mean photon number; this is the vacuum state.

\subsection{The issue of Entanglement breaking}
 
It is known that the beamsplitter channel is not entanglement
breaking \cite{holevo08ebt}. It should thus be possible, as it is obligatory, to demonstrate
that this channel cannot be represented using a set of rank one Kraus
operators. We begin by noting that in the limiting case
$\kappa=0$, all our Kraus operators $B_{\ell}(0)$ are of rank
one. Indeed, $(B_{\ell}(0))_{mn} = \delta_{m0} \delta_{n \ell}$. This
singular limit corresponds to the quantum-limited ${\cal A}_1$ channel which is known to be
entanglement breaking. 
We consider therefore the nontrivial case $\kappa \neq 0$. It is
manifestly clear that rank
$B_{\ell}(\kappa) = \infty$ for all $\ell$ (for $\kappa \neq 0$). If we represent this channel using
another set of Kraus operators $\{ B^{\,'}_{r}(\kappa) \}$, then these new operators
should necessarily be in the support of the set of operators $\{ B_{\ell}(\kappa)\} $. Thus a necessary
condition that one is able to represent the channel $\{ B_{\ell}(\kappa) \}$ using
rank one Kraus operators is that there be (sufficient number of) rank one operators in the
support of $\{ B_{\ell}(\kappa) \}$. It turns out that there is not
even one rank one operator
in this support. Indeed, a much stronger result is true. 
\begin{theorem}: 
There exists no finite rank operator in the support of the set
$\{ B_{\ell}(\kappa)\} ,\, \kappa \neq 0  $.  
\end{theorem}
Proof follows immediately from the structure of the $B_{\ell}(\kappa)$'s \,: $B_0(\kappa)$
is diagonal, and the $mn^{\text{th}}$ entry of $B_{\ell}(\kappa)$ is
nonzero iff $n = m+\ell$. Any matrix in the linear span of $ \{B_{\ell}(\kappa) \}$ is of
the form $M = \sum_{\ell} c_{\ell}B_{\ell}(\kappa)$, and is upper
diagonal. Let $N$ be the smallest  $\ell$ for which the $c$-number
coefficient $c_{\ell} \neq 0$. Let
$\tilde{M}$ be the matrix obtained from the upper-diagonal $M$ by deleting the first $N$
columns. Clearly, rank $\tilde{M} =$ rank $M$. Further, the diagonal entries
of the upper triangular $\tilde{M}$ are all nonzero, being the nonzero entries of
$B_N(\kappa)$. Now, the rank of an upper triangular matrix is not less than
that of its diagonal part. Thus, rank $\tilde{M}$ is not less than
rank $B_{N}(\kappa) = \infty$, thus completing the proof.

\subsection{Semigroup property}
It is clear from (\ref{bs7}) that successive actions of
two beamsplitter channels with parameter values ${\kappa}_1, {\kappa}_2$ is a single beamsplitter 
channel whose parameter ${\kappa}$ equals the product ${\kappa}_1{\kappa}_2$ of the 
individual channel parameters \,:
\begin{eqnarray}
{\cal C}_1(\kappa_1)\,:  ~~\chi_W(\xi) &\rightarrow &\chi{\,'}_W(\xi) = \chi_W({\kappa}_1  \,\xi) ~
\exp{[-(1-{\kappa}_1^{2}) |\xi|^2/2]}, \nonumber \\
{\cal C}_1(\kappa_2)\,:  ~~ \chi{\, '}_W(\xi) &\rightarrow &\chi{\,''}_W(\xi) = \chi{\, '}_W({\kappa}_2  \,\xi) ~
\exp{[-(1-{\kappa}_{2}^{2}) |\xi|^2/2]} \nonumber \\
&=& \chi_W({\kappa}_1{\kappa}_2  \,\xi) ~\exp{[-(1-{{\kappa}}_{1}^{2} {{\kappa}}_{2}^{2}
  )|\xi|^2/2]}. 
\label{bs21}
\end{eqnarray} 
It is instructive to see how this semigroup property emerges in the 
Kraus representation. Let $\{B_{\ell_1}(\kappa_1)\}$ and
$\{B_{\ell_2}(\kappa_2) \}$ be the Kraus operators of the two
channels. The product of two Kraus operators
$B_{{\ell}_1}{({\kappa_1})}$, $B_{{\ell}_2}{({\kappa_2})}$, one
from each set, is 
\begin{align}
B_{{\ell}_1}({\kappa}_1) B_{{\ell}_2}({\kappa}_2) = &\sum_{m =0}^{\infty}\,
\sqrt{{}^{{\ell}_1+{\ell}_2} C_{\ell_1}} \left(\sqrt{1-\kappa_1^2}\right)^{\ell_1} \left(\sqrt{1-\kappa_2^2}\right)^{\ell_2}
 \nonumber \\ 
&\times  \sqrt{{}^{m + {\ell}_1+{\ell}_2} C_{{\ell}_1+{\ell}_2}}
\, ({\kappa}_1{\kappa}_2)^{m} \,{\kappa}_2^{{\ell}_1}  |m 
  \rangle\langle m + {\ell}_1 + {\ell}_2|. 
\label{bs22}
\end{align}
Thus the action of the product channel on the input operator 
$|r \rangle \langle r+\delta|$ is
\begin{eqnarray}
&&\sum_{{\ell}_1,{\ell}_2} B_{{\ell}_1}({\kappa}_1) B_{{\ell}_2}({\kappa}_2) |r \rangle \langle r+\delta|
B_{{\ell}_2}({\kappa}_2)^{\dagger} \, B_{{\ell}_1}({\kappa}_1)^{\dagger} \nonumber \\
&&= \sum_{{\ell}_1,{\ell}_2,m,n} \sqrt{{}^{{\ell}_1+{\ell}_2} C_{\ell_1}}  \left(\sqrt{1-\kappa_1^2}\right)^{\ell_1} \left(\sqrt{1-\kappa_2^2}\right)^{\ell_2}
\sqrt{{}^{m + {\ell}_1+{\ell}_2} C_{{\ell}_1+{\ell}_2}}
\,({\kappa}_1{\kappa}_2)^{m} {\kappa}_2^{{\ell}_1}  
  \nonumber \\
&&\,\,~~~\,\,\,\,\times  \sqrt{{}^{{\ell}_1+{\ell}_2} C_{\ell_1}} \, \left(\sqrt{1-\kappa_1^2}\right)^{\ell_1} \left(\sqrt{1-\kappa_2^2}\right)^{\ell_2}
\sqrt{{}^{n + {\ell}_1+{\ell}_2} C_{{\ell}_1+{\ell}_2}}
\,({\kappa}_1{\kappa}_2)^{n} {\kappa}_2^{{\ell}_1} 
\nonumber\\ 
&& ~~~~~~\times |m
  \rangle\langle m + {\ell}_1 + {\ell}_2 |r \rangle \langle
  r+\delta| n+ {\ell}_1 + {\ell}_2
  \rangle\langle n |.
\label{bs23}
\end{eqnarray}
Denoting ${\ell}_1+{\ell}_2 = {\ell}$, the expression on the RHS becomes 
\begin{align}
\text{RHS  }= \sum_{{\ell}=0}^{r} \sum_{{\ell}_1=0}^{{\ell}} \sum_{m,n=0} ^{\infty}
&{}^{{\ell}}C_{{\ell}_1} {\kappa}_2^{2{\ell}_1}
{({1-\kappa_1^2})}^{{\ell}_1} 
{({1-\kappa_2^2})}^{({\ell} -{\ell}_1)}
({\kappa}_1{\kappa}_2)^{m+n} \nonumber\\
&\times \sqrt{{}^{{\ell}+m} C_{\ell} {}^{{\ell}+n} C_{\ell}} 
\, \delta_{r,m+\ell} \, \delta_{r+\delta, n+\ell} \, 
|m \rangle \langle n |.
\label{bs24}
\end{align}
The sum over ${\ell}_1$ is the binomial expansion of
$[({{1-\kappa_1^2})}{\kappa}_2^2+{({1-\kappa_2^2})}]^{\ell} =
(1-\kappa_1^2\kappa_2^2)^{\ell}$ and, in addition, we have the
constraints 
$m + {\ell} =r$ and $n + {\ell} = r+ \delta$. With this the
expression (\ref{bs24}) reduces to 
\begin{align}
\text{RHS }=\sum_{{\ell}=0}^{r} \, (1-\kappa_1^2\kappa_2^2)^{\ell} \, \sqrt{ {}^{r} C_{\ell}\,
  {}^{r+\delta} C_{\ell}} \, ({\kappa}_1{\kappa}_2)^{2r-2{\ell} + \delta} |r-{\ell} \rangle
\langle r-{\ell} + \delta|. 
\label{bs25}
\end{align} 
Comparing Eqs. \eqref{bs12} and (\ref{bs25}) we find that the expression in
(\ref{bs25}) is precisely the action of a quantum-limited
attenuator channel with parameter ${\kappa}_1{\kappa}_2$\,:
\begin{align}
\label{bs26}
\sum_{{\ell}_1,{\ell}_2 =0}^{\infty} B_{{\ell}_1}({\kappa}_1) B_{{\ell}_2}({\kappa}_2) |r \rangle \langle r+\delta|
B^{\dagger}_{{\ell}_2}({\kappa}_2) B^{\dagger}_{{\ell}_1}({\kappa}_1)
= \sum_{\ell=0}^{\infty} \,B_{\ell}({\kappa}_1{\kappa}_2) |r \rangle
\langle r+\delta|  B_{\ell}^{\dagger}({\kappa}_1{\kappa}_2).
\end{align}  
\noindent
An identical result can be similarly obtained for the behaviour of $|r
+ \delta \rangle \langle r|$, and thus we have proved the semigroup
property 
\begin{align}
{\cal C}_1(\kappa_1)\circ {\cal C}_1(\kappa_2) = {\cal C}_1(\kappa_1\kappa_2).
\end{align}

\noindent 
{\bf Remark on interrupted evolution and Zeno-like effect}\,:\\
As seen from (\ref{bs1}) the parameter $\kappa$ specifying the
channel ${\cal C}_1(\kappa)$ equals $\cos{\theta}$, where $\theta$ is a
measure of the two-mode rotation effected by the `beamsplitter'
coupling the system mode to an ancilla mode assumed to be in the
vacuum state initially. The associated two-mode unitary operator
$U(\theta) = \exp [-\theta(a^{\dagger}b - b^{\dagger}a)]$ may be
viewed as effecting evolution for `duration' $\theta$ under the
Hamiltonian $-i (a^{\dagger}b - b^{\dagger}a)$. It is clear that
attenuation increases monotonically as $\theta$ varies from $0$ to
$\pi/2$, with total attenuation achieved at $\theta=\pi/2$. 

\begin{figure}
\includegraphics{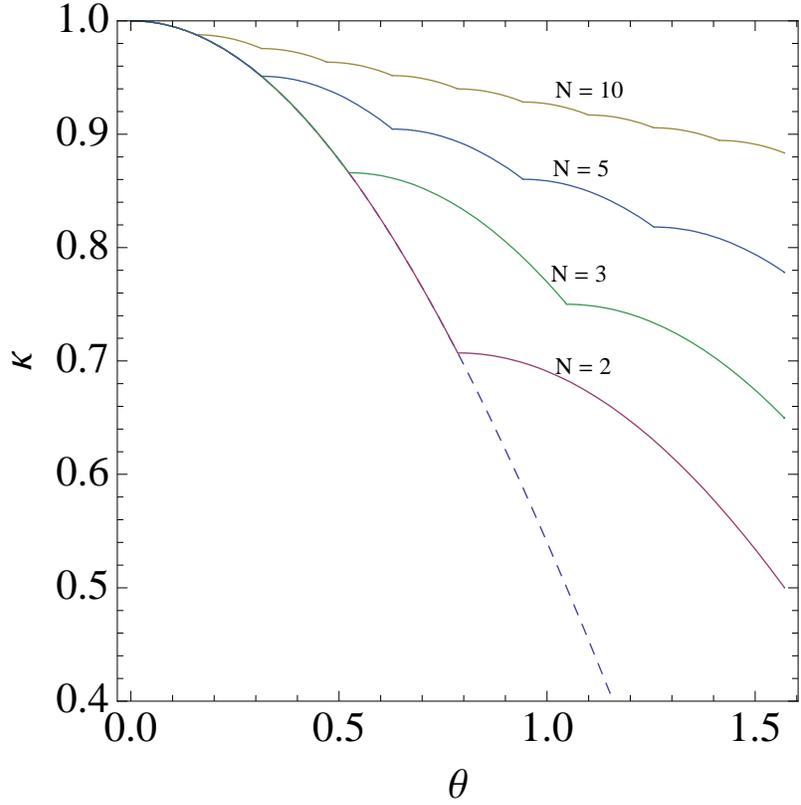}
\caption{Showing the variation of the attenuation factor $\kappa$ as a
  function of the evolution parameter $\theta$, with the evolution
  interrupted periodically once every $N^{-1}\pi/2$ in $\theta$. The
  broken curve represents the uninterrupted evolution for $0 \leq
  \theta \leq \pi/2$, total attenuation $(\kappa =0)$ being achieved
  at $\theta =\pi/2$ in this case. The other curves correspond to $N = 2,
  3, 5,$ and $10$. The Zeno tendency of $\kappa$, for large $N$, to
  become linear in $\theta$, with slope $\sim N^{-1}$ may be noted. \label{fig1}  }
\end{figure}

Two-mode evolution for duration $N^{-1} \pi/2$ followed by tracing
away of the ancilla mode results in the channel ${\cal
  C}_1(\kappa_{N,1})$ where $\kappa_{N,1} \equiv \cos
(N^{-1}\pi/2)$. Now suppose that we have interrupted evolution in the
sense that this process leading to ${\cal
  C}_1(\kappa_{N,1})$  is repeated $\ell$ times. By the semigroup
property the net result is a quantum-limited attenuator ${\cal
  C}_1(\kappa_{N,\ell})$ where $\kappa_{N,\ell} = (\kappa_{N,1})^{\ell}
= (\cos(N^{-1}\pi/2))^{\ell}$. The behaviour of the attenuation factor
$\kappa_{N,\ell}$ is depicted in Fig. {\ref{fig1}}. That the effect of
interruption is to slow down attenuation is transparent. For large $N$
we have $\kappa_{N,\ell} \approx 1- \frac{\pi^2}{4N} \left(
\frac{\ell}{N} \right)$
reminiscent of quantum Zeno effect \cite{zeno}. 

\section{Amplifier channel \, ${\cal C}_2(\kappa), \, \kappa \geq 1$ }
The two-mode  metaplectic unitary operator describing a single-mode
quantum-limited amplifier channel 
corresponds to the following  symplectic
transformation on the mode operators \cite{caruso06}\,: 
\begin{eqnarray}
S &=& ~
\left(
\begin{matrix}
\cosh{\nu} &0 & \sinh \nu & 0 \\
0& \cosh {\nu} & 0 & -\sinh{\nu} \\
\sinh \nu & 0& \cosh \nu & 0\\
0& -\sinh{\nu} & 0 & \cosh{\nu} 
\end{matrix}
\right).
\label{am1}
\end{eqnarray}
As in the earlier two cases of ${\cal D}(\kappa)$ and ${\cal
  C}_1(\kappa)$, the position and momentum variables do not mix under
the action of ${\cal C}_2(\kappa)$.  
The position variables transform as
\begin{equation}
\left ( \begin{matrix}
  {q}_{1}^{}  \\
  {q}_{2}^{} 
 \end{matrix}
\right) \rightarrow
\left ( \begin{matrix}
  {q}_{1}^{\,'}  \\
  {q}_{2}^{\,'} 
 \end{matrix}
\right) = 
M \, \left ( \begin{matrix}
  q_1  \\
  q_2 
 \end{matrix}
\right)= 
\left(
\begin{matrix} 
\cosh \nu & -\sinh \nu \\
-\sinh \nu & \cosh \nu 
\end{matrix}
\right) ~
\left(
\begin{matrix}
q_1\\
q_2
\end{matrix}
\right),
\label{am2}
\end{equation}
and the momentum variables transform according to $M^{-1}$. 
Thus the parameter $\kappa$ in ${\cal C}_2(\kappa)$ is related to the two-mode
squeeze parameter $\nu$ through $\kappa= \cosh{\nu}$. The function
$F(z_1, z_2,\eta_1, \eta_2 )$ in \eqref{ph11} is  readily computed to
be 
\begin{eqnarray}
F(z_1, z_2,\eta_1, \eta_2 )=
\kappa^{-1} \exp \left\{ \kappa^{-1} (\eta_{1}z_1 + \eta_{2} z_2)+
  (\sqrt{1-\kappa^{-2}})(\eta_1 \eta_2 - z_1 z_2)   \right\}.
\label{am4}
\end{eqnarray}
As in the earlier cases of ${\cal D}(\kappa)$ and ${\cal C}_1(\kappa)$,
the differentiation on $F(z_1, z_2,\eta_1, \eta_2 )$ can be performed to
obtain the matrix elements of the unitary operator corresponding to
the symplectic $S$ in \eqref{am1}. We obtain, after some algebra
patterned after the earlier two cases, 
\begin{eqnarray}
C_{n_1n_2}^{m_1m_2}&&= \frac{\kappa^{-1}}{\sqrt{n_1! n_2! m_1! m_2!} } ~n_1! m_2!  \nonumber \\ 
&&\times \sum_{r =  0}^{n_2} \, \sum_{j = 0}^{m_1} {}^{n_2}C_{r}
{}^{m_1}C_{j}\,(-1)^{r}\,  
(\sqrt{1- \kappa^{-2}})^{r + m_1
-j}\, (\kappa^{-1})^{n_2 +j -r}  \delta_{n_1, r+j} \, \delta_{m_2,
  n_2 + m_1 -r -j} \, .~~~
\label{am5}
\end{eqnarray}  
The Kraus operators are obtained from ${C}^{m_1 m_2}_{n_1 n_2}$ by setting $n_2=0$, and $m_2=\ell$. 
Setting  $n_2=0$ $\Rightarrow$ $r=0$, and we have 
\begin{eqnarray}
A_{\ell}(\kappa)= \kappa^{-1} \sum_{m=0}^{\infty} \sqrt{{}^{m+\ell}
  C_{\ell}} \left(\sqrt{1-\kappa^{-2}}\right)^{\ell} (\kappa^{-1})^{m}  
| m +\ell \rangle \langle m |, \,\,\,\,\, \ell =0,1,2, \cdots
\label{am6}
\end{eqnarray}
as the Kraus operators of the quantum-limited amplifier channel ${\cal
  C}_2(\kappa), 
\, \kappa >1$ \cite{carmichael10}.  

\subsection{Duality between the attenuator family ${\cal C}_1(\cdot)$ and the amplifier family ${\cal C}_2(\cdot)$ }
The Kraus operators $A_{\ell}(\kappa),\, \kappa>1$ of the amplifier channel ${\cal C}_2(\kappa)$ have an 
interesting dual relationship
to the Kraus operators $B_{\ell}(\kappa^{-1}), \,\kappa >1$ of the attenuator channel ${\cal C}_1(\kappa^{-1})$. While 
$\sum_{\ell=0}^{\infty} {A}_{\ell}^{\dagger}(\kappa)
{A}_{\ell}(\kappa) = 1\!\!1, ~\kappa >1$ and $\sum_{\ell=0}^{\infty}
{B}_{\ell}^{\dagger}(\kappa^{\,'}) {B}_{\ell}(\kappa^{\,'})= 1\!\!1, ~\kappa^{\,'}<1  
$, consistent with the trace-preserving property of ${\cal
  C}_2(\kappa)$ and ${\cal C}_1(\kappa^{\,'})$, we have 
\begin{align}
 &\sum_{\ell=0}^{\infty} {A}_{\ell}(\kappa) {A}_{\ell}^{\dagger}(\kappa) = \kappa^{-2} 1\!\!1, \nonumber \\
&\sum_{\ell=0}^{\infty} {B}_{\ell}(\kappa^{\,'})
{B}_{\ell}^{\dagger}(\kappa^{\,'}) = 
(\kappa^{\,'})^{-2} 1\!\!1.
\end{align}
Thus the (trace-preserving) families ${\cal C}_1$ and ${\cal C}_2$ are
not unital. But they are `almost unital', for the failure to be unital
is by just a scalar factor. This shows that the family $\{\kappa
A_{\ell}(\kappa)^{\dagger} ,\, \kappa >1\}$ and the family $\{
{\kappa^{\,'}}^{-1} B_{\ell}(\kappa^{\,'})^{\dagger}, \,\kappa^{\,'} <1 \}$ too
describe trace-preserving CP maps, and we may ask what these `new'
channels stand for.  

The meaning of these channels may be easily seen by considering the
adjoints $A_{\ell}(\kappa)^{\dagger},\, \kappa>1 $ of the Kraus
operators of the amplifier channel\,:   
\begin{align}A_{\ell}(\kappa)^{\dagger} &= \kappa^{-1}
  \sum_{m=0}^{\infty} \sqrt{{}^{m+\ell}C_{\ell}}
  \left(\sqrt{1-\kappa^{-2}} \right)^{\ell} \kappa^{-m} |m \rangle
  \langle m+\ell| \nonumber \\ 
 & = \kappa^{-1} B_{\ell}(\kappa^{-1}) 
\end{align}
Thus $\{ \kappa A_{\ell}(\kappa)^{\dagger} \},\, \kappa>1 $ are the
Kraus operators of the beamsplitter channel ${\cal C}_1({\kappa^{\,'}})$
with ${\kappa^{\,'}} = \kappa^{-1} <1$. Similarly it can be seen that $\{{\kappa^{\,'}}
B_{\ell}({\kappa^{\,'}})^{\dagger} \}, \, {\kappa^{\,'}}<1$ represents the
amplifier channel ${\cal C}_2(\kappa)$ with $\kappa =
({\kappa^{\,'}})^{-1}>1$.  Thus we have
\begin{theorem}
 The amplifier family ${\cal C}_2(\kappa)$ and the attenuator family ${\cal
   C}_1(\kappa^{-1}), \, \kappa>1$ are mutually dual: their Kraus operators are connected
 through the adjoint operation.
\end{theorem}
%
%

\subsection{Properties of the Kraus operators}
We now explore the properties of the Kraus operators of ${\cal
  C}_2(\kappa)$ exhibited in
Eq. (\ref{am6}), relating the action of these operators in the
Fock basis  to the defining or expected transformation property of the
quasiprobability in phase
space. We begin by establishing that 
the amplifier channel simply scales 
the $Q$ function. Then we show, through consideration of the
cumulants, that there is no fixed point for the amplifier
channel.  The
structure of the Kraus operators is invoked to comment on the fact
that the quantum-limited amplifier channel is not an entanglement breaking channel.
Finally, the semigroup structure 
of the quantum-limited amplifier channels is brought out in the Kraus representation. 

Under the action of the amplifier channel ${\cal C}_2(\kappa)$ the Weyl-ordered
characteristic function transforms as follows, and this may be
identified with the very definition of the channel\,:  
\begin{align}
\chi_W(\xi) \rightarrow \chi^{\,'}_W(\xi) = \chi_W(\kappa\,\xi) ~
\exp{[-(\kappa^2 -1 ) |\xi|^2/2]}. 
\label{am9}
\end{align}
Given a Weyl-ordered characteristic function $\chi_W(\xi)$, the corresponding 
antinormal ordered characteristic function corresponding to the $Q$-distribution is \cite{cahill692} 
\begin{align}
\chi_A(\xi) = \chi_W(\xi)~ \exp{[-|\xi|^2/2]}.
\label{am10}
\end{align}
Therefore the channel action Eq. \eqref{am9}, written in terms of $\chi_A(\xi)$, reads 
\begin{align}
\chi_A(\xi) \rightarrow \chi^{\,'}_A(\xi) =\chi_A(\kappa \,\xi).
\label{am11}
\end{align}
That is, $\chi_A(\xi)$ simply scales under the
action of the amplifier channel, a fact that should be profitably
compared with the scaling behaviour \eqref{bs9} for the attenuator
channel. Since $\chi_A(\xi)$ and the $Q$-
function form a Fourier transform pair,  the action of the amplifier
channel is fully described as a scaling transformation of the $Q$-function\,:
$Q(\alpha) \rightarrow Q^{\,'}(\alpha) = \kappa^{-2}
Q(\kappa^{-1}\alpha), \, \kappa >1$ \cite{agarwal09}. 

It is instructive to see in some detail how our Kraus operators $
A_{\ell}(\kappa)$ bring out this behaviour. Given a state 
\begin{eqnarray}
{\rho}=\sum_{n,m=0}^{\infty}|n \rangle \langle n | {\rho} |m \rangle \langle m|=
\sum_{n,m=0}^{\infty} {\rho}_{nm} |n \rangle \langle m|,
\label{am13}
\end{eqnarray} 
its corresponding $Q$ function is \cite{cahill692}  
\begin{eqnarray}
Q_{\rho}(\alpha) = \langle \alpha | {\rho}| \alpha \rangle=
\exp[-{|\alpha|}^2]\sum_{n,m=0}^{\infty} \frac{({\alpha}^*)^n}{\sqrt{n!}}
\frac{(\alpha)^m}{\sqrt{m!}} {\rho}_{nm}.
\label{am14}
\end{eqnarray}
To see the action of the linear map ${\cal C}_2(\kappa)$ on an arbitrary ${\rho}$, it is sufficient
to exhibit its action on the operators $|n \rangle \langle m|$, 
for all $n, m \geq 0$. We have
\begin{eqnarray}
&&{|n \rangle \langle m|} \rightarrow \sum_{\ell=0}^{\infty} {A}_{\ell}(\kappa) |n \rangle \langle m|
{A}_{\ell}^{\dagger}(\kappa) \nonumber \\ 
&&={\kappa}^{-2} \frac{{(\kappa)}^{-(n+m)}}{\sqrt{n!m!}} \sum_{\ell=0}^{\infty}\frac{ (1-\kappa^{-2})^{\ell}}{{\ell}!}
{\sqrt{(n+{\ell})!}} {\sqrt{(m+{\ell})!}}|n+{\ell}\rangle \langle m+{\ell} |.
\label{am15}
\end{eqnarray}
Thus, under the action of the channel ${\cal C}_2(\kappa)$, ${\rho}$ goes to
\begin{eqnarray}
{\rho}^{\,'}= {\kappa}^{-2}\sum_{n,m=0}^{\infty}{\rho}_{nm}
\frac{{\kappa}^{-(n+m)}}{\sqrt{n!m!}} ~\sum_{{\ell}=0}^{\infty}\frac{(1-\kappa^{-2})^{{\ell}}}{{\ell}!}
{\sqrt{(n+{\ell})!}} {\sqrt{(m+{\ell})!}}\,|n+{\ell}\rangle \langle m+{\ell} |.
\label{am16}
\end{eqnarray}
The $Q$ function of the resultant or output state ${\rho}^{\,'}$ is 
\begin{eqnarray}
\langle \alpha|{\rho}^{\,'}|\alpha \rangle &=&
{\kappa}^{-2} \exp[-{|\alpha|}^2]\sum_{n,m=0}^{\infty}{\rho}_{nm}\frac{{\kappa}^{-(n+m)}}{\sqrt{n!m!}}
({\alpha}^*)^n (\alpha)^m \left(\sum_{{\ell}=0}^{\infty} 
\frac{(1-\kappa^{-2})^{}\ell}{{\ell}!}{|\alpha|}^{2{\ell}}\right) \nonumber \\
&=& \kappa^{-2} \exp[-{|\kappa^{-1} \alpha|}^2]\sum_{n,m=0}^{\infty}\frac{(\kappa^{-1}{\alpha}^*)^n}
{\sqrt{n!}}\frac{(\kappa^{-1}{\alpha})^m}{\sqrt{m!}} {\rho}_{nm} \nonumber\\
&=& \kappa^{-2} Q(\kappa^{-1}\alpha).
\label{am17}
\end{eqnarray}
We thus conclude
\begin{theorem}
The scaling  $Q_{\rho}(\alpha)$ $\rightarrow$ 
${Q}_{\rho^{\,'}}(\alpha)= \kappa^{-2} Q_{\rho} (\kappa^{-1} \alpha)$, $0< \kappa^{-1} < 1$, is a completely positive
map whose Kraus decomposition is given by  $\{ A_{{\ell}}(\kappa) \}$.
\label{amth1}
\end{theorem}
\noindent
This result may be compared with Theorem 6 for the ${\cal C}_1(\cdot)$
family of channels.

The amplifier channel has the following property in respect of 
nonclassicality of the output states \,:  
\begin{corollary}
The amplifier channel cannot generate nonclassicality.
\label{ampcol1}
\end{corollary}
\noindent
{\em Proof}\,: By Eq. (\ref{am9}), the normal ordered characteristic
function transforms as follows
\begin{align}
{\cal C}_2(\kappa)\,: \chi_N(\xi) \rightarrow \chi^{\,'}_N(\xi)=
\chi_W(\kappa\xi)  \exp{[-(\kappa^2 
  -2)   |\xi|^2/2]}. 
\label{am18}
\end{align}
This may be rewritten in the suggestive form 
\begin{align}
 \chi_N(\xi) \rightarrow \chi^{\,'}_N(\xi) = \chi_N(\kappa\xi) \exp [-(\kappa^2-1)|\xi|^2]. 
\end{align}
Fourier transforming, we see that the diagonal weight
$\phi^{\,}(\alpha)$ of the output state is the convolution of the
(scaled) input diagonal weight with a Gaussian (corresponding to the
last factor), and hence it is pointwise nonnegative whenever the input
diagonal weight $\phi(\alpha)$ is pointwise nonnegative.  \\

\noindent
{\bf Remark}\,: We are not claiming that the amplifier channel cannot
destroy nonclassicality [compare the structure of Corollary 2 with
that of Corollary 1 following Theorem 6]. Indeed, it is easy to show that
nonclassicality of every Gaussian state will be destroyed by any
${\cal C}_2(\kappa)$ with $\kappa \geq \sqrt{2}$ \cite{allegra10,
  carmichael10, agarwal09,lutkenhaus95}. It is also 
easy to show that there are states whose nonclassicality will survive
${\cal C}_2(\kappa)$ even for arbitrarily large $\kappa$
\cite{lutkenhaus95, carmichael10, agarwal09}. To see this, note first of all, that any state
$\rho$ whose $Q$-function $Q(\alpha) = \langle \alpha | \rho | \alpha
\rangle$ vanishes for some $\alpha$ is necessarily nonclassical. The
assertion simply follows from the fact that under the scaling
$Q(\alpha) \rightarrow \kappa^{-2} Q(\kappa^{-1}\alpha)$ a  zero
$\alpha_0$ of $Q(\alpha)$ goes to a zero at $\kappa\alpha_0$.

%

\subsection{Fixed points}
By Eq. (\ref{am9}) we have, under the action of the channel ${\cal C}
_2(\kappa)$, the following behaviour for the moment generating function\,:
\begin{eqnarray}
\Gamma(\xi) \rightarrow \Gamma^{\,'}(\xi) =\text{log }[\chi_W( \kappa\xi)] \, -\frac{1}{2} |\xi|^2 (\kappa^2-1).
\end{eqnarray}
The cumulants of order $>2$ of the output characteristic function are
\begin{eqnarray}
\gamma^{\,'}_{m_1 m_2}&=&  (\kappa)^{m_1+m_2} \left( \frac{\partial}{\partial(it_1)} \right)^{m_1}  
\left( \frac{\partial}{\partial(it_2)}
\right)^{m_2}   \, \text{log }[\chi_W(t)] , ~~~(\kappa\xi=t)\nonumber \\
&=&  (\kappa)^{m_1+m_2} \,\gamma_{m_1m_2},
\end{eqnarray}
where $\gamma_{m_1m_2}$ are the cumulants of the input state. 
Thus, for any non-Gaussian input state, the
higher order cumulants  grow monotonically with repeated use of the channel.
Thus, leaving out the case $\kappa=1$ which corresponds to the identity
channel, there is no non-Gaussian state that is 
preserved. To see if there is any fixed point among the Gaussian
states, it is sufficient to consider only thermal states as input 
\eqref{phc28a}. By Eq. (\ref{am15}), the output state is 
\begin{align}
{\rho}_{\text{out}}=(1-x) \kappa^{-2} \sum_{n=0}^{\infty} \kappa^{-2n}(\kappa^2-1 + x)^n
|n \rangle \langle n|,
\label{am21}
\end{align}
where we used the binomial expansion to perform a sum. 
Comparing Eqs. (\ref{phc28a}) and (\ref{am21}), we have 
\begin{align}
{\cal C}_2(\kappa)\,: x \rightarrow x^{\,'}=\kappa^{-2}(\kappa^2-1 +
x) \text{ or  } a_0 \rightarrow a_0^{\,'} =\kappa^2 a_0 + \kappa^2-1.   
\label{am22}
\end{align}
It is clear that the output thermal parameter $a_0^{\,'}$ is strictly greater than
$a_0$. Hence there is no thermal state that is left invariant under the action
of the amplifier channel. Collecting these facts together, we conclude 
\begin{theorem}
There exists no state which is a fixed point of the quantum-limited amplifier
channel. 
\end{theorem} 

As a simple illustration, consider the action of the channel
on a Fock state. We have by Eq. (\ref{am15})    
\begin{align}
{\cal C}_2(\kappa)\,:\, |n \rangle \langle n| \rightarrow
\kappa^{-2} \sum_{\ell=0}^{\infty} {}^{n+\ell} C_{\ell}\,  
(1-\kappa^{-2})^{\ell} \kappa^{-2n} |n+\ell \rangle \langle n+\ell|. 
\label{am20}
\end{align}
Thus an input Fock state $| n \rangle \langle n|$ 
is taken to a convex combination of all Fock
states with photon number {\em greater than or equal} to $n$. It may be noted
that this behaviour is complementary to that of the
beamsplitter channel where the output was a convex combination of all
Fock states {\em upto} $|n\rangle \langle n|$. And we find, analogous
to the beamsplitter case, 
that any state which is in the support of a finite number of Fock states
 is not preserved by the channel.

As a second example, consider the phase averaged coherent state
\eqref{phc32a} as the input. By 
Eq. (\ref{am20}) the output is
\begin{align}
{\rho}_{\text{out}} = \sum_{j,\ell} {}^{j+\ell}
C_{\ell}\,\frac{\lambda^j e^{-\lambda}}{j!} 
(1- \kappa^{-2})^{\ell}
\kappa^{-2j}  |j+\ell \rangle \langle j+\ell|.
\label{am23}
\end{align} 
It is clear that we cannot solve consistently for parameters
$\kappa>1$ and $\lambda$ such that the output PND is also
Poissonian. In view of Corollary \ref{ampcol1}, we may conclude that
an input Poissonian PND generically results in a super-Poissonian PND.\\

\noindent 
{\bf Remark on entanglement breaking}\,: It is well known that the
quantum-limited 
amplifier channel is not entanglement breaking \cite{holevo08ebt}. It may
be pointed out in passing that this fact follows also from the structure of
our Kraus operators $\{A_{\ell}(\kappa) \}$. Since these operators coincide with
the transpose of the beamsplitter channel Kraus operators $\{B_{\ell}(\kappa^{-1}) \}$,
apart from a $\ell$-independent multiplicative factor, there exists
{\em no
finite rank operator} in the support of the set of operators $\{A_{\ell}(\kappa)
\}$. In particular, there are no rank one operators in the support of
$\{A_{\ell}(\kappa) \}$. Hence, ${\cal C}_2(\kappa)$ is not an
entanglement breaking channel.

\subsection{Semigroup property}

It follows from the very definition of the amplifier channel that the
composition of two quantum-limited amplifier channels
 with parameters $\kappa_1$ and $\kappa_2$ is also such an amplifier channel with
 parameter $\kappa_1\kappa_2 >1$\,:
\begin{align}
{\cal C}_2(\kappa_2) \circ {\cal C}_2(\kappa_1): \, \chi_W(\xi) \rightarrow \chi^{\,'}_W(\xi) =\chi_W(\kappa_1 \kappa_2 \,\xi) ~
\exp{[-(\kappa_{1}^{2} \kappa_{2}^{2}  -1 ) |\xi|^2/2]}.
\end{align}
That is, 
\begin{align}
{\cal C}_2(\kappa_2) \circ {\cal C}_2(\kappa_1) = {\cal C}_2(\kappa_1
\kappa_2) = {\cal C}_2(\kappa_1) \circ {\cal C}_2(\kappa_2). 
\label{am24}
\end{align}
It will be instructive to examine how this fact emerges from the structure of
the Kraus operators. Let the set 
$\{ A_{\ell_1}(\kappa_1)\}$ be the Kraus operators of the first amplifier and
let $\{ A_{\ell_2}(\kappa_2)\}$ be that of the second. Then the product of
a pair of Kraus operators, one from each set, is  
\begin{align}
A_{\ell_1}(\kappa_1) A_{\ell_2}(\kappa_2) = (\kappa_1 \kappa_2)^{-1} \sqrt{{}^{\ell_1+\ell_2}C_{\ell_1}}
&\sum_{n=0}^{\infty} \sqrt{{}^{n+\ell_1+\ell_2}C_{\ell_1+\ell_2}}
\left(\sqrt{1-\kappa_1^{-2}}\right)^{\ell_1}
\left(\sqrt{1-\kappa_2^{-2}} \right)^{\ell_2} \nonumber \\ 
~~~~~~&~\times (\kappa_1\kappa_2)^{-n} \kappa_1^{-\ell_2} |n+\ell_1+\ell_2 \rangle \langle n|. 
\label{am26}
\end{align}  
Thus, under the successive action of these two amplifier channels the operator
$|j \rangle \langle j+\delta|$ goes to
\begin{align}
&\sum_{{\ell}_1,{\ell}_2} A_{{\ell}_1}(\kappa_1) A_{{\ell}_2}(\kappa_2) |j \rangle \langle j+\delta|
A_{{\ell}_2}(\kappa_2)^{\dagger} A_{{\ell}_1}(\kappa_1)^{\dagger} \nonumber \\
&=(\kappa_1 \kappa_2)^{-2} \sum_{{\ell}_1,{\ell}_2}\sum_{n=0}^{\infty}\sum_{m=0}^{\infty} 
{}^{{\ell}_1+{\ell}_2}C_{{\ell}_1}\,  \left({1-\kappa_1^{-2}}\right)^{\ell_1} \left({ 1-\kappa_2^{-2}}\right)^{\ell_2} (\kappa_1\kappa_2)^{-(n+m)}
\kappa_1^{-2{\ell}_2} \nonumber \\ 
& ~~~~~~~~\times
\,\sqrt{{}^{n+{\ell}_1+{\ell}_2}C_{{\ell}_1+\ell_2}\,{}^{m+{\ell}_1+{\ell}_2}C_{{\ell}_1+\ell_2}} 
|n+{\ell}_1+{\ell}_2 \rangle \langle n| j \rangle \langle j+\delta| m \rangle
\langle m +{\ell}_1 +{\ell}_2|.  
\end{align}
Denoting ${\ell}_1+{\ell}_2 = {\ell}$, the right hand side of the above expression
reduces to  
\begin{align}
&(\kappa_1 \kappa_2)^{-2} \,  \sum_{{\ell}=0}^{\infty} \sum_{{\ell}_1=0}^{{\ell}}
\sum_{n=0}^{\infty}\sum_{m=0}^{\infty} \,
{}^{{\ell}}C_{{\ell}_1}  \, (1-\kappa^{-2}_1)^{{\ell}_1} \, 
(\kappa_1^{-2} (1-\kappa_2^{-2}))^{({\ell}- {\ell}_1)}
 \, (\kappa_1\kappa_2)^{-(n+m)} \nonumber \\ 
 &~~~~\times \sqrt{{}^{n+{\ell}}C_{{\ell}}\, {}^{m+{\ell}}C_{{\ell}}} \,
\, \delta_{m,j+\delta} \, \delta_{n,j}\,
|n+{\ell}\rangle \langle n+\delta +{\ell}| 
\label{am26a}
\end{align}
The summation over the index $\ell_1$ is a binomial expansion \,:
\begin{align*}
\sum_{\ell_1=0}^{\ell} \, {}^{{\ell}}C_{{\ell}_1}\,
(1-\kappa^{-2}_1)^{{\ell}_1} \, (\kappa_1^{-2}
(1-\kappa_2^{-2}))^{({\ell}- {\ell}_1)} = (1-\kappa_1^{-2}\kappa_2^{-2})^{\ell}.
\end{align*}
Thus the expression in \eqref{am26a} reduces to 
\begin{align}
 (\kappa_1 \kappa_2)^{-2}  \sum_{{\ell}=0}^{\infty} \,
(1-\kappa_1^{-2}\kappa_2^{-2} )^{\ell} \,
(\kappa_1\kappa_2)^{-(j+j+\delta)} \,
\sqrt{{}^{j+\ell}C_{{\ell}}\,{}^{j+\ell+\delta}C_{{\ell}}} \, |j+{\ell}
\rangle \langle j+{\ell}+\delta|.
\label{am27}
\end{align}
Comparing Eqs. \eqref{am15} and \eqref{am27}, we see that the latter
is the Kraus representation for a single quantum-limited amplifier
channel. That is, 
\begin{align}
\sum_{{\ell}_1,{\ell}_2} A_{{\ell}_1}(\kappa_1) A_{{\ell}_2}(\kappa_2) |j \rangle \langle j+\delta|
A_{{\ell}_2}(\kappa_2)^{\dagger} A_{{\ell}_1}(\kappa_1)^{\dagger} 
=  \sum_{\ell}  A_{\ell}(\kappa_1\kappa_2)|j \rangle \langle j+\delta|
A_{{\ell}}(\kappa_1\kappa_2)^{\dagger}.
\label{am28}
\end{align}
A similar behaviour holds for $|j+\delta \rangle \langle j|$ as well. 
And this is what we set out to demonstrate. \\

\noindent
{\bf Remark on interrupted evolution and Zeno-like effect}\,:\\
As seen from (\ref{am1}) the parameter $\kappa$ of ${\cal
  C}_2(\kappa)$ equals $\cosh{\mu}$, where $\mu$ is a measure of the
two-mode squeezing effected with the help of an ancilla initially in
the vacuum state. The relevant two-mode unitary squeeze operator
$U(\mu) = \exp [-\mu ( a^{\dagger}b^{\dagger} - ab)]$ can be viewed as
evolution for a duration $\mu$ under the Hamiltonian $-i(
a^{\dagger}b^{\dagger} - ab)$, the amplification factor $\kappa = \cosh{\mu}$ increasing
monotonically with increasing $\mu$.

\begin{figure}
\includegraphics{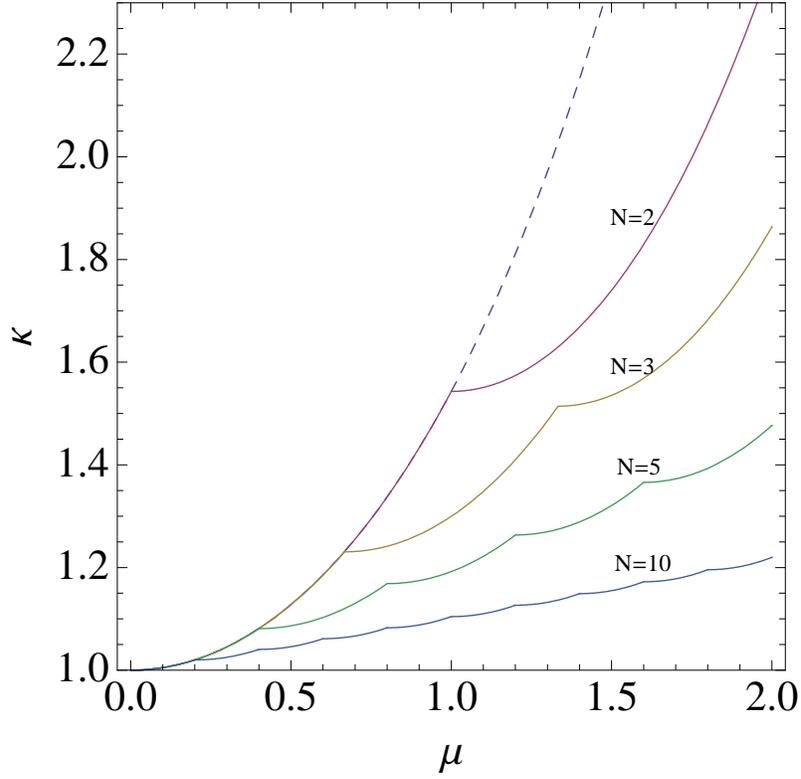}
\caption{Showing the evolution of the amplification factor $\kappa$ as
  a function of the parameter $\mu$, with the evolution interrupted
  every $2N^{-1}$ in $\mu$. The broken curve represents uninterrupted
  evolution for $0 \leq \mu \leq 2$. The other curves correspond to $N
  = 2, 3, 5$ and $10$, the total duration adding to 2 for each $N$. The Zeno tendency of $\kappa$, for large $N$, to
  become linear in $\mu$ with slope $\sim N^{-1}$ should be noted. \label{fig2}}
\end{figure} 

For convenience let us fix the total duration in $\mu$ to some value,
say 2. In place of this single evolution, let us consider a sequence
of $N$ interrupted evolutions, by tracing away the ancilla after every
$2N^{-1}$ duration in $\mu$, the total duration adding upto 2. The
semigroup property of these interrupted evolutions after a total
duration of 2 in $\mu$ will be a quantum-limited amplifier ${\cal
  C}_2(\kappa)$, with $\kappa = [\cosh{(2/N)}]^N$, which should be
compared with $\kappa = \cosh{2}$. The effect of interruption in slowing down
amplification is clear. The behaviour of this interrupted
amplification is shown in Fig. {\ref{fig2}}. After $\ell$ such
evolutions the amplification factor will be $\kappa =
(\cosh{2/N})^{\ell}$, which for large $N$ has the behaviour $\kappa
\approx 1 + \frac{2}{N} \left( \frac{\ell}{N} \right)$, reminiscent of
quantum Zeno effect \cite{zeno} as in the case of quantum-limited
attenuation in Section IV.

\section{The Singular case ${\cal A}_2$} 
We now consider briefly ${\cal A}_2$, the last of the quantum 
limited Bosonic Gaussian channels. The two-mode metaplectic unitary operator
representing ${\cal A}_2$ produces a symplectic transformation on the
quadrature variables which does not mix the position variables with
the momentum variables \cite{caruso06}\,:
\begin{eqnarray}
\left ( \begin{matrix}
  {q}_{1}^{}  \\
  {q}_{2}^{} 
 \end{matrix}
\right)
&\rightarrow&
\left ( \begin{matrix}
  {q}_{1}^{\,'}  \\
  {q}_{2}^{\,'} 
 \end{matrix}
\right) = 
M \, \left( \begin{matrix}
  q_1  \\
  q_2 
 \end{matrix}
\right), \nonumber \\
\left ( \begin{matrix}
  {p}_{1}^{}  \\
  {p}_{2}^{} 
 \end{matrix}
\right)
&\rightarrow&
\left ( \begin{matrix}
  {p}_{1}^{\,'}  \\
  {p}_{2}^{\,'} 
 \end{matrix}
\right) = 
(M^{-1})^T \, \left( \begin{matrix}
  p_1  \\
  p_2 
 \end{matrix}
\right), \nonumber \\
M &=& \left(
\begin{matrix} 
0 &1 \\
1 &-1 
\end{matrix}
\right).
\label{sing2}
\end{eqnarray}
Therefore, our general scheme applies to this case as well. Unlike in
the earlier cases of ${\cal D}(\kappa)$, ${\cal C}_1(\kappa)$, and
${\cal C}_2(\kappa)$, in the present case 
it turns out to be more convenient to evaluate the matrix 
elements of $U^{(ab)}$ in a mixed basis\,: 
\begin{align}\label{sing3}
C_{n_1n_2}^{m_1q} &=  \langle m_1| \langle  q | U^{(ab)} |
n_1 \rangle |n_2 \rangle. 
\end{align}
Here $|q \rangle$ labels the position basis of the ancilla mode.  
With this mixed choice, the Kraus operators are labelled by a continuous index
`$q$', and are given by
\begin{align}
V_q = \langle q|U^{(ab)}|0\rangle = \sum_{m_1,n_1} C_{n_1
  0}^{m_1q} | m_1 \rangle \langle n_1|,
\label{sing4}
\end{align}
where 
\begin{align}\label{sing5}
C_{n_1 0}^{m_1q} &= \int dq_1 \langle m_1| q_1 \rangle \langle q_1|
\langle q | U^{(ab)} | n_1 \rangle |0\rangle  \nonumber \\
&=  \int dq_1 \langle m_1| q_1 \rangle \langle q,q_1-q| n_1,0\rangle.
\end{align} 
Here we have used, as in the earlier cases, the action of the unitary operator in the
position eigenstates of the two-mode system. Employing the position space wavefunctions of the
Fock states, we have
\begin{align}\label{sing6}
C_{n_1 0}^{m_1q} = \frac{\pi^{-3/4}}{\sqrt{2^{n_1+m_1} n_1!m_1!}}
 H_{n_1}(q) e^{-\frac{q^2}{2}} \int dq_1 H_{m_1}(q_1) e^{-\frac{(q_1-q)^2}{2}-\frac{{q}_{1}^{2}}{2}}. 
\end{align}  
The above integral is easily evaluated \cite{gradstein2}, and we have  
\begin{align}\label{sing7}
C_{n_1 0}^{m_1q} &=  \frac{\pi^{-1/4}}{\sqrt{2^{n_1+m_1} n_1!m_1!}}\,
  q^{m_1} H_{n_1}(q) \exp[-3q^2/4]  \nonumber \\
 &= \langle m_1 | q/\sqrt{2}) \langle q | n_1
   \rangle, 
\end{align} 
where $ | q/\sqrt{2} )$ is the coherent state
$|\alpha \rangle$ for $\alpha =q/\sqrt{2}$, and the purpose of
the round bracket being to distinguish the same from the position
eigenket $| q/\sqrt{2}\rangle$. With this notation
the Kraus operators are   
\begin{align}\label{sing8}
V_q =  | q/\sqrt{2} ) \, \langle q |.
\end{align} 
That the trace-preserving condition on the Kraus operators is
satisfied emerges from the fact that the position kets are complete\,:
$\int dq\, 
V_q^{\dagger} V_q = \int dq |q \rangle \langle q| = 1\!\!1$ . 

To connect these Kraus operators $V_q$ to the action of
the channel in the phase space picture, we examine the behaviour of an arbitrary 
pure state $|\psi\rangle$ under passage through the channel. We have  
\begin{align}\label{sing9}
{\cal A}_2\,: \rho = |\psi\rangle \langle \psi|  \rightarrow \rho^{\,'}&= \int dq ~  |
  q/\sqrt{2} ) \,  \langle q | 
\psi\rangle  \langle \psi|q \rangle \,   ( q/\sqrt{2} |   \nonumber \\
&= \int dq ~ |\psi(q)|^2   \,  | q/\sqrt{2} )\,  (q/\sqrt{2} |  \nonumber \\
&= \int dq\, dp ~ |\psi(q)|^2 \delta(p) \,  | [q+ip]/\sqrt{2} )~  (
  [q+ip]/\sqrt{2} |  .
\end{align} 
The last expression is already in the `diagonal' form in the coherent
states basis, with $
|\psi(q)|^2 \delta(p)$, $\alpha = (q+ip)/\sqrt{2}$ forming the diagonal weight function
$\phi(\alpha)$. It follows by convexity that for an arbitrary input state $\rho$
the output of the channel is given by 
\begin{align}
\rho^{\,'} = \pi^{-1}\int d^2\alpha \, \phi(\alpha) \, |\alpha \rangle \langle
\alpha|, ~~ \phi(\alpha) = \langle q|\rho|q\rangle \, \delta(p).
\label{sing9a}
\end{align} 
It is seen
that this transformation is the same as $ \chi_W({\xi})
\rightarrow \chi_W \left( \frac{(1\!\!1 + \sigma_3)}{2}  ~ \xi \right)
\exp[-|\xi|^2/2]$, the expected behaviour of the characteristic
function under passage through ${\cal A}_2$ \cite{giovannetti10}.  

The above results can be alternatively understood through the action of the
channel in the Fock basis. Under passage through the channel,
\begin{align}\label{sing10}
|n\rangle\langle m| &\rightarrow \int dq V_q |n\rangle\langle m|
V_q^{\dagger} \nonumber \\
&= \int dq \,  | q/\sqrt{2} )\,  \langle q |
n\rangle\langle m|   q \rangle \, ( q/\sqrt{2} |
\nonumber \\
&= \int dq \, \frac{\pi^{-1/2}}{\sqrt{2^{n+m} n! m!}} \, H_n(q) H_m(q) \, 
e^{-q^2}  | q/\sqrt{2} ) \, ( q/\sqrt{2}|, 
\end{align}
for all $n,m$. The outcome for an arbitrary input state $\rho$ follows
by linearity, and we have  
\begin{theorem}
The channel ${\cal A}_2$ is both nonclassicality breaking and
entanglement breaking. 
\end{theorem}
\noindent
{\em Proof}\,: 
We note from Eq. \eqref{sing8} that the Kraus operators are already
in rank one form, thereby
showing that the channel is entanglement breaking. And from
Eq. \eqref{sing9a} we see that the output of the channel, for every
input state $\rho$, supports a diagonal representation with
nonnegative weight $\langle q|\rho|q\rangle \, \delta(p) \geq 0$, for
all $\alpha = (q+ip)/\sqrt{2}$, showing
that the output is classical for all input states. \\

\noindent
{\bf Remark on fixed points}\,: 
It can be seen directly from the action of the channel on the
characteristic function that the momentum variable of the output is
set to zero,  and
then multiplied by a Gaussian in that variable. Thus any  cumulant of
order $>2$ in this variable is 
 set to zero by the channel action. Moreover it is easily seen that
the  second moments  are not preserved, thus showing that there is
no state that is invariant under the action of this channel.

\section{Single Quadrature classical noise  channel ${\cal B}_1(a), \,a\geq 0$}
The channel ${\cal B}_1(a)$, whose action is to simply inject Gaussian
noise of magnitude $a$ into one quadrature of the oscillator, is
neither quantum-limited nor extremal. It can be realized in the form 
\begin{align}
{\cal B}_1(a) \,: ~ \rho \rightarrow \rho^{\,'} = \frac{1}{\sqrt{\pi
    a}} \int dq \, \exp [-q^2/a] \, D(q/\sqrt{2})\, \rho \, D(q/\sqrt{2})^{\dagger},
\label{one1}
\end{align}
where $D(\alpha)$'s are the unitary displacement operators. ${\cal
  B}_1(a)$ is thus a case of the so-called random unitary channels
\cite{kraus}, a convex sum of unitary channels. The
continuum \begin{equation} Z_q \equiv (\pi
a)^{-1/4} \exp[-^2/2a]\, D(q/\sqrt{2}) \label{one1a}\end{equation} are the Kraus
operators of this realization. The quantum-limited end of ${\cal
  B}_1(a)$ is obviously the identity channel, corresponding to $a
\rightarrow 0$ [$\lim_{a \rightarrow 0} \sqrt{\pi a}^{-1} \exp
[-q^2/a]= \delta(q),$ and $Z_{q=0}=$ identity]. One may assume $a=1$ without loss of
generality. 

The reason we present a brief treatment of this channel here is just
to demonstrate that this case too subjects itself to our general
scheme presented in Section II. A further reason why we treat this
noisy channel here is this. In Section IX we shall treat every
noisy channel as the composite of two quantum-limited channels, the
case of ${\cal B}_1(a)$ constituting the only exception wherein this
cannot be done. 

The two-mode metaplectic unitary operator
representing ${\cal B}_1$ produces a symplectic transformation on the
quadrature variables which, as in the earlier cases of ${\cal
  D}(\kappa), \, {\cal C}_1(\kappa), \, {\cal C}_2(\kappa)$, and
${\cal A}_2$, does not mix the position variables with
the momentum variables \cite{caruso06}\,:
\begin{eqnarray}
\left ( \begin{matrix}
  {q}_{1}^{}  \\
  {q}_{2}^{} 
 \end{matrix}
\right)
&\rightarrow&
\left ( \begin{matrix}
  {q}_{1}^{\,'}  \\
  {q}_{2}^{\,'} 
 \end{matrix}
\right) = 
M \, \left( \begin{matrix}
  q_1  \\
  q_2 
 \end{matrix}
\right), \nonumber \\
M &=& \left(
\begin{matrix} 
1 &-1 \\
0 &1 
\end{matrix}
\right).
\label{sing13}
\end{eqnarray}
And $p_1, p_2$ transform according to $(M^{-1})^T$. 

As in the immediate previous case ${\cal A}_2$, the matrix
elements of $U^{(ab)}$ are 
\begin{align}\label{sing14}
C_{n_1n_2}^{m_1q} &=  \langle m_1| \langle  q | U^{(ab)} |
n_1 \rangle |n_2 \rangle ,
\end{align}
where $|q \rangle$'s are the position eigenvectors. In view of this the
Kraus operators are labelled by a continuous index 
`$q$' and are given by
\begin{align}\label{sing15}
 \langle q|U^{(ab)}|0\rangle = \sum_{m_1,n_1} C_{n_1
  0}^{m_1q} | m_1 \rangle \langle n_1|,
\end{align}
where 
\begin{align}\label{sing16}
C_{n_1 0}^{m_1q} &= \int dq_1 \langle m_1| q_1 \rangle \langle q_1|
\langle q | U^{(ab)} | n_1 \rangle |0\rangle  \nonumber \\
&=  \int dq_1 \langle m_1| q_1 \rangle \langle q_1-q,q| n_1,0\rangle.
\end{align} 
Here we made the two-mode metaplectic unitary operator act on the
position basis.

\noindent
To evaluate the Kraus operator, it is sufficient to evaluate
the matrix elements
\begin{align}
{C}^{n_1 q}_{m_1 0} = \frac{\pi^{-3/4}}{\sqrt{2^{n_1+m_1} n_1!m_1!}}
 e^{-\frac{q^2}{2}} \int dq_1 H_{n_1}(q_1-q)  H_{m_1}(q_1)
 e^{-\frac{q_1^2}{2}} e^{-\frac{(q_1-q)^2}{2}}.
\label{sing17}
\end{align}
The above integral can be readily performed \cite{gradstein3}, and we obtain
\begin{align}\label{sing18}
{C}^{n_1 q}_{m_1 0} &= \pi^{-1/4} e^{-\frac{q^2}{2}} \left[
  e^{-\frac{q^2}{4}} \sqrt{\frac{m_1!}{n_1!}} \left(
    \frac{-q}{\sqrt{2}}\right)^{n_1-m_1} L^{n_1-m_1}_{m_1}(q^2/2)
\right] \nonumber \\
&\equiv  \pi^{-1/4} e^{-\frac{q^2}{2}}  \langle m_1 | D(q/\sqrt{2})
|n_1 \rangle = \, Z_q. 
\end{align}
We have thus recovered \eqref{one1a}, but staying entirely within our
general scheme.

\section{Extremal Gaussian channels}
It follows from the very identification of channels with
trace-preserving CP maps that channels form a convex set, and since a
convex set is fully characterized by its extremal elements, it is of interest
to know if there are any extremal elements among the quantum-limited Bosonic Gaussian
channels. The present Section is devoted to this issue.

A convenient characterization of extremality in terms of Kraus
representation is due to Choi \cite{choimap75}\,: A trace-preserving CP
map $\Omega$ is extremal if and only if $\Omega$ supports a Kraus
representation ${\rho} \rightarrow {\rho}^{\,'} = \sum_{n}
W_{n}\,{\rho}\, W_{n}^{\dagger}$,\, $\sum_{n}W_{n}^{\dagger}\,
W_{n}=1\!\!1$, such that the set of operators $\{ W_{m}^{\dagger}\,
W_{n}\}$, with $m$ and $n$ independently running over their range, are
linearly independent. While Choi's original result was formulated in
the finite dimensional case, it is to be expected that this result
will generalize to the infinite-dimensional case with suitable
technical regularization. That this is the case is indicated, for
instance, by Theorem 2.4 of the work of Tsui \cite{tsui96}. Since we
are dealing with concrete physical situations, we may assume in what
follows that such technical requirements are indeed satisfied.

With these preliminary remarks in place, we are now ready to test the
quantum-limited Gaussian channels for extremality. We begin with the
transpose or phase conjugation channel.

\subsection{The phase conjugation channel ${\cal D}(\kappa)$}
The Kraus operators $\{T_{m}(\kappa) \}$ for ${\cal D}(\kappa)$ have
been presented in (\ref{phc12}). We may readily express the products 
$\{T_{m}^{\dagger}(\kappa)\,T_{n}(\kappa) \}, \, 0 \leq m,n < \infty$
in the convenient form  
\begin{eqnarray}
T_{\ell + \delta}^{\dagger}(\kappa)\,T_{\ell}(\kappa)&=& \sum_{j=0}^{\infty}
d(\delta)_{\ell j}\,|j+\delta \rangle \langle j|, \nonumber \\
T_{\ell}^{\dagger}(\kappa) T_{\ell + \delta}(\kappa)&=&
\sum_{j=0}^{\infty} d(\delta)_{\ell j}\,|j \rangle \langle j+ \delta|,\nonumber \\
d(\delta)_{\ell j}&=&
\sqrt{{}^{\ell}C_j\,{}^{\ell+\delta}C_{j+\delta}} ~
[1+\kappa^2]^{-j-\delta/2-1} \, 
[(1+\kappa^{-2})^{-1}]^{\ell-j} \,\,\,\, {\rm for}\,\,\, j \leq
\ell \nonumber \\
&=& 0\,\,\,\,{\rm for}\,\,\, j > \ell\,. 
\label{epc}
\end{eqnarray}
The two cases involving $|\ell + \delta \rangle \langle
\ell|$ and $|\ell \rangle \langle \ell + \delta|,\, \delta = 0, 1, 2,
\cdots  $ correspond respectively to $m-n \geq 0$ and $m-n \leq
0$. It is clear from \eqref{epc} 
that the set $\{T_{m}^{\dagger}(\kappa)\,T_{n}(\kappa) \}, \, 0 \leq m,n < \infty$ breaks into
nonintersecting subsets labelled by $m-n=0,\, \pm1,\, \pm2 \cdots$. The
index $\ell = {\rm min}\{m,n \}$ labels the $ T_{m}^{\dagger}
(\kappa)\, T_{n}(\kappa)$'s within a subset. That subsets corresponding
to two different values of $m-n$ are mutually orthogonal is
manifest. Thus, we are left to examine only linear independence within each
subset, and this is easily accomplished as follows.

For a given value of $m-n$ (i.e. fixed value of $\delta$) arrange the
$c$-number coefficients $d(\delta)_{\ell j}$ into a
matrix $d(\delta)=((d(\delta)_{\ell j}))$. This matrix is lower
diagonal, and its diagonals are nonzero [indeed, all the entries on
  and below the diagonal are non zero]. Thus $d(\delta)$ is
nonsingular. Therefore the linear independence of
$T_{\delta}^{\dagger}(\kappa)T_{0}(\kappa)$,
$T_{1+ \delta}^{\dagger}(\kappa)T_{1}(\kappa)$,
$T_{2+ \delta}^{\dagger}(\kappa)T_{2}(\kappa)$, $\cdots$ follows from
the linear independence of the mutually orthogonal $| \delta \rangle
\langle 0 |$, $|1+ \delta \rangle \langle 1 |$, $|2+ \delta \rangle
\langle 2|$, $\cdots$ and, similarly,  the linear independence of 
$T_{0}^{\dagger}(\kappa)T_{\delta}(\kappa)$,
$T_{1}^{\dagger}(\kappa)T_{1+\delta}(\kappa)$,
$T_{2}^{\dagger}(\kappa)T_{2+ \delta}(\kappa)$, $\cdots$
follows from the linear independence of 
$|0 \rangle\langle \delta |$, $|1 \rangle \langle 1+ \delta |$, $|2 \rangle
\langle 2 + \delta|$, $\cdots$.

This completes proof of the linear independence of the set
$\{T_{m}^{\dagger}(\kappa)\,T_{n}(\kappa) \},  \, 0 \leq m,n < \infty$ and hence proves
extremality of ${\cal D}(\kappa)$ in the sense of Choi, for all
$\kappa \neq 1$. \\

\noindent 
{\bf Remark on the Doubly Stochastic Case ${\cal D}(1)$}\,:
The channel ${\cal D}(1)$ is exceptional within the phase conjugation
family of quantum-limited channels in that it is both trace-preserving
and unital, 
i.e. it is doubly stochastic. While all random unitary channels
(convex sums of unitary channels) \cite{bhatia} are manifestly doubly
stochastic, the converse is not true. Early (finite-dimensional)
counter examples to the converse can be found in Ref. \cite{landau93, tregub}. It
is this phenomenon that underlies a conjecture of Winter et al
\cite{winter}. The channel ${\cal D}(1)$ is a counter example from the
infinite-dimensional Gaussian domain, and the only doubly stochastic case
among quantum-limited Gaussian channels. That ${\cal D}(1)$ is not
random unitary readily follows from its extremality which we have
established above. Not every non-extremal doubly stochastic map is
random unitary, but an extremal doubly stochastic map can have not
even one unitary operator in the support of its Kraus operators, for
if it had it will be a convex sum of that unitary channel and another
doubly stochastic map. The noisy classical noise channels of families
${\cal B}_1$ and ${\cal B}_2$ are obviously doubly stochastic and
obviously random unitary.

\subsection{The attenuation/beamsplitter channel ${\cal
    C}_{1}(\kappa)$}
The beamsplitter channel ${\cal C}_1 (\kappa)$ is described by the
Kraus operators $\{B_{m}(\kappa) \}$ given in (\ref{bs6}). As in the
earlier case of ${\cal D}(\kappa)$ we can compute the products $\{
B_{m}^{\dagger}(\kappa)\, B_{n}(\kappa)\}$ in the form
\begin{eqnarray}
B_{\ell + \delta}^{\dagger}(\kappa)B_{\ell}(\kappa)&=& \sum_{j=0}^{\infty}
f(\delta)_{\ell j}\, |j+\delta \rangle \langle j|, \nonumber \\
B_{\ell}^{\dagger}(\kappa) B_{\ell + \delta}(\kappa)&=&
\sum_{j=0}^{\infty} f(\delta)_{\ell j}\, |j \rangle \langle
j+ \delta|,
\nonumber \\ 
f(\delta)_{\ell j}&=& \sqrt{{}^{j}C_{\ell} ~
  {}^{j+\delta}C_{\ell+\delta} } \, (1-\kappa^2)^{\ell + \delta/2} \,
\kappa^{2(j-\ell)} \,\,\,\, {\rm for}\,\,\, j \geq \ell 
\nonumber \\
&=& 0\,\,\,\,{\rm for}\,\,\, j < \ell.
\label{ebs1}
\end{eqnarray} 
Again the two cases $|j+\delta \rangle \langle j|$ and $|j \rangle
\langle j+\delta|$ correspond respectively to $m-n \geq 0$ and $m-n
\leq 0$.

The situation is similar to the earlier case of ${\cal
  D}(\kappa)$. The set $\{B_{m}^{\dagger}(\kappa) B_{n}(\kappa)\},  \,
0 \leq m,n < \infty$ 
fibrates into nonintersecting subsets, and these are labelled by
$m-n=0,\,\pm1,\,\cdots$. The index $\ell ={\rm min}\{m,n \}$ acts as
the label within a given subset. The subsets being mutually orthogonal,
it only remains to examine linear independence within each subset.

As in the case of ${\cal D}(\kappa)$ we arrange the $c$-number
coefficients $f(\delta)_{\ell j}$ into matrices
$f(\delta)=((f(\delta)_{\ell j}))$, one matrix for each subset $\delta$. These matrices are upper diagonal. None of the
diagonal elements vanishes, and so the matrices are nonsingular, proving
the linear independence of the sets $\{B_{\ell +
  \delta}^{\dagger}(\kappa) B_{\ell}(\kappa) \}_{\ell}$ and
$\{B_{\ell}^{\dagger}(\kappa) B_{\ell + \delta}(\kappa) \}_{\ell}$ and hence
the extremality of ${\cal C}_1(\kappa)$, for all $\kappa \geq 0$.

\subsection{The amplifier channel ${\cal C}_2 (\kappa)$}
The amplifier channel ${\cal C}_{2}(\kappa)$ described by the Kraus
operators $\{A_{m}(\kappa) \}$ presented in (\ref{am6}) turns out to be a
little more subtle in respect of our present purpose. There is
considerable similarity with the two earlier cases, though. As in the
case of ${\cal D}(\kappa)$ and ${\cal C}_1 (\kappa)$, let us begin by
expressing the products $\{A_{m}^{\dagger}(\kappa)\, A_{n}(\kappa) \}$
in the form
\begin{eqnarray}
A_{\ell + \delta}^{\dagger}(\kappa)\, A_{\ell}(\kappa)&=& \sum_{j=0}^{\infty}
h(\delta)_{\ell j}|j \rangle \langle j+ \delta|, \nonumber \\
A_{\ell}^{\dagger}(\kappa) A_{\ell + \delta}(\kappa)&=&
\sum_{j=0}^{\infty} h(\delta)_{\ell j}|j+\delta \rangle \langle
j|,\nonumber \\ 
h(\delta)_{\ell j}&=&  \kappa^{-2} \sqrt{{}^{\ell + j
    +\delta}C_{\ell + \delta}\,  {}^{\ell +j +\delta}C_{j +
    \delta}}\, (\sqrt{1-\kappa^{-2}})^{2 \ell + \delta}\,(\kappa^{-1})^{2j+\delta}.
\label{eam}
\end{eqnarray} 
The two cases correspond respectively to $m-n\geq 0$ and $m-n \leq 0$.
The set of operators $\{A_m^{\dagger}(\kappa) A_n(\kappa) \}$ manifestly separate into
nonintersecting subsets, determined by $m-n$; the index $\ell$ acts as
a label within each subset; the subsets are mutually orthogonal; and
it only remains to determine the linear independence within each
subset. Up to this point the situation is similar to the earlier two
cases.

As in the earlier two cases, let us arrange the $c$-number coefficients $h(\delta)_{\ell j}$ 
into matrices $h(\delta)$, one matrix for each $\ell$. And now arises the
distinction\,: whereas the invertibility of $d(\delta)$ and
$f(\delta)$ was manifest, being lower or upper diagonal
matrices, this is not so in respect of the present case. We are
therefore led to demonstrate the nonsingularity of $h(\delta)$ 
in a somewhat different manner. 

The multiplicative scalar
$(\kappa^{-1})^{2+\delta}(\sqrt{1-\kappa^{-2}})^{\delta}$ can be
dropped from $h(\delta)$ for our present purpose. We note further that $h(\delta)$ can be
simplified by left and right multiplication by diagonal matrices $L$, $M$\,:
\begin{align}
h(\delta)&=L \, \tilde{h} R \nonumber \\
L_{rs}= \frac{(1-\kappa^{-2})^r}{\sqrt{r! (r + \delta)!}} \delta_{rs}
&,~~ R_{rs}= \frac{\kappa^{-2r}}{\sqrt{r! (r + \delta)!}} \delta_{rs}.
\end{align} 
Since $L$, $R$ are nonsingular, invertibility of $h(\delta)$ is
equivalent to that of $\tilde{h}(\delta)$. The new matrix $\tilde{h}$ is
symmetric and has a simple structure\,: $\tilde{h}(\delta)_{rs}=
(r+ s + \delta)!$. To demonstrate the nonsingularity of
$\tilde{h}(\delta)$, we begin by writing out its entries in detail\,:
\begin{eqnarray}
\tilde{h}(\delta)&=& \left( \begin{array}{ccccccllcl}
\delta! && (\delta +1)! && (\delta+2)! && \cdots & \cdots & (\delta+k)! &\cdots \\
(\delta+1)! && (\delta +2)! && (\delta+3)! && \cdots & \cdots & \cdots &\cdots \\
(\delta+2)! && (\delta +3)! && (\delta+4)! && \cdots & \cdots & \cdots &\cdots \\
\vdots&&\vdots&&\vdots&&\vdots&\vdots&\cdots&\cdots \\
(\delta+\ell)!&&\vdots&&\vdots&&\vdots&\vdots&(\delta +k+\ell)!&\cdots \\
\vdots&&\vdots&&\vdots&&\vdots&\vdots&\cdots&\cdots 
\end{array} \right)
\end{eqnarray}
Dividing the first row of $\tilde{h}(\delta)$ by $\delta!$, the second row by
$(\delta+1)!$, the third row by $(\delta +2)!$, and so on, we obtain
\begin{eqnarray}
\left( \begin{array}{ccccccccccc}
1&&(\delta+1)&&(\delta +1)(\delta+2) &&(\delta +1)(\delta+2)(\delta+3)  
&& \cdots&&\cdots \\
1 && (\delta+2)&& (\delta+2)(\delta +3)&&
(\delta+2)(\delta +3)(\delta+4)&& \cdots&&\cdots \\
1 &&(\delta +3)&&(\delta+3)(\delta+4)
&&(\delta+3)(\delta+4)(\delta+5) &&\cdots &&\cdots\\
\vdots&&\vdots&&\vdots&&\vdots&& &&
\end{array} \right).
\end{eqnarray}
Now performing the row transformations $R_1 \rightarrow R_1-R_0$, $R_2
\rightarrow R_2-R_1$, and 
so forth we obtain
\begin{eqnarray}
\left( \begin{array}{cccccc}
1&(\delta+1)&(\delta +1)(\delta+2)&(\delta +1)(\delta+2)(\delta+3)&\cdots
&\cdots \\
0&1& 2(\delta+2)&3(\delta+2)(\delta+3)& \cdots & \cdots \\
0&1&2(\delta+3)&3(\delta+3)(\delta+4)& \cdots &\cdots \\
0&1&2(\delta+4)&3(\delta+4)(\delta+5)& \cdots &\cdots \\
\vdots&\vdots&\vdots&\vdots& & 
\end{array} \right).
\end{eqnarray}
We can repeatedly do this kind of row transformations, 
starting with row $R_2$ in the next iteration and
row $R_3$ in the subsequent iteration and so on. The matrix
$\tilde{h}(\delta)$ gets finally transformed to the form
\begin{eqnarray}
\left(\begin{array}{cccccc}
1& \cdots & \cdots & \cdots & \cdots & \cdots \\
0& 1&\cdots & \cdots & \cdots & \cdots \\
0&0&2!& \cdots & \cdots & \cdots \\
0&0&0&3!&\cdots & \cdots \\
0&0&0&0&4!&\cdots \\
\vdots&\vdots&\vdots&\vdots&\vdots&\ddots
\end{array}\right),
\end{eqnarray} 
which is upper triangular with non-vanishing diagonal entries, and 
thus has `full rank'. We may thus conclude that the coefficient
matrix $h(\delta)$ is invertible, showing that the set $\{
{A}_{\ell+\delta}^{\dagger} A_{\ell}\}_{\ell}$ [as well as $\{
{A}_{\ell}^{\dagger} A_{\ell+\delta}\}_{\ell}$] is linearly
independent, for each $\delta$.

This completes proof of  linear independence of the set
$\{{A}_{m}^{\dagger}(\kappa)\, A_{n}(\kappa) \}, 0 \leq m,n < \infty$. Extremality
of the amplifier channel ${\cal C}_2(\kappa)$ is thus established,,
for all $\kappa >1$.

\subsection{The Singular case ${\cal A}_2$}
Computation of the product $V_{{q^{\,'}}}^{\dagger} V_q  $ is nearly trivial
in this case. We have from \eqref{sing8}\,: 
\begin{align} \label{sing11}
V_{q^{\,'}}^{\dagger} V_q   =  ( q^{\,'}/\sqrt{2}
|   q/\sqrt{2} ) \,  |{q^{\,'}} \rangle \langle q |.
\end{align}
The inner product $(q^{\,'}/\sqrt{2}
| q/\sqrt{2} ) = \exp[ -q^2/4
  -{q^{\,'}}^2/4+qq^{\,'}/2 ]$ is nonzero for
arbitrary 
$q$, ${q^{\,'}}$. Since $ |{q^{\,'}} \rangle \langle q |$ form a linearly
independent set of operators, we conclude that the channel ${\cal A}_2$ is
extremal. 

We have considered in this Section the quantum-limited channels ${\cal
D}(\kappa)$, ${\cal C}_1(\kappa)$, ${\cal C}_2(\kappa)$, and ${\cal
A}_2$ in that order. Since these are the only quantum-limited Bosonic
Gaussian channels, the main conclusion of our study can be stated in
the following concise form\,: 
\begin{theorem}
All quantum-limited Bosonic Gaussian channels are extremal. 
\end{theorem}

\noindent 
{\bf Remark on ${\cal A}_1$}\,:  
\noindent We have not treated separately the quantum-limited channel ${\cal
A}_1$, for this may be viewed as a particular case of ${\cal C}_1(\kappa)$,
corresponding to $\cos(\theta)=0$ in \eqref{bs1} or, equivalently,
$\kappa=0$.  Thus, the assertion above
covers ${\cal A}_1$ as well.

\section{Noisy channels as composites of quantum-limited channels}
Our considerations so far have been in respect of quantum-limited
channels. We turn our attention now to the case of noisy channels. It
turns out that every noisy channel, except ${\cal B}_1(a)$ which
corresponds to injection of classical noise in just one quadrature,
can be realised (in a non-unique way) as composition of two quantum
limited channels, so that the Kraus operators are products of those of
the constituent quantum-limited channels. 

We have noted in Sections IV and V that composition of two quantum
limited attenuator (or amplifier) channels is again a quantum-limited
attenuator (or amplifier) channel. This special semigroup property
however does not obtain under composition for other quantum-limited
channels. In general, composition of two quantum-limited channels 
results in a channel with additional classical noise. For this reason we restore in
this Section the original notation ${\cal D}(\kappa;0)$, ${\cal
  C}_1(\kappa;0)$, and ${\cal C}_2(\kappa;0)$ in order to make room for
this additional classical noise. 

\subsection{The composite ${\cal C}_2(\kappa_2;0) \circ {\cal C}_1(\kappa_1;0), ~
  \kappa_2 \geq 1, ~ \kappa_1 \leq 1$ \label{comp1}}
It is clear from the very definition of these channels through their action
on the characteristic function that the composite ${\cal
  C}_2(\kappa_2;0) \circ {\cal C}_1(\kappa_1;0) $ is a noisy amplifier or
attenuator depending on the numerical value of $\kappa_2\kappa_1$\,:
it equals ${\cal C}_1(\kappa_2\kappa_1; 2(\kappa_2^2-1))$ for
$\kappa_2\kappa_1 \leq 1$, and ${\cal C}_2(\kappa_2\kappa_1; 2\kappa_2^2(1-\kappa_1^2))$ for
$\kappa_2\kappa_1 \geq 1 $, as may be readily read off from Table \ref{tab1}.

\begin{table}
\begin{center}
\begin{tabular}{|c|c|c|c|c|}
\hline
\hline
$X ,  Y \rightarrow$  & ${\cal D}(\kappa_1;0)$  & ${\cal C}_1(\kappa_1;0)$ &
${\cal C}_2 (\kappa_1;0)$ & ${\cal A}_2(0)$ \\
$\downarrow$~~~~~~~~ &&&&\\
\hline 
  & ${\cal C}_1(\kappa_2\kappa_1;2\kappa_2^2(1+\kappa_1^2))$,
& &  & \\
${\cal D}(\kappa_2;0)$ &for $\kappa_2\kappa_1 \leq 1$.&${\cal D}(\kappa_2\kappa_1; 2\kappa_2^2(1- \kappa_1^2)$&${\cal
  D}(\kappa_2\kappa_1;0 )$&${\cal A}_2(2\kappa_2^2) $\\
&${\cal C}_2(\kappa_2 \kappa_1;2(1+\kappa_2^2))$, &&&\\
&for $\kappa_2\kappa_1 \geq1 $.&&&\\
\hline
  &  &  & ${\cal
   C}_1(\kappa_2\kappa_1;2\kappa_2^2(\kappa_1^2-1))$,&\\
 ${\cal C}_1(\kappa_2;0)$ & ${\cal D}(\kappa_2\kappa_1;0)$&${\cal
   C}_1(\kappa_2\kappa_1;0)$&for $\kappa_2\kappa_1 \leq 1$.& ${\cal A}_2(0)$\\ 
&&&${\cal C}_2(\kappa_2\kappa_1;2(1-\kappa_2^2))$,&\\ 
&&&for $\kappa_2 \kappa_1 \geq 1 $&\\
\hline
&
& ${\cal C}_1(\kappa_2\kappa_1;2(\kappa_2^2-1))$,  &  &\\
${\cal C}_2(\kappa_2;0)$ & ${\cal
  D}(\kappa_2\kappa_1;2(\kappa_2^2-1))$&for $\kappa_2\kappa_1 \leq
1$.&${\cal C}_2(\kappa_2 
\kappa_1;0)$&${\cal A}_2(2(\kappa_2^2-1)) $\\ 
&&${\cal C}_2(\kappa_2\kappa_1; 2\kappa_2^2(1-\kappa_1^2))$, &&\\
&&for $\kappa_2 \kappa_1 \geq 1$.&&\\
\hline
${\cal A}_2(0)$ &${\cal A}_2(\sqrt{\kappa_1^2+2} -1)$&${\cal
  A}_2(\sqrt{2-\kappa_1^2}-1)$&${\cal A}_2(\kappa_1-1)$&${\cal
  A}_2(\sqrt{2}-1)$\\ 
\hline 
\hline
\end{tabular}
\end{center}
\caption{Showing the composition $X \circ Y$ of quantum-limited channels
  $X,Y$ {\em assumed to be in their respective canonical forms
    simultaneously}. The composition results, in several cases, in
  noisy channels thereby enabling description of noisy Gaussian
  channels, including the classical noise channel ${\cal B}_2(a)$, in terms of {\em
    discrete sets} of linearly independent Kraus operators.  \label{tab1}}
\end{table}

The Kraus operators for the composite is given by the set
$\{A_m(\kappa_2) B_{n}(\kappa_1)\} $ with $m$,$n$ running
independently over the range $0 \leq m,n < \infty$. That {\em these
  Kraus operators are linearly independent} may be seen as
follows. Since $A_{m}(\kappa_2)$ is the same as
$B_m(\kappa_2^{-1})^{\dagger}$ except for an $m$-independent
multiplicative constant, linear independence of the set
$\{A_m(\kappa_2) B_n(\kappa_1)\}$ is the same as linear independence
of $\{B_m(\kappa^{-1}_{2})^{\dagger} B_n(\kappa_1)\}$. It will be
recalled that in proving linear independence of the set
$\{B_m(\kappa)^{\dagger}B_n(\kappa) \} $ in Section V in the context
of extremality of ${\cal C}_2(\kappa)$, we used only the structure of
the $B_{\ell}(\cdot)$'s in respect of the expansion coefficients in
the basis $\{|m\rangle \langle n|\}$ being zero or nonzero, and {\em not
  the actual numerical values of the nonzero coefficients}. Hence the same
argument should be expected to apply to proof of linear independence
of $\{B_m(\kappa_2^{-1})^{\dagger}B_n(\kappa_1) \}$ $\sim$ $\{A_m(\kappa_2)
B_n(\kappa_1)\} $ as well. That this is indeed the case is seen by
computing the products $A_m(\kappa_2) B_n(\kappa_1)$ in the form 
\begin{align}
A_{\ell + \delta}(\kappa_2) B_{\ell}(\kappa_1) &= \sum_{j=0}^{\infty}
g_1(\delta)_{\ell j} | j + \delta \rangle \langle j|, \nonumber \\
A_{\ell}(\kappa_2) B_{\ell+\delta}(\kappa_1) &= \sum_{j=0}^{\infty}
\tilde{g}_1(\delta)_{\ell j} | j \rangle \langle j+\delta|,\nonumber \\
g_1(\delta)_{\ell j} &= \kappa_2^{-1} \, \sqrt{{}^{j+\delta}C_{\ell +
    \delta} \, {}^{j}C_{\ell}} \, \left(\sqrt{1-\kappa_1^2}\right)^{\ell} \,
\left(\sqrt{1-\kappa_2^{-2}}\right)^{\ell+ \delta} \,
(\kappa_2^{-1}\kappa_1)^{j-\ell}, \text{  for } j\geq \ell, \nonumber\\
&= 0, \text{  for } j < \ell;\nonumber \\ 
\tilde{g}_1(\delta)_{\ell j} &= \kappa_2^{-1} \, \sqrt{{}^{j+\delta}C_{\ell +
    \delta} \, {}^{j}C_{\ell}} \, \left(\sqrt{1-\kappa_2^{-2}}\right)^{\ell} \,
\left(\sqrt{1-\kappa_1^{2}}\right)^{\ell + \delta} \,
(\kappa_2^{-1}\kappa_1)^{j-\ell}, \text{  for } j\geq \ell,
\nonumber\\ 
&= 0, \text{  for } j < \ell.
\end{align}
Comparing this with \eqref{ebs1}, it is now clear that $A_m(\kappa_2)
B_n(\kappa_1)$'s are {\em linearly independent} for exactly the same reason
by which $B_m(\kappa)^{\dagger} B_n(\kappa)$'s were linearly independent
rendering the quantum-limited attenuator ${\cal C}_1(\kappa;0)$ extremal. 

\subsection{ The composite ${\cal C}_1(\kappa_2;0) \circ {\cal C}_2(\kappa_1;0), ~
  \kappa_2 \leq 1, ~ \kappa_1 \geq 1$ \label{comp2}}
Again the composite ${\cal C}_1(\kappa_2;0) \circ {\cal C}_2(\kappa_1;0)$ is 
a noisy amplifier or attenuator depending on the numerical value of
$\kappa_2\kappa_1$, and the details may be read off from Table \ref{tab1}. The
Kraus operators for the composite ${\cal C}_1(\kappa_2;0) \circ {\cal
C}_2(\kappa_1;0)$ are given by $\{B_m(\kappa_2) A_n(\kappa_1) \}$, $0
\leq m,n < \infty$. To establish their {\em linear independence} we
compute $B_m(\kappa_2) A_n(\kappa_1)$ in the form
\begin{align}
B_{\ell + \delta}(\kappa_2) A_{\ell}(\kappa_1) &= \sum_{j=0}^{\infty}
g_2(\delta)_{\ell j} |j \rangle \langle j+ \delta|, \nonumber \\
B_{\ell}(\kappa_2) A_{\ell+\delta}(\kappa_1) &= \sum_{j=0}^{\infty}
\tilde{g}_2(\delta)_{\ell j} |j +\delta \rangle \langle j|, \nonumber \\ 
g_2(\delta)_{\ell j}&= \kappa_1^{-1}\,\sqrt{{}^{j+\ell+\delta}C_{\ell + \delta} \,
  {}^{j+\delta+\ell}C_{\ell}} \, \kappa_1^{-(j+\delta)}  \left(\sqrt{1-\kappa^{-2}_1}\right)^{\ell}\,
\kappa_2^{j} \left(\sqrt{1-\kappa^2_2}\right)^{\ell+ \delta},\nonumber \\
\tilde{g}_2(\delta)_{\ell j}&= \kappa_1^{-1}\,\sqrt{{}^{j+\ell+\delta}C_{\ell + \delta} \,
  {}^{j+\delta+\ell}C_{\ell}} \, \kappa_1^{-j}  \left(\sqrt{1-\kappa^{-2}_1}\right)^{\ell+\delta}\,
\kappa_2^{j+\delta} \left(\sqrt{1-\kappa^2_2}\right)^{\ell}.
\end{align}
Comparing with \eqref{eam} we see that the Kraus operators
$B_m(\kappa_2)A_n(\kappa_1)$ of ${\cal C}_1(\kappa_2;0) \circ {\cal
C}_2(\kappa_1;0)$ are linearly independent for the same reason by which
the quantum-limited amplifier channel was extremal. 

\subsection{The composite $ {\cal D}(\kappa_2) \circ {\cal D}(\kappa_1), ~ \kappa_2,
  \kappa_1 > 0$ \label{comp3}}
Similar to the earlier two cases, the composite ${\cal D}(\kappa_2;0)
\circ {\cal D}(\kappa_1;0)$ is a noisy amplifier or attenuator depending
on the numerical value of $\kappa_2\kappa_1$, and the details can be read off
from Table \ref{tab1}. It may be noted, again from Table \ref{tab1},  that this case tends to be more noisy
that the earlier two cases. 

The Kraus operators for this composite are given by
$\{T_m(\kappa_2)T_n(\kappa_1)\}, ~ 0 \leq\, m,n < \infty $. To exhibit
the {\em linear independence} of these Kraus operators we compute the
products $T_m(\kappa_2)T_n(\kappa_1)$ in the form 
\begin{align}
T_{\ell + \delta}(\kappa_2)T_{\ell}(\kappa_1) &= \sum_{j=0}^{\infty}
g_3(\delta)_{\ell j} |j+\delta \rangle \langle j|, \nonumber \\ 
T_{\ell}(\kappa_2)T_{\ell + \delta}(\kappa_1) &= \sum_{j=0}^{\infty}
\tilde{g}_3(\delta)_{\ell j} |j+\delta \rangle \langle j|, \nonumber \\  
g_3(\delta)_{\ell j} &= \left(\sqrt{1+\kappa_1^2}\right)^{-1} 
\left(\sqrt{1+\kappa_2^2}\right)^{-1}\,
\sqrt{{}^{\ell}C_{j}\, {}^{\ell+\delta}C_{j}} \,
\left[\sqrt{(1+\kappa_2^2)(1+\kappa_1^{-2})}  \right]^{-(\ell -j)}
\,\nonumber \\
&\times \left[\sqrt{(1+\kappa_1^2)(1+\kappa_2^{-2})}\right]^{-j} \,
\left(\sqrt{1+\kappa_2^{-2}}\right)^{-\delta}, ~~ \text{for } j\leq \ell, \nonumber \\ 
&=0, ~~\text{for } j>\ell, \nonumber \\ 
\tilde{g}_3(\delta)_{\ell j} &= \left(\sqrt{1+\kappa_1^2}\right)^{-1}
\left(\sqrt{1+\kappa_2^2}\right)^{-1}\, 
\sqrt{{}^{\ell}C_{j}\, {}^{\ell+\delta}C_{j}} \,
\left[\sqrt{(1+\kappa_2^2)(1+\kappa_1^{-2})}\right]^{-(\ell -j)}
\,\nonumber \\ 
&\times \left[\sqrt{(1+\kappa_1^2)(1+\kappa_2^{-2})}\right]^{-j} \,
\left(\sqrt{1+\kappa_1^2}\right)^{-\delta}, ~~ \text{for } j\leq \ell, \nonumber \\ 
&=0, ~~\text{for } j>\ell. 
\end{align} 
Comparing with \eqref{epc} it is readily seen that these Kraus operators are
linearly independent for exactly the same reason by which the quantum
limited transpose channel was extremal. 

\subsection{The composite  ${\cal D}(\kappa_2;0) \circ {\cal C}_1(\kappa_1;0), ~ 
  \kappa_2 > 0, \, 0 \leq \kappa_1  \leq 1$ \label{comp4}}
Kraus operators of this composite, which always corresponds to a noisy
transpose channel (see Table \ref{tab1}), are $\{T_m(\kappa_2) B_n(\kappa_1) \}, ~ 0 \leq m,n
< \infty$. We have 
\begin{align}
T_m(\kappa_2)B_n(\kappa_1) &= \sum_{j=0}^{\infty} \xi^{j}_{mn} |m-j \rangle
\langle n+j|, \nonumber \\
\xi^j_{mn} &= \left(\sqrt{1+\kappa_2^2}\right)^{-1} \sqrt{{}^mC_j\,
  {}^{n+j}C_j} \, \left(\sqrt{1+\kappa_2^2}\right)^{-j}\, 
\left(\sqrt{1+\kappa_2^{-2}}\right)^{-(m-j)}\nonumber \\ 
&~~~~~\times \kappa_1^j \left(\sqrt{1-\kappa_1^2}\right)^{n}, \text{ for }
j \leq m; \nonumber \\
&=0, \, \text{ for } j>m.
\label{c4}
\end{align}  
It is immediately clear that $T_m(\kappa_2) B_n(\kappa_1)$ and
$T_{m^{\,'}}(\kappa_2) B_{n^{\,'}}(\kappa_1)$ are (trace-)orthogonal
unless $m+n = m^{\,'} + n^{\,'}$.

We may therefore divide the Kraus operators into orthogonal subsets
determined by $m+n =$ constant $\equiv N$. Then linear independence
will have to be established just within each subset $\Omega_N = \{
T_0(\kappa_2) B_N(\kappa_1), \, T_1(\kappa_2) B_{N-1}(\kappa_1),
\cdots, T_N(\kappa_2) B_0(\kappa_1)  \} $. It is seen from \eqref{c4} that
$T_0(\kappa_2) B_N(\kappa_1)$ is a multiple of $|0 \rangle \langle
N|$, $T_1(\kappa_2) B_{N-1}(\kappa_1)$ is a linear combination of $|0
\rangle \langle N|$ and $|1 \rangle \langle N-1|$, $T_2(\kappa_2)
B_{N-2}(\kappa_1)$ is a linear combination of $|0 \rangle \langle
N|, \,|1 \rangle \langle N-1|$, and $|2 \rangle \langle N-2| $, and so
on. Thus linear independence within $\Omega_N$ follows as an
immediate consequence of the fact that $\xi^0_{mn} \neq 0 \,\, \forall \, 
m,n$. 

\subsection{The composite ${\cal C}_1(\kappa_2;0) \circ {\cal D}(\kappa_1;0), ~ 
  \kappa_1 > 0, 0 \leq \kappa_2  \leq 1$ \label{comp5}}
This composite channel corresponds to a {\em quantum-limited} transpose
channel (see Table \ref{tab1}) which we have already considered in much detail in Section
III. The Kraus operators $\{B_m(\kappa_2) T_n(\kappa_1) \}, ~ 0 \leq m,n
< \infty $ (which as a set should be equivalent to
$\{T_{\ell}(\kappa_2\kappa_1) \}, \, 0 \leq \ell < \infty$), are 
\begin{align}
B_m(\kappa_2) T_n(\kappa_1) &= \sum_{j=m}^{n}\, \xi^j_{mn} |j-m
\rangle \langle n-j|, \nonumber \\
\xi^j_{mn} &= \sqrt{{}^{j}C_{m}\,
  {}^{n}C_{j}}\, \left(\sqrt{1-\kappa_2^2}\right)^{m}\, \kappa_2^{j-m}\,
\left(\sqrt{1+\kappa_1^2}\right)^{-(n-j+1)}\nonumber \\ 
& \times
\,\left(\sqrt{1+\kappa_1^{-2}}\right)^{-j}, \text{ for } n\geq m ;
\nonumber \\
&= 0, \text{ for } n<m.  
\end{align}
Thus $B_m(\kappa_2) T_n(\kappa_1) =0$ if $m>n$. Further, all
$B_m(\kappa_2) T_n(\kappa_1)$ with $m=n$ correspond to a multiple of
the vacuum projector $|0 \rangle \langle 0|$, showing that these
Kraus operators are not linearly independent. Even so, this is also a valid
representation of the quantum-limited channel ${\cal C}_1(\kappa_2;0)
\circ {\cal D}(\kappa_1;0) = {\cal D}(\kappa_2\kappa_1;0)$ .

\subsection{The composite ${\cal C}_2(\kappa_2;0) \circ {\cal D}(\kappa_1;0), ~ 
  \ \kappa_2  \geq 1 , \kappa_1 > 0$ \label{comp6}} 
This composite channel corresponds, for all $\kappa_1, \kappa_2$,  to
a noisy transpose channel, 
similar to the case of ${\cal D}(\kappa_2;0)\circ {\cal C}_1(\kappa_1;0)$
considered earlier. The Kraus operators $\{A_m(\kappa_2) T_n(\kappa_1)
\}, ~ 0 \leq m,n < \infty$ have the form 
\begin{align}
A_m(\kappa_2) T_n(\kappa_1) &= \sum_{j=0}^n \, \xi^j_{mn} |j+m \rangle
\langle n-j|, \nonumber \\
\xi^j_{mn} &= \kappa_2^{-1}\left(\sqrt{1+\kappa_1^2}\right)^{-1} 
\, \sqrt{{}^{m+j}C_{j}
  \, {}^{n}C_{j}} \, \left(\sqrt{1-\kappa_2^{-2}}\right)^{m} \, \kappa_2^{-j}\,
\left(\sqrt{1+\kappa_1^2}\right)^{-(n-j)} \nonumber \\ &\times 
\left( \sqrt{1+\kappa_1^{-2}}\right)^{-j}, \text{ for } j\leq n;
\nonumber \\
&=0, \text{ for } j>n. 
\end{align}
Linear independence of the Kraus operators can be established in a
manner similar to the earlier case of ${\cal D}(\kappa_2;0)\circ {\cal
  C}_1(\kappa_1;0)$. 

\subsection{The composite ${\cal D}(\kappa_2;0)\circ {\cal C}_2(\kappa_1;0), ~ \kappa_2
  >0, \kappa_1 \geq 1$ \label{comp7}}
This composite is a {\em quantum-limited} transpose channel (see Table \ref{tab1}),
with Kraus operators $\{T_m(\kappa_2) A_n(\kappa_1) \}, ~ 0 \leq m,n <
\infty$. Similar to the earlier quantum-limited case of ${\cal
  C}_1(\kappa_1;0) \circ {\cal D}(\kappa_2;0)$, these Kraus operators too
are not linearly independent; they do represent the quantum-limited
channel ${\cal D}(\kappa_2\kappa_1;0)$, though.  

We conclude this Section with some further observations. It is seen
from Table \ref{tab1} that to realise the noisy attenuator channel ${\cal
  C}_1(\kappa;a),$ for the full parameter range $ 0 \leq \kappa \leq 1, \, a>0$, where $a$ is a
measure of the additional classical noise above the quantum limit, as
the composite ${\cal C}_2(\kappa_2;0)\circ {\cal C}_1(\kappa_1;0)$, 
 we have to solve $\kappa_2\kappa_1 = \kappa$ and
$2(\kappa_2^2-1)=a$ for $0 \leq \kappa_1 \leq 1, \, \kappa_2 \geq 1$,
and we have as solution $\kappa_2 = \sqrt{1+ a/2}>1, \, \kappa_1 =
\kappa/\kappa_2 < \kappa \leq 1$. It should be appreciated that this
realisation contains as 
special cases the noisy ${\cal A}_1(a)$ for $\kappa=0$ (i.e.,
$\kappa_1 =0$), and the
classical noise channel ${\cal B}_2(a)$ for $\kappa=1$. 

Similarly, the noisy amplifier ${\cal C}_2(\kappa;a), \, \kappa \geq 1,
a>0$ can be realised through the same composite if we solve can
$\kappa_2\kappa_1=\kappa$ and $2\kappa_2^2(1- \kappa_1^2) =a$ for $0
\leq \kappa_1 \leq 1, \kappa_2 \geq 1$. We have the solution
$\kappa_2 = \sqrt{\kappa^2 + a/2} > \kappa \geq 1$ and $\kappa_1 =
\kappa/\kappa_2 < 1$. Again the classical noise channel ${\cal
  B}_2(a)$ is contained as the special case $\kappa=1$. 

Finally, to realise the noisy transpose channel ${\cal D}(\kappa;a),
\, \kappa \geq0,\, a>0$ as the composite ${\cal D}(\kappa_2;0) \circ {\cal
  C}_1(\kappa_1;0), \, 0 \leq \kappa_1 \leq 1,\, \kappa_2>0$, of quantum-limited
channels we solve $\kappa=\kappa_2\kappa_1$ and $a = 2 \kappa_2^2(1- \kappa_1^2)$
 to obtain $\kappa_2 = \sqrt{\kappa^2 + a/2} > \kappa , \, \kappa_1 =
 \kappa/\kappa_2 < 1$. The same can also be realised as the
 composite ${\cal C}_2(\kappa_2;0) \circ  {\cal D}(\kappa_1;0), \, \kappa_2
 \geq 1, \kappa_1 >0$ by solving $\kappa = \kappa_1\kappa_2$ and $a =
 2(\kappa_2^2-1)$, and we have $\kappa_2 = \sqrt{1+a/2} >1, \, \kappa_1
 = \kappa/\kappa_2 \geq 0$.  

We may summarise some aspects of our consideration thus far in this
Section in the following manner.
\begin{theorem}
All the nonsingular noisy channels can be realised as the composition
of a pair of quantum-limited channels. Equivalently, and as a consequence,
each of the nonsingular Gaussian channel has an operator-sum
representation in terms of a {\em discrete set of linearly independent
} Kraus operators. 
\end{theorem}
The above assertion includes in particular the case of the {\em classical
noise channel} ${\cal B}_2(a)$. Further, the case ${\cal A}_1(a)$ is not
exempted from our consideration above, for it is just a special case,
corresponding to $\kappa=0$ of the noisy attenuator ${\cal
  C}_1(\kappa;a)$. 

It is seen
from Table \ref{tab1} that the noisy singular case ${\cal A}_2(a)$ can be
realised as a composite of two quantum-limited channels\,: either
following or preceding the quantum-limited ${\cal A}_2(0)$ by quantum
limited ${\cal C}_1(\cdot;0)$, ${\cal C}_2(\cdot;0)$, or ${\cal
  D}(\cdot;0)$. Consequently, the Kraus operators will be indexed by
one discrete and {\em one real} variable. Thus, the single quadrature
noise channel ${\cal B}_1(a)$ is the only singular case that does not
submit itself to our consideration above, in the sense that there
seems to be no way of realising it as composite of a pair of quantum
limited channels.

While we have obtained in the Section Kraus representations for noisy
channels with the aid of pairs of quantum-limited channels it is,
of course, possible to obtain Kraus representation using unitary
dilation of Section 2, with the ancilla in a thermal state rather than
the vacuum state. But we believe our present approach has the
advantage of leading to Kraus operators of extremely simple structure,
in addition to the advantage of connecting the noisy case to the
quantum-limited case.

We conclude this section with three remarks, two of them are in
respect of Table \ref{tab1} while the third one is in the context of error
correctability. \\

\noindent
{\bf Remark 1}\,: We have already noted in Sections III and V that
the quantum-limited family ${\cal D}(\cdot)$ is self dual ${\cal
  D}(\kappa) \sim {\cal D}(\kappa^{-1})$, whereas the families ${\cal C}_1(\cdot)$
and ${\cal C}_2(\cdot)$ are dual to one another \,: ${\cal C}_1(\kappa)
\sim {\cal C}_2(\kappa^{-1})$. The reader will recall that these
duality relations are a consequence of the failure to be unital of
these (trace-preserving) maps by just a multiplicative
scalar. Remnants of these duality relations may be readily observed in
Table \ref{tab1}. The composite ${\cal D}(\cdot) \circ {\cal C}_1(\cdot)$ is
quantum-limited, the dual fact being that ${\cal C}_2(\cdot) \circ
{\cal D}(\cdot)$ is quantum-limited. The fact that ${\cal C}_2(\cdot) \circ {\cal C}_2(\cdot)$
is quantum-limited is dual to the fact that ${\cal C}_1(\cdot) \circ
{\cal C}_1(\cdot)$ is quantum-limited. The fact that ${\cal D}(\cdot)
\circ {\cal C}_1(\cdot)$ is a noisy conjugator has as its dual the
fact that ${\cal C}_2(\cdot) \circ {\cal D}(\cdot)$ is a noisy
conjugator. Finally, ${\cal D}(\cdot) \circ {\cal D}(\cdot)$, ${\cal
  C}_1(\cdot) \circ {\cal C}_2(\cdot)$, and ${\cal C}_2(\cdot) \circ
{\cal C}_1(\cdot)$ are self duals. Since the failure of the quantum
limited ${\cal A}_2$ to be unital is nontrivial, ${\cal A}_2$ does not
figure in any such duality relation. \\

\noindent
{\bf Remark 2}\,: In Table \ref{tab1} we have considered the composition of
pairs of  quantum-limited
channels under the assumption that the two channels are simultaneously
in their canonical forms. In order to help
the reader appreciate this remark it should 
first be emphasised that a typical composite of this kind, say ${\cal
  D}(\kappa_2;0) \circ {\cal C}_1(\kappa_1;0)$, {\em should not} stand
for composition of two quantum-limited channels which are already in
their respective canonical forms, but rather to two channels
picked one from either $Sp(2,R)$ orbit or double coset. That
is ${\cal D}(\kappa_2;0) \circ {\cal C}_1(\kappa_1;0)$, for 
instance should stand for $((U(S_1) \circ {\cal D}(\kappa_2;0)\circ
U(S_2))\circ(U(S_1^{\,'}) \circ {\cal C}_1(\kappa_1;0) \circ
(U(S_2^{\,'})) $, for arbitrary metaplectic unitaries corresponding to
$S_1, S_2, S_1^{\,'}, S_2^{\,'} \in \, Sp(2,R)$. The fact that two
channels cannot in general be taken to their respective canonical
forms simultaneously, i.e., with $S_1^{\,'} = S_1$, $S_2^{\,'} = S_2$,
brings out the nontriviality of the assumption underlying Table
1. When this assumption is lifted, Table \ref{tab1} gets much enriched into
Table 2 as shown in the Appendix. \\

\noindent 
{\bf Remark 3}\,: We have noted in Sections III to VI that the
Kraus operators $\{W_{\ell} \}$ of quantum-limited Gaussian channels
possess the property that the associated nonnegative operators
$W_{\ell}^{\dagger} W_{\ell}$ are simultaneously diagonal (in the Fock
basis for ${\cal D}(\cdot)$, ${\cal C}_1(\cdot)$, and ${\cal
  C}_2(\cdot)$ and in the position basis for ${\cal A}_2$). In the
present Section we presented for each nonsingular noisy channel a
discrete set of Kraus operators, say $W_{mn}$, indexed by a pair of
integer variables $m,n$ and it can be readily verified in each case
that the associated nonnegative operators $W_{mn}^{\dagger} W_{mn}$
are simultaneously diagonal (in the Fock basis). For the noisy channel
${\cal A}_2(a)$, it may be verified in the realization ${\cal
  D}(\cdot) \circ {\cal A}_2$ and ${\cal
  C}_2(\cdot) \circ {\cal A}_2$ [and not in ${\cal
  A}_2 \circ {\cal D}(\cdot)$, ${\cal
  A}_2 \circ {\cal C}_1(\cdot)$, or ${\cal
  A}_2 \circ {\cal C}_2(\cdot)$ ] that the relevant nonnegative
operators are simultaneously diagonal in the position basis. Finally,
the single quadrature classical noise channel ${\cal B}_1(a)$ being
random unitary, the associated nonnegative operators are all multiples
of unity. We may thus state
\begin{theorem} 
For every Gaussian channel it is possible to obtain a Kraus
representation such that the nonnegative operators associated with the
Kraus operators are all simultaneously diagonal. 
\end{theorem}
The above observation leads to the following remark on error
correction. \\

\noindent 
{\bf Remark on Error correction}\,:
The fact that
$W_{\ell}^{\dagger}\,W_{\ell}$'s are simultaneously
diagonal in the Fock basis for all $\ell$
for the channels $D(\kappa;0), \, C_1(\kappa;0), \, C_2(\kappa;0)$ and
their composites, imply in the view of the work of \cite{gregoratti03}
that the Fock states could be used  
to reliably transmit classical information through this channel. In
such a case the channel is viewed as a generalised measurement. In
other words, any classical information encoded in Fock states and 
passed through these channels can be reliably retrieved by a  
restoring channel.


\section{Conclusion}
We have obtained operator-sum representations for all single-mode
bosonic Gaussian channels presented in their respective canonical
forms. Evidently, the operator-sum representation of a channel not in
the caonical form follows by adjoining of appropriate unitary Gaussian
evolutions before and after the channel. The
Kraus operators were obtained from the matrix elements of the two-mode
metaplectic unitary operator which 
effects the channel action on a single mode. The two-mode symplectic
transformation in each case did not mix the
position and momentum variables and this fact proved valuable for our
study. The Kraus operators for the quantum-limited channels except the
singular case were found to have a simple and sparse structure in the Fock
basis. 

It was shown that the phase conjugation channels ${\cal D}(\kappa)$ and
${\cal D}(\kappa^{-1})$ are dual to one another, and the attenuator
and the amplifier families ${\cal C}_1(\kappa)$ and ${\cal
  C}_2(\kappa^{-1}), \, \kappa <1$ are mutually dual. The channels ${\cal
  D}(\kappa), \, {\cal C}_1(\kappa),$ and ${\cal C}_2(\kappa)$ were found
to be almost unital; in the sense that the unit operator was taken to
a scalar times the unit operator. The channel ${\cal D}(1)$ was found
to be bistochastic but not random unitary. The unitary bistochastic
channels being the classical 
noise channels ${\cal B}_1$ and ${\cal B}_2$  and the trivial
identity channel ${\cal C}_1(1) =  {\cal C}_2(1) $.

 In the case of the
phase conjugation channel, the action in phase space was brought out
explicitly through the action of the Kraus operators on the Fock
basis. The 
attenuator channel resulted in the scaling of the diagonal weight
function $\phi(\alpha)$ and the amplifier channel resulted in the
scaling of the Husimi $Q$-function as expected. Further, the output of
the channel 
with respect to classicality/nonclassicality was studied. It was found
that the phase conjugation channel ${\cal D}(\kappa)$ and the singular
channel ${\cal A}_2$ are classicality breaking while the
attenuator channel ${\cal C}_1(\kappa)$ and the amplifier channel
${\cal C}_2(\kappa)$ do not generate nonclassicality. 

The action of the channel in the Fock basis gave an insight into the
fixed points of the channel. The action in the Fock basis together with the action of the
channel in phase space led us to conclude that there is a unique
thermal state which is an invariant state for ${\cal D}(\kappa), \,
\kappa<1$. Further it was shown that the vacuum state is the only invariant
state for the attenuator channel ${\cal C}_1(\kappa)$, and that there is no finite energy
state that is invariant for either the amplifier channel ${\cal
  C}_2(\kappa)$ and the singular channel ${\cal A}_2$. 

Using Choi's theorem, it was shown that all
quantum-limited bosonic Gaussian channels are extremal. The Kraus
operators of the phase
conjugation channel was brought to a rank one form, thereby explicitly
bringing out the entanglement breaking nature of the phase conjugation
channel. It was further shown that there is no finite rank operator in
the support of the Kraus operators of either the amplifier or the attenuator channel, and this
explicitly demonstrates that the quantum-limited attenuator and the amplifier families of
channels are not entanglement breaking. 
The Kraus operators of the singular channel ${\cal A}_2$ was also
obtained in the rank one form thereby manifestly showing that this
 channel is entanglement breaking.

It was shown that every noisy Gaussian channel (except the singular
case ${\cal B}_1(a)$), can be obtained as the composition of a pair
of quantum-limited channels as shown in Table \ref{tab1}. This in turn implies
that apart from those compositions that involve ${\cal A}_2$, there is
a discrete set of Kraus operators for all the noisy Gaussian
channels. Further, the nonnegative operators
$\{W_{\ell}^{\dagger} W_{\ell}\}$  were found to be simultaneously diagonal
for all Gaussian channels. This throws light on the error
correctability of these channels.

In bringing out the semigroup structure of the amplifier and the
attenuator families of quantum-limited channels, it was shown that
interrupted evolution slows down both amplification and attenuation in
a manner characteristic of the quantum Zeno effect.\\

\noindent
{\bf Acknowledgements}\,: This work originated from an inspiring seminar on
Gaussian channels which Raul Garcia-Patron Sanchez presented at the
Institute of Mathematical Sciences, and the authors would like to
thank him for stimulating their interest in the problems studied in
this paper. They would like to thank K. R. Parthasarathy and
V. S. Sunder for insightful 
comments, from mathematicians' point of view, on the subtleties
involved in the extension of Choi's theorem on extremality of unital
[or trace-preserving] CP maps
to the infinite dimensional case.\\

\noindent
{\bf Appendix \,: Composition of a General pair of Quantum
  Limited Channels} \\
Given two quantum-limited Gaussian channels whose $(X,Y)$ matrices in
the sense of Section 1 are respectively $(X_1,Y_1)$,  $(X_2,Y_2)$,
with $|\rm{det}\,X_j| = \kappa_j^2$, we have for the composite channel
$(X,Y)$ 
\begin{equation*}
X = X_1 X_2, ~~~ Y = X_2^T Y_1 X_2 + Y_2.
\end{equation*}
If $(X_1^0,Y_1^0)$, $(X_2^0,Y_2^0)$ are the canonical forms of
$(X_1,Y_1)$, $(X_2,Y_2)$ in the sense of \eqref{int3}, the most
general $(X_1,Y_1)$, $(X_2,Y_2)$ should necessarily have the form
$(S_1 X_1^0 S_2, \, S_2^T Y_1^0 S_2)$, $(S_3 X_2^0 S_4, \, S_4^T Y_2^0
S_4)$, with $S_1, S_2, S_3, S_4 \, \in Sp(2,R)$, so that 
\begin{align*}   
X &= S_1 X_1^0 S_2 S_3 X_2^0 S_4, \nonumber \\
Y &= S_4^T X_2^0 S_3^T S_2^T Y_1^0 S_2 S_3 X_2^0 S_4 + S_4^T Y_2^0 S_4.
\end{align*}
Our problem now is to classify the orbits or cosets under the unitary
equivalence $(X,Y) \sim (\tilde{S_1} X \tilde{S_2},\, \tilde{S_2}^T Y
\tilde{S_2}) $, $\tilde{S_1}, \tilde{S_2} \in Sp(2,R)$. Basically we
have to determine the determinants of $X, Y$ in terms of the canonical
parameters $\kappa_1, \kappa_2$ of the constituent channels. While
det$\,X$ is independent of $S_1, S_2, S_3, S_4$, det$\,Y$ needs a
careful consideration. It is this situation that $Y \geq 0$ is the sum
of two (positive) terms that breaks our analysis into two distinct
cases. 

Let us first consider the case in which $X_2^0$ is nonsingular, so
that both the terms of $Y$ are nonsingular\,: this case obviously
corresponds to the first twelve entries of Table \ref{tab2}. With the choice
$\tilde{S_2} = S_4^{-1}$ the second term of $Y$ becomes a multiple of
the identity, and the first term of $Y$ becomes a multiple of $S_3^T
S_2^T S_2S_3 \in Sp(2,R) $, a positive symplectic matrix. We can now do a
rotation $\tilde{S_2} = R \in SO(2) \subset Sp(2,R)$ without affecting
the second term in $Y$, so that the first term of $Y$ becomes a
multiple of diag$(\lambda, \lambda^{-1})$, $\lambda, \lambda^{-1}$
being eigenvalues of $S_3^T S_2^T S_2S_3$. 
\begin{table}
\begin{center}
\begin{tabular}{|c|c|}
\hline
${\cal C}_1(\kappa_2;0) \circ {\cal C}_1(\kappa_1;0)$ &  ${\cal
  C}_1\left(\kappa_2\kappa_1;
\sqrt{(1-\kappa_2^2\kappa_1^2)^2+\kappa_2^2(1-\kappa_1^2)(1-\kappa_2^2)(\lambda
  - \lambda^{-1})^2} - (1-\kappa_2^2\kappa_1^2)\right)$ \\
\hline
${\cal C}_{2}(\kappa_2;0) \circ {\cal C}_{2}(\kappa_1;0)$ &  ${\cal
  C}_{2}\left(\kappa_2 \kappa_1;\sqrt{(\kappa_1^2 \kappa_2^2 -1)^2
    + \kappa_2^2(\kappa_1^2 -1)(\kappa_2^2 -1)(\lambda -
    \lambda^{-1})^2}- (\kappa^2_{2}\kappa_1^2 -1) \right)$ \\
\hline
${\cal C}_{2}(\kappa_2;0) \circ {\cal C}_{1}(\kappa_1;0)$ &  ${\cal
  C}_{1}\left(\kappa_2 \kappa_1;\sqrt{(2\kappa_2^2 -\kappa_1^2\kappa_2^2 -1)^2
    + \kappa_2^2(1-\kappa_1^2)(\kappa_2^2 -1)(\lambda -
    \lambda^{-1})^2}- (1-\kappa^2_{2}\kappa_1^2) \right)$, \\
&~~ for $\kappa_2\kappa_1 \leq 1$; \\
&  ${\cal C}_{2}\left(\kappa_2 \kappa_1;\sqrt{(2\kappa_2^2 -\kappa_1^2\kappa_2^2 -1)^2
    + \kappa_2^2(1-\kappa_1^2)(\kappa_2^2 -1)(\lambda -
    \lambda^{-1})^2}- (\kappa^2_{2}\kappa_1^2 - 1) \right)$, \\
 &~~ for  $\kappa_2\kappa_1 \geq 1$ \\
\hline
${\cal C}_{1}(\kappa_2;0) \circ {\cal C}_{2}(\kappa_1;0)$ & ${\cal
  C}_{1}\left(\kappa_2 \kappa_1;\sqrt{(\kappa_2^2\kappa_1^2 -2\kappa_2^2 +1)^2
    + \kappa_2^2(1-\kappa_2^2)(\kappa_1^2 -1)(\lambda -
    \lambda^{-1})^2}- (1-\kappa^2_{2}\kappa_1^2) \right)$, \\
&~~ for $\kappa_2\kappa_1 \leq 1$; \\
& ${\cal C}_{2}\left(\kappa_2 \kappa_1;\sqrt{(\kappa_2^2\kappa_1^1 -2\kappa_2^2+1)^2
    + \kappa_2^2(1-\kappa_2^2)(\kappa_1^2 -1)(\lambda -
    \lambda^{-1})^2}- (\kappa^2_{2}\kappa_1^2 - 1) \right)$, \\
 &~~ for  $\kappa_2\kappa_1 \geq 1$. \\
\hline
${\cal D}(\kappa_2;0) \circ {\cal D}(\kappa_1;0)$ & ${\cal
  C}_{1}\left(\kappa_2 \kappa_1;\sqrt{(\kappa_2^2\kappa_1^2 +2 \kappa_2^2 +1)^2
    + \kappa_2^2(1+\kappa_1^2)(1+\kappa_2^2)(\lambda -
    \lambda^{-1})^2}- (1-\kappa^2_{2}\kappa_1^2) \right)$, \\
&~~ for $\kappa_2\kappa_1 \leq 1$; \\
& ${\cal C}_{2}\left(\kappa_2 \kappa_1;\sqrt{(\kappa_2^2\kappa_1^2 +2\kappa_2^2 +1)^2
    + \kappa_2^2(1+\kappa_1^2)(1+\kappa_2^2)(\lambda -
    \lambda^{-1})^2}- (\kappa^2_{2}\kappa_1^2 - 1) \right)$, \\
 &~~ for  $\kappa_2\kappa_1 \geq 1$. \\
\hline
${\cal D}(\kappa_2;0) \circ {\cal C}_1(\kappa_1;0)$ & ${\cal
  D}\left(\kappa_2 \kappa_1;\sqrt{(2 \kappa_2^2-\kappa_2^2\kappa_1^2  +1)^2
    + \kappa_2^2(1-\kappa_1^2)(1+\kappa_2^2)(\lambda -
    \lambda^{-1})^2}- (1+\kappa^2_{2}\kappa_1^2) \right)$ \\
\hline
${\cal D}(\kappa_2;0) \circ {\cal C}_2(\kappa_1;0)$ & ${\cal
  D}\left(\kappa_2 \kappa_1;\sqrt{(1+\kappa_2^2\kappa_1^2)^2
    + \kappa_2^2(\kappa_1^2-1)(1+\kappa_2^2)(\lambda -
    \lambda^{-1})^2}- (1+\kappa^2_{2}\kappa_1^2) \right)$ \\
\hline
${\cal C}_2(\kappa_2;0) \circ {\cal D}(\kappa_1;0)$ & ${\cal
  D}\left(\kappa_2 \kappa_1;\sqrt{(2 \kappa_2^2+\kappa_2^2\kappa_1^2  -1)^2
    + \kappa_2^2(1+\kappa_1^2)(\kappa_2^2-1)(\lambda -
    \lambda^{-1})^2}- (1+\kappa^2_{2}\kappa_1^2) \right)$ \\
\hline
${\cal C}_1(\kappa_2;0) \circ {\cal D}(\kappa_1;0)$ & ${\cal
  D}\left(\kappa_2 \kappa_1;\sqrt{(1+\kappa_2^2\kappa_1^2 )^2
    + \kappa_2^2(\kappa_1^2+1)(1-\kappa_2^2)(\lambda -
    \lambda^{-1})^2}- (1+\kappa^2_{2}\kappa_1^2) \right)$ \\
\hline
 ${\cal C}_1(\kappa_2) \circ  {\cal A}_2(0)  $ & ${\cal A}_2
\left(\sqrt{1+\kappa_2^2(1-\kappa_2^2)(\lambda-\lambda^{-1})^2} -1
\right)$  \\
\hline
${\cal C}_2(\kappa_2) \circ  {\cal A}_2(0)  $ & ${\cal A}_2
\left(\sqrt{(2\kappa_2^2-1)^2+\kappa_2^2(\kappa_2^2-1)(\lambda-\lambda^{-1})^2} -1
\right)$ \\
\hline 
${\cal D}(\kappa_2) \circ  {\cal A}_2(0)  $ & ${\cal A}_2
\left(\sqrt{(2\kappa_2^2+1)^2 + \kappa_2^2(1+\kappa_2^2)(\lambda-\lambda^{-1})^2} -1
\right)$  \\
\hline
${\cal A}_2(0) \circ {\cal C}_1(\kappa_1)$ & ${\cal A}_2
(\sqrt{1+(\lambda \cos^2{\theta}+ \lambda^{-1}\sin^2{\theta}
    )(1-\kappa_1^2)} -1  )$ \\
\hline
${\cal A}_2(0) \circ {\cal C}_2(\kappa_1)$ & ${\cal A}_2
(\sqrt{1+(\lambda \cos^2{\theta}+ \lambda^{-1}\sin^2{\theta}
    )(\kappa_1^2-1)} -1  )$  \\
\hline 
${\cal A}_2(0) \circ {\cal D}(\kappa_1)$ & ${\cal A}_2
(\sqrt{1+(\lambda \cos^2{\theta}+ \lambda^{-1}\sin^2{\theta}
    )(1+\kappa_1^2)} -1  )$ \\
\hline
${\cal A}_2(0) \circ  {\cal A}_2(0)  $ & ${\cal A}_2\left( \sqrt{1+\lambda \cos^2{\theta}+ \lambda^{-1}
    \sin^2{\theta}  } -1 \right)$, or 
\\&${\cal C}_1 \left(0; \sqrt{1+\lambda \cos^2{\theta}+ \lambda^{-1} \sin^2{\theta}  } -1
\right)$. \\
\hline 
\end{tabular}
\end{center}
\caption{Showing composition of two quantum-limited Gaussian channels
  which are not necessarily in their respective canonical forms. It
  may be seen that the case $\lambda =1$ coresponds to Table
  \ref{tab1}. In the case of the first twelve entries the noise is
  always greater than what obtains in Table \ref{tab1}. But in the
 remaining four cases the noise can be either more or less, depending
 on the value of $\theta$. \label{tab2}}
\end{table}
Det$\,Y$ can now be easily
evaluated. Removal of the mandatory quantum-limited noise, as dictated by the
value of det$\,X$ [i.e., $|1-\kappa^2|$, $1+ \kappa^2$, or 1 depending
on det$\,X$ being positive, negative or zero, $\kappa^2$ equalling det$|X|$
], then yields the classical noise indicated in Table \ref{tab2} by the
second argument of the composite channel. This procedure is the one
used for the first twelve entries of Table \ref{tab2}. 

The case of singular $X_2^0$ is somewhat different. With the removal
of $S_4$ with the choice $\tilde{S_2} = S_4^{-1}$, as in the earlier
case, in the first term of $Y$ the projection $X_2^0 = \Bigg( \begin{matrix}
  1&0 \\ 0&0 \end{matrix}\Bigg)$ picks out the 1,1 element of the
matrix $S_3^T S_2^T S_2S_3$. With the eigenvalues of this matrix again
denoted $(\lambda, \lambda^{-1})$ and assuming that $\theta$ is the
rotation needed to diagonalize it, the 1,1 element equals $\lambda
\cos^2{\theta} + \lambda^{-1} \sin^2{\theta}$ (a combination appearing
in the last four entries of Table \ref{tab2}). 
\noindent 
With this, the second term in $Y$ is a multiple of identity whereas
the first term is a multiple of $\lambda
\cos^2{\theta} + \lambda^{-1} \sin^2{\theta}$ times the projector $\left( \begin{matrix}
  1&0 \\ 0&0 \end{matrix}\right)$; the determinant of $Y$ can be
readily computed, and removal of the quantum-limited noise gives the
numerical value of the second argument of the composite channel. 

Finally, the two cases in the last entry of Table \ref{tab2}, the case of the
composite ${\cal A}_2(0) \circ {\cal A}_2(0)$, correspond to the two
possible situations that may arise with $X$ when both $X_1^0$ and
$X_2^0$ are of rank one\,: $X$ could either be rank one or it could be a
null matrix. Correspondingly, the composite  ${\cal A}_2(0) \circ
{\cal A}_2(0)$ should be viewed as a member of the ${\cal C}_1(0;a) =
{\cal A}_1(a)$ family or the ${\cal A}_2(a)$ family, the extra
classical noise being `the same' in both cases. 

We conclude with the following observations in respect of Table \ref{tab2}. It
is evident that in all cases $\lambda =1$ corresponds to the situation
in which both constituents of the composite are already in their
canonical forms, and the reader can verify that Table \ref{tab2}
reduces to Table \ref{tab1} in this case. And thus we see that $\lambda \neq \lambda^{-1}$ results,
in the first sixteen entries of Table \ref{tab2}, in classical noise which is
{\em always more in magnitude} than the case $\lambda =1$. The last
four entries of Table \ref{tab2} are distinguished in this regard\,: since the
classical noise depends also on the choice of $\theta$, it can be {\em
  either more or less} than what obtains in the $\lambda =1$ case.




\begin{thebibliography}{}


\bibitem{caves94} C. M. Caves, P. D. Drummond, Rev. Mod. Phys. {\bf 66},
  481 (1994). 

\bibitem{adesso07}  S. L. Braunstein, and P. van Loock,
  Rev. Mod. Phys. {\bf 77}, 513--577 (2005); G. Adesso and F. Illuminati, J. Phys. A:
  Math. Theor. {\bf 40}, 7821 (2007); X.-B. Wang, T. Hiroshima,
  A. Tomita, M. Hayashi, Phys. Rep. {\bf 448}, 1 (2007); K.
  Hammerer   A. S. S\o rensen and E. S. Polzik, Rev. Mod. Phys. {\bf 82}, 1041â1093 (2010) .



\bibitem{rev2} J. DiGuglielmo, B. Hage, A. Franzen, J. Fiur\'{a}
  \v{s}ek, and R. Schnabel, \pra {\bf 76}, 012323 (2007); J. Laurat,
  G. Keller, J. A. O.-Huguenin, C. Fabre, T. Coudreau, A. Serafini,
  G. Adesso, and F. Illuminati, J. Opt. B: Quantum
  semiclass. Opt. {\bf 7}, S577 (2005); V. D'Auria, S. Fornaro,
  A. Porzio, S. Solimeno, S. Olivares, and M. G. A. Paris, \prl {\bf
    102}, 020502 (2009); D. Buono, G. Nocerino, V. D'Auria, A. Porzio,
  S. Olivares, and M. G. A. Paris, J. Opt. Soc. Am. B, vol.27, No.6
  (2010).
 
\bibitem{polzik} C. Schori, J. L. S\"{o}rensen, and E. S. Polzik,
Phys. Rev. A {\bf 66}, 033802 (2002); J. Eisert, M. B. Plenio,
S. Bose, and J. Hartley, Phys. Rev. Lett. {\bf 93}, 190402 (2004). 


  



\bibitem{tele}  S. L. Braunstein, and H. J. Kimble, \prl {\bf 80},
  869 (1998); S. L. Braunstein, and H. J. Kimble, \pra {\bf 61},
  042302 (2000). 

\bibitem{crypto} F. Grosshans and P. Grangier,  Phys. Rev. Lett. {\bf 88},
057902 (2002); Ch. Silberhorn, T. C. Ralph, N. L\"{u}tkenhaus, and
G. Leuchs, Phys. Rev. Lett. {\bf 89}, 167901 (2002); M. Navascu\'{e}s,
J. Bae, J. I. Cirac, M. Lewenstein, A. Sanpera, 
and A. Ac\'{i}n, Phys. Rev. Lett. {\bf 94}, 010502 (2005). 


\bibitem{hall94} M. J. W. Hall, \pra {\bf 50}, 3295 (1994). 


\bibitem{preskill01} J. Harrington, and J. Preskill, 
\pra  {\bf 64}, 062301 (2001).

\bibitem{hirota99} A. S. Holevo, M. Sohma and O. Hirota, \pra {\bf 59}, 1820 (1999).

\bibitem{giovannetti042}  V. Giovannetti, S. Guha, S. Lloyd, L. Maccone, J. H. Shapiro, and 
H. P. Yuen, \prl {\bf 92}, 027902 (2004).

\bibitem{holevo01} A. S. Holevo and R. F. Werner, \pra {\bf 63}, 032312 (2001).  

\bibitem{mmwolf07} M. M. Wolf, D. P. -Garc\'{i}a, and G. Giedke, \prl {\bf 98}, 130501 (2007).







\bibitem{wolf05} J. Eisert and M.M. Wolf, Quantum Information with 
Continous Variables of Atoms and Light, pages 23-42 (Imperial College Press, London, 2007




\bibitem{caruso061} F. Caruso and V. Giovannetti, \pra {\bf 74}, 062307 (2006).

\bibitem{giovannetti041} V. Giovannetti, S. Guha, S. Lloyd, L. Maccone and J. H. Shapiro,
\pra {\bf 70}, 032315 (2004).


\bibitem{giovannetti043} V. Giovannetti, S. Lloyd, L. Maccone, J. H. Shapiro, and B. J. Yen, 
\pra {\bf 70}, 022328 (2004).


\bibitem{caruso06} F. Caruso, V. Giovannetti and A. S. Holevo, New. J. Phys. 
{\bf 8}, 310 (2006). 


\bibitem{holevo07} A. S. Holevo, Probl. Inf. Trans. {\bf 43}, 1-11 (2007) 
(Preprint quant-ph/0607051).

\bibitem{mmwolf08} M. M. Wolf, \prl {\bf 100}, 070505 (2008).

\bibitem{caruso08} F. Caruso, J. Eisert, V. Giovannetti and A. S. Holevo, New. J. Phys. 
{\bf 10}, 083030 (2008).

\bibitem{caruso10} F. Caruso, J. Eisert, V. Giovannetti, and
  A. S. Holevo, arXiv:1009.1108.








\bibitem{simon87} R. Simon, E. C. G. Sudarshan, and N. Mukunda, 
 Phys. Rev. A {\bf 36}, 3868 (1987).

\bibitem{simon88}  R. Simon, E. C. G. Sudarshan, and N. Mukunda, \pra
  {\bf 37}, 3028 (1988).   

\bibitem{dutta94} R. Simon, N. Mukunda, and B. Dutta, 
Phys. Rev. A {\bf 49}, 1567  (1994).

\bibitem{krp} K. R. Parthasarathy, Commun. Stoc. Anal. vol.4, No.2,
  143 (2010).



\bibitem{demoen79} B. Demoen, P. Vanheuverzwijn, A. Verbeure,
  Rep. Math. Phys. vol.{\bf 15}, pp. 27-39 (1979).

 
\bibitem{lindblad00} G. Lindblad, J. Phys. A {\bf 33}, 5059 (2000).

\bibitem{hellwig} K. Hellwig and K. Kraus, Comm. Math. Phys. {\bf 16}, 142 (1970).

\bibitem{matthews61} E. C. G. Sudarshan, P. M. Matthews and J. Rau, 
Phys. Rev. {\bf 121}, 920 (1961).

\bibitem{choimap75} M. D. Choi, Lin. Alg. Appl. {\bf 10}, 285 (1975).

\bibitem{stinespring} W. F. Stinespring, Proc. Amer. Math. Soc. {\bf 6},
211-216 (1955).

\bibitem{peres96} A. Peres, Phys. Rev. Lett. {\bf 77}, 1413 (1996).


\bibitem{horodecki96} M. Horodecki, P. Horodecki, and R. Horodecki,
  Phys. Lett. A {\bf 223}, 1 (1996).    

\bibitem{simon00} R. Simon, Phys. Rev. Lett. \textbf{84}, 2726 (2000).



\bibitem{horodecki03}  M. Horodecki, P. W. Shor, M. B. Ruskai,
  Rep. Math. Phys. {\bf 15}, 629 (2003). 


\bibitem{holevo08ebt}  A. S. Holevo, Probl. Inf. Trans. {\bf 44}, 3-18 (2008) 
(Preprint quant-ph/0802:0235).

\bibitem{shirokov05} A. S. Holevo, M. E. Shirokov, and R. F. Werner, 
Russian Math. Surveys, vol. 60, N2, (2005).




\bibitem{winter} J. A. Smolin, F. Verstraete, and A. Winter, \pra
{\bf 72}, 052317 (2005); R. F. Werner, Open Problems in Quantum
Information Theory Online at: http://www.imaph.tu-bs.de/qi/problems.

\bibitem{gregoratti03} M. Gregoratti and R. F. Werner,
  J. Mod. Opt. {\bf 50}, 915-933 (2003); M. Gregoratti and
  R. F. Werner, J. Math. Phys. {\bf 45}, 2600 (2004).



\bibitem{hayden-king} P. Hayden and C. King, Quant. Inf. Comp. {\bf
  5}(2), 156 (2005).



\bibitem{agarwal09} G. S. Agarwal, S. Chaturvedi, and A. Rai,
  arXiv: 0912.5134. 

\bibitem{carmichael10} H. Nha, G. J. Milburn, and H. J. Carmichael,  
  New J. Phys. \textbf{12}, 103010 (2010).

\bibitem{allegra10} M. Allegra, P. Giorda, 
and M. G. A. Paris, \prl  {\bf 105},100503 (2010).


\bibitem{leonhardt94} U. Leonhardt, \pra {\bf 49}, 1231 (1994).

\bibitem{cahill692} K. E. Cahill and R. J. Glauber, Phys. Rev. {\bf
  177}, 1857 (1969); K. E. Cahill and R. J. Glauber, Phys. Rev. {\bf
    177}, 1882 (1969).




\bibitem{gradstein1} I. S. Gradshteyn and I. M. Ryzhik, 
{\em Table of Integrals, Series, and Products}, 
Academic Press (2001), p.802, {\bf 7.414} 3, 4.





\bibitem{sudarshan63} E. C. G. Sudarshan, \prl {\bf 10}, 277 (1963);  
R. J. Glauber, Phys. Rev. {\bf 131}, 2766 (1963).


\bibitem{schack90} R. Schack and A. Schenzle, \pra {\bf 41}, 3847 (1990).

\bibitem{solong08} J. Solomon Ivan, M. Sanjay Kumar, and R. Simon,
  quant-ph/0812.2800.



\bibitem{browne03} D. E. Browne, J. Eisert, S, Scheel, and M. B. Plenio,
\pra {\bf 67}, 062320 (2003).


\bibitem{mary97}  R. Simon, M. Selvadoray, Arvind, N. Mukunda,
  quant-ph/9709030.





\bibitem{landau93} L .J. Landau, and R. F. Streater, Lin. Alg. Appl.
{\bf 193}, 107-127 (1993).



\bibitem{saleh89} R. A. Campos, B. E. A. Saleh, M. C. Teich, \pra {\bf
40}, 1371 (1989); M. S. Kim, W. Son, V. Bu\v{z}ek, and P. L. Knight,
  \pra {\bf 65}, 032323 (2002); B. Yurke, S. L. McCall, and
  J. R. Klauder, \pra {\bf 33}, 4033 (1986).



\bibitem{nair} H. P. Yuen and R. Nair, Phys. Rev. A {\bf 80}, 023816
  (2009). 

\bibitem{zeno} B. Mishra and E. C. G. Sudarshan, J. Math. Phys. {\bf
    18}, 756 (1977); P. Facchi, H. Nakazato, and S. Pascazio, \prl
  {\bf 86}, 2699 (2001). 

\bibitem{imoto04} Y. -X. Liu, \c{S}. K. \"{O}zdemir, A. Miranowicz, and
  N. Imoto, Phys. Rev. A {\bf 70}, 042308 (2004); I. L. Chuang, D. W. Leung, and
Y. Yamamoto, \pra {\bf 56}, 1114 (1997). 




\bibitem{lutkenhaus95} N. L\"{u}tkenhaus and S. M. Barnett, Phys. Rev. A
  {\bf 51}, 3340 (1995). 


\bibitem{gradstein2} The integral {\bf 7.374} 6 on on p.796 of
  Ref.\cite{gradstein1}. 




\bibitem{giovannetti10} V. Giovannetti, A. S. Holevo, S. Lloyd,
  and L. Maccone, J. Phys. A:Math. Theor. {\bf 43}, 415305 (2010).



\bibitem{kraus}K. Kraus, States, Effects, and Operations: Fundamental 
Notions in Quantum Theory, Lecture Notes in Physics vol. 190,
Springer-Verlag, Berlin (1983).


\bibitem{gradstein3} The integral {\bf 7.377} on p.797 of Ref.\cite{gradstein1}.

 
\bibitem{tsui96} S. -K. Tsui, Proc. Am. Math. Soc. {\bf 124}, 437
  (1996).


\bibitem{bhatia} R. Bhatia, {\em Matrix Analysis}, Springer
  (2004).



\bibitem{tregub} S. L. Tregub, Sov. Math. {\bf 30}, 105 (1986);
B. Kummerer and H. Maassen, Commun. Math. Phys. {\bf 109}, 1 (1987). 

















\end{thebibliography}
\end{document}